\documentclass[journal,12pt,onecolumn,draftclsnofoot]{IEEEtran}
\usepackage{graphicx}
\usepackage{epsf}
\usepackage{mathtools}
\usepackage{amssymb}
\usepackage{amsmath, amsfonts}
\usepackage{latexsym}
\usepackage{multirow}
\usepackage{multicol}
\usepackage[table]{xcolor}
\usepackage{color}

\usepackage{tabularx}
\usepackage{lipsum}

\usepackage{algorithm}
\usepackage{algorithmic}

\usepackage{mathrsfs}

\usepackage{fixltx2e}

\usepackage[mathscr]{euscript}

\usepackage{commath}
\usepackage{MnSymbol}

\usepackage{capt-of}
\usepackage{caption}
\usepackage{subcaption}

\usepackage{mathtools} 
\DeclarePairedDelimiterX\setc[2]{[}{]}{\,#1 \;\delimsize\vert\; #2\,}
\DeclarePairedDelimiterX\parth[2]{(}{)}{\,#1 \;\delimsize\vert\; #2\,}

\DeclarePairedDelimiter{\ceil}{\lceil}{\rceil}
\DeclarePairedDelimiter{\floor}{\lfloor}{\rfloor}

\PassOptionsToPackage{hyphens}{url}
\usepackage[unicode=true, bookmarks=true, bookmarksnumbered=true, bookmarksopen=true, bookmarksopenlevel=1, breaklinks=false, pdfborder={0 0 0}, pdfborderstyle={}, backref=false, colorlinks=false]{hyperref}

\usepackage{epstopdf}

\usepackage[utf8]{inputenc}
\usepackage[english]{babel}

\usepackage{theorem}
{
\theorembodyfont{\rmfamily}
\newtheorem{assumption}{Assumption}
\newtheorem{remark}{Remark}
\newtheorem{lemma}{Lemma}
\newtheorem{proposition}{Proposition}
\newtheorem{definition}{Definition}
}

\definecolor{orange}{RGB}{255,127,0}
\definecolor{blue}{RGB}{0,0,255}
\definecolor{red}{RGB}{255,0,0}
\definecolor{green}{RGB}{50,160,50}
\definecolor{grey}{RGB}{125,120,125}
\definecolor{yellow}{RGB}{210,210,0}

\begin{document}
{
\title{{\fontsize{20}{2}\selectfont Spatiotemporal Analysis on Broadcast Performance of\\\vspace{-0.1 in}DSRC with External Interference in 5.9 GHz Band}}

\author
{
Seungmo Kim, \textit{Member}, \textit{IEEE}, and Mehdi Bennis, \textit{Senior Member}, \textit{IEEE}

\thanks{S. Kim is with Department of Electrical and Computer Engineering, Georgia Southern University in Statesboro, GA, USA (e-mail: seungmokim@georgiasouthern.edu). M. Bennis is with Centre for Wireless Communications, University of Oulu in Finland (e-mail: mehdi.bennis@oulu.fi).

An initial version of analysis presented in this paper was presented in IEEE Globecom 2018 \cite{globecom18}.}
}

\maketitle

\vspace{-0.6 in}
\begin{abstract}
Coexistence between the dedicated short-range communications (DSRC) and other wireless technologies needs thorough study since the United States legislative bodies still remain undecided on the shared use of the 5.9 GHz band (5.850-5.925 GHz). If the band is decided to be shared among multiple technologies, the DSRC is expected to experience a performance degradation even in safety-critical application. As such, it is a natural question how much the performance degradation will be. However, it is not trivial to precisely model the behaviors of a vehicular-to-everything (V2X) network since it requires to formulate both spatial and temporal aspects in concert while the network topology keeps dynamic due to the nodes' mobility. Moreover, DSRC relies on broadcast of basic safety messages (BSMs) for supporting safety-critical applications. Most prior work uses performance metrics such as packet delivery rate (PDR) and packet delay, which cannot accurately capture the performance of DSRC broadcasts. To this end, this paper (i) provides a comprehensive analysis framework formulating both spatial and temporal factors in concert and (ii) proposes a new performance metric that can more accurately characterize the broadcast performance of a DSRC network. Based on the new metric, the results present (i) the fundamental performance of a DSRC network under inter-RAT interference from Wi-Fi and C-V2X and (ii) provide suggestions on optimal selection of networking parameters.
\end{abstract}


\IEEEpeerreviewmaketitle

\section{Introduction}\label{sec_introduction}
In 1999, the Federal Communications Commission (FCC) allocated the 5.9 GHz band (5.850-5.925 GHz) for intelligent transportation system (ITS) applications based on dedicated short-range communications (DSRC) and adopted basic technical rules for the DSRC operations \cite{fcc_dsrc}. The DSRC is now at the stake of sharing the 5.9 GHz band with the following two other radio technologies (RATs).

The first RAT is \textit{Wi-Fi}. As suggested by the Congress in September 2015, the FCC, in its latest public notice \cite{fcc1668a1}, now seeks to refresh the record of its pending 5.9 GHz rulemaking to provide potential sharing solutions between proposed Unlicensed National Information Infrastructure (U-NII) devices and DSRC operations in the 5.9 GHz band. The focus of the FCC's solicitation \cite{fcc1668a1} is two-fold: (i) prototype of interference-avoiding devices for testing; (ii) test plans to evaluate electromagnetic compatibility of unlicensed devices and DSRC.

More recently, the \textit{cellular V2X (C-V2X)} is seeking to operate in the 5.9 GHz band as well \cite{5gaa}. At present, only DSRC is permitted to operate in the ITS band in United States (US), while the 5G Automotive Association (5GAA) has requested a waiver to the FCC to allow C-V2X operations in the band \cite{5gaa}. The key problem here is that C-V2X and DSRC are not compatible with each other. It means that if some vehicles use DSRC and others use C-V2X, these vehicles will be unable to communicate with each other, which leads to a scenario where the true potential of V2X communications cannot be attained.

Nevertheless, DSRC still holds significance as the key technology enabling safety-critical applications. Europe recently mandated DSRC as the sole technology operating in the 5.9 GHz band \cite{europe}. Also, in the US, 50 state transport departments request reserving the 5.9 GHz band for transport safety \cite{dot}. However, state departments of transportation (DOTs) are experiencing confusion and inefficiency in enhancement/expansion of connected vehicle technologies, due to the FCC's indecisiveness on ``shared use'' of the 5.9 GHz band with other wireless systems--\textit{i.e.}, Wi-Fi and cellular vehicle-to-everything communications (C-V2X) \cite{dot}. Also, at the US DOT, Phase I of a three-phase testing plan has been completed, which investigated the DSRC’s interoperability with the other wireless systems \cite{fcc_phase1}. However, Phase II has not yet even started, which is supposed to involve basic field tests to assess the efficacy of findings from Phase I \cite{aashto}. Until these real-world tests are completed, one will not know conclusively to what extent, or whether at all, DSRC and the other technologies can operate together without interference, which has significant influence on the vehicle safety and mobility.

To this end, this paper provides an extensive spatiotemporal analysis on DSRC's broadcast of basic safety messages (BSMs). The key application of this study will be the feasibility of DSRC in 5.9 GHz band in all the possible interoperability scenarios that the FCC is considering.

\section{Related Work}\label{sec_related}
\subsection{Performance of DSRC}\label{sec_related_dsrc}
\subsubsection{Temporal Analysis on BSM Broadcast in DSRC}
An analysis was provided on the performance of DSRC's broadcast of BSMs \cite{elsevier14}. However, it presented a limited generality due to a few critical shortcomings: (i) no randomness on the position of vehicles--\textit{i.e.}, no statistics on the spatial perspectives; (ii) an over-simplified spatial model--\textit{i.e.}, one-dimensional, linear spatial analysis; and (iii) only numerical solutions without closed-form expressions.

There was an effort to modify the IEEE 802.11 MAC protocol for higher packet reliability in DSRC through ``retransmissions'' \cite{vnc10}. However, we need to thoroughly consider the abundance of BSMs relative to a vehicle's movement. For instance, a vehicle can only move 26.8224 meters per second (mps) even at the speed of 60 miles per hour (mph). At the broadcast rate of 10 Hz, which is the nominal consensus in the field, each of the 10 BSMs during the last second can cover 2.6822 m. Realistically, in an environment where a vehicle can move at 60 mph, it is not likely that a drastic change occurs within such a short distance as 2.6822 m. So, we doubt that adding a retransmission capability to DSRC is practical in the logic of cost and benefit.

A recent study presented a temporal analysis on DSRC beaconing \cite{eurasip19}. But the stochastic geometry was not considered, which may degrade the applicability of its findings. Further, the temporal analysis lacked details: \textit{i.e.}, the exact timing ``within a beacon period'' and explicit formulation of the `probability of a packet expiration.' It is of particular significance because a packet expiration occurs under a set of particular parameters--\textit{e.g.}, an inter-broadcast interval shorter than a value of backoff time.

\subsubsection{Markov Process for Modeling BSM Broadcast in DSRC}
Markov process has been considered as a useful stochastic tool to model the behavior of an IEEE 802.11-based system \cite{bianchi}. 

DSRC's broadcast of BSM within a beaconing period was also attempted to be modeled based on Markov process in recent literature. A semi-Markov process model was proposed to characterize the medium contention and back off behavior for a tagged vehicle and influences from other vehicles \cite{tcom13}. However, it lacks consideration of ``packet expiration,'' which is one of the critical features in characterizing a DSRC Tx. In DSRC, a node is allowed decrement its backoff counter only when the medium is found idle, which is in contrast to an IEEE 802.11 distributed coordinated function (DCF) where a backoff counter is decremented with the probability of 1 \cite{bianchi}. As such, the chance that a BSM is transmitted within a beaconing period is very significantly determined by the number of busy slots that the BSM experiences. Leaving it not discussed limits the applicability of an analysis.

The impact of a packet expiration has been considered in another study \cite{haenggi16}, wherein a spatiotemporal analysis is provided based on the Markov process and Poisson point process (PPP). However, the system model is too simplistic to be assured to be applied in general scenarios: a one-dimensional system model and no explicit mathematical expressions on BSM expiration and collisions, which degrades the applicability of the results.

\begin{figure}[t]
\vspace{-0.3 in}
\centering
\includegraphics[width = 0.8\linewidth]{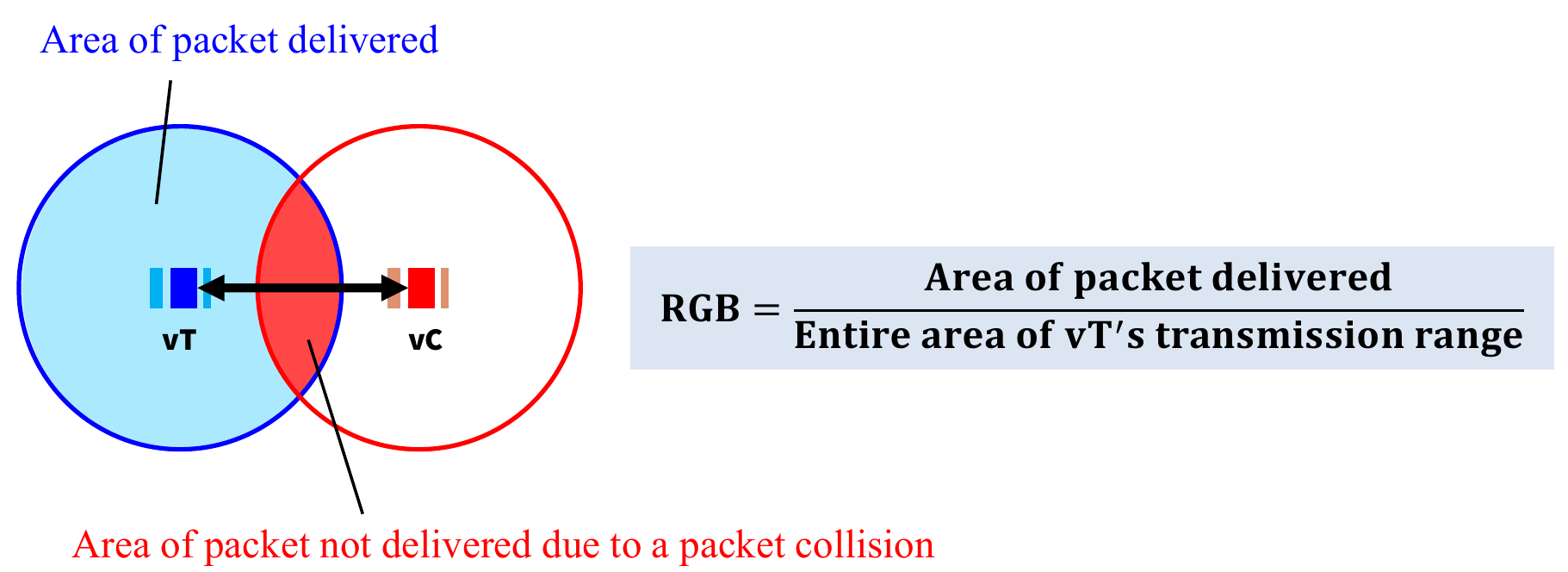}
\caption{Geometric analysis of a broadcast-based vehicular network \cite{globecom18}}
\label{fig_rgb_definition}
\vspace{-0.3 in}
\end{figure}

\subsubsection{Metrics in Measurement of BSM Broadcast Performance}
Currently used metrics are not always able to accurately measure the broadcast performance of a DSRC network \cite{globecom18}. The main objective of DSRC's broadcast of basic safety messages (BSMs) is to support safety-critical applications. Prior relevant work such as \cite{infocom_6}-\cite{infocom_10} relied on typical metrics--\textit{i.e.}, packet delivery rate (PDR) and packet delay/latency, which capture only the ``temporal'' aspect of BSM broadcast in a DSRC network.

An advanced metric that addresses this problem is proposed in \cite{irt_vanet06}\cite{irt_elsevier16}, namely inter-reception time (IRT). Typical packet delay/latency was measured only among successfully received packets, which di not capture the impact of packet losses and collisions on the latency. Defined as the time elapsed between two successive successful reception of packets broadcast by a specific transmitter (Tx), the IRT can more accurately display the performance of a BSM-enabled safety-critical application.

Nevertheless, IRT still is another metric that displays the temporal behaviors of a DSRC network. Here is an example scenario that cannot be captured by the IRT. For instance, when a BSM is transmitted from a Tx vehicle, $\mathsf{vT}$, all the vehicles located in the transmission range of $\mathsf{vT}$ become potential receivers (Rx's). Suppose that the BSM is collided by another packet transmitted from vehicle $\mathsf{vC}$, which is located on the `right' of $\mathsf{vT}$. Then, the vehicles that are located at the `left' side of $\mathsf{vT}$ are still able to receive the BSM successfully. This spatial insight cannot be captured by any of the classical metrics: \textit{i.e.}, PDR, packet delay/latency, and IRT.

As a solution, a metric evaluating such a spatial aspect was proposed in a recent literature \cite{globecom18}--namely, the ratio of geometry for reception of broadcast ($\mathsf{RGB}$) \footnote{While the abbreviation remains the same, the full spelled name has changed for a better intuition from the original one--namely, the \textit{reception geometry for broadcast} \cite{globecom18}.}. However, it lacks in-depth analysis on the temporal behaviors of DSRC, which still leaves the necessity of a comprehensive analysis encompassing both spatial and temporal perspectives.

\subsection{Coexistence of DSRC with Other Wireless Technologies}\label{sec_related_models}
Notice that some parts of the ``spatial'' analysis have previously been presented by the author \cite{globecom18}. Building atop the previous work, the key improvements presented in this paper are identified as (i) bridging the spatial analysis to a temporal analysis and (ii) derivation of closed-form expressions expressing the intertwined impacts between the spatial and temporal aspects on the broadcast behaviors of DSRC.

There is a body of prior work discussing coexistence between DSRC system and IEEE 802.11ac \cite{infocom_12}-\cite{gaurang17}. Especially in \cite{gaurang17}, a method of allocating channels for DSRC and 802.11ac was suggested and evaluated via experiments and simulations. However, the discrepancy between their experiment results and simulation results is too significant to neglect, which degrades the credibility of the study.

A recently proposed coexistence method between C-V2X and VANET was based on an ``on and off'' mechanism \cite{bian18}. However, allocation of different chunks of time resource is not applicable to coexistence with an IEEE 802.11-based system, which is asynchronous. In other words, it is not an efficient approach to make an asynchronous system to keenly turns on or off at a certain time instant due to all different time clocks among different nodes.

A latest work focused to address the interference among DSRC-based vehicles \cite{milcom19}. It suggests to ``prioritize'' a packet transmission according to the level of danger, which is measured by the inter-vehicle distance.

General spectrum contention control methods would be worth mentioning \cite{elsevier14_30}\cite{elsevier14_31}. However, these studies suffer from a critically unrealistic assumption: \textit{i.e.} all the vehicles are in each other’s transmission range and hence no hidden nodes. The impact of hidden nodes is one of the most dominant factors determining the broadcast performance of DSRC \cite{elsevier14}\cite{globecom18}. Moreover, in \cite{elsevier14_30}, no mobility of a node is assumed, which is even further from reality. As such, these studies cannot be regarded to suggest reliable ideas.

An intelligence-based approach has been proposed lately \cite{bennis_tcom19}-\cite{bennis_tvt19}. While they provides a generalized analysis framework for V2X networking, it did not provide enough detail on the feasibility of the 5.9 GHz band for DSRC under external interference from other RATs.

\subsection{Contributions}\label{sec_related_contributions}
Motivated from the limitations of the aforementioned related work, this paper presents the following contributions:
\begin{enumerate}
\item \textit{A spatiotemporal analysis framework:} This paper provides a comprehensive analysis framework that embraces both spatial and temporal factors determining the broadcast performance of a DSRC system.
\begin{itemize}
\item The main advantage of the proposed method is that it can precisely evaluate the reciprocal impacts between the spatial and temporal factors in concert. It is certainly an improvement from the typical metrics that can measure either of the two aspects separately. The advantage is highlighted at its ability to accurately capture the broadcast nature of DSRC basic safety message dissemination: some Rx's are still able to receive a packet from a Tx even if other Rx's are not.
\item For the spatial aspects of the analysis, this paper adopts PPP, which results in the following key efficiencies:
\begin{itemize}
\item Specifically, it formulates coexistence of multiple RATs as a `superposition' of heterogeneous PPPs defined in distinct two-dimensional spaces representing each RAT.
\item Closed-form characterization of technical details in the 5.9 GHz coexistence--\textit{e.g.}, carrier-sense range as \textit{thinning} and hidden-node interference as \textit{superposition} of a PPP. The key advantage is that it is straightforward: addition of multiple RATs can be expressed as a \textit{sum of the intensities} of the PPPs.
\end{itemize}
\end{itemize}

\item \textit{Analysis on the coexistence among multiple RATs at 5.9 GHz:} In addition to a more accurate, realistic modeling of a vehicular network, we consider the impact of the secondary Wi-Fi network on the performance of the vehicular network.
\begin{itemize}
\item A generalized framework for multiple interfering RATs, which can accommodate the current discussions regarding the 5.9 GHz band.
\end{itemize}

\item \textit{Analytical and numerical results suggesting appropriate behaviors of DSRC with and without presence of external interference:} This applies the proposed metric (namely, $\mathsf{RGB}$) to characterization of the coexistence of DSRC with other RATs--\textit{i.e.}, Wi-Fi and C-V2X. The results suggest adequate selection of the contention window (CW) values to improve the performance of a BSM broadcast.
\end{enumerate}

\begin{figure}[t]
\vspace{-0.2 in}
\hspace{-0.2 in}
\minipage{0.38\textwidth}
\centering
\includegraphics[width = \linewidth]{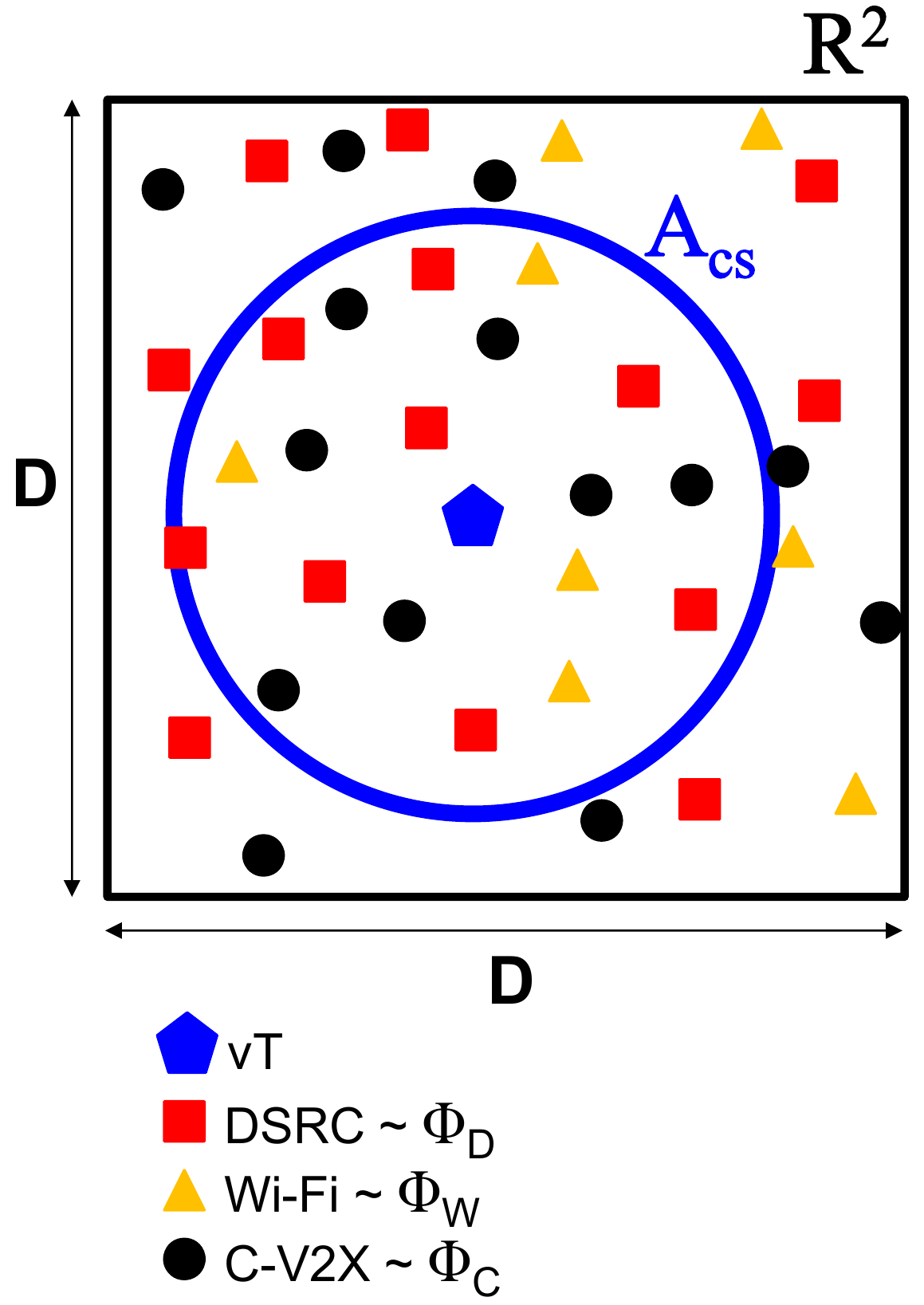}
\caption{Geometry of the system}
\label{fig_geometry}
\endminipage\hfill
\minipage{0.6\textwidth}
\centering
\captionsetup{type=table}
\scriptsize
\caption{Summary of Key Notation}
\begin{tabular}{c | l}
\hline
\textbf{Notation} & \textbf{Description}\\
\hline
EXP & Packet expiration\\
SYNC & Collision by synchronized transmission\\
HN & Collision by hidden node\\
$\lambda_{\left(\cdot\right)}$ & The intensity of a PPP\\
$\mathtt{x}_{\left( \cdot \right)}=\left(x_{\left( \cdot \right)},y_{\left( \cdot \right)}\right)$ & Position of a node \\
$\mathsf{l}\left(\mathtt{x}_{\left( m \right)}, \mathtt{x}_{\left( n \right)}\right)$ & Distance between nodes $m$ and $n$\\
$\mathsf{vR}$ & Rx vehicle\\
$\mathsf{vT}$ & Tx vehicle\\
$\mathsf{vC}$ & Vehicle transmitting a colliding BSM\\
$\left(\cdot\right)_{\boldsymbol{\mathsf{D}}}$ & A variable for DSRC\\
$N_{\left(\cdot\right)}$ & Number of packets\\
$n_{\left(\cdot\right)}$ & Number of nodes\\
$l_{bcn}$ & Length of a beacon in slots\\
$L_{bcn}$ & Length of a beaconing period in slots\\
CW & Contention window\\
$\tau$ & Probability of a BSM transmission within a slot\\
$\mathsf{P}_{b}$ & Probability of a slot found busy\\
$\mathsf{P}_{\text{start}}$ & Probability of a BSM transmission within a beaconing period\\
\hline
\end{tabular}
\label{table_notation}
\endminipage
\vspace{-0.2 in}
\end{figure}

\section{System Model}\label{sec_model}
For formulation of the DSRC broadcast performance, this paper establishes the following four key assumptions.

\subsubsection{Circular Space for Node Distribution}
A generalized ``circular'' environment instead of an example road segment, for the most generic form of analysis as done in a related literature \cite{access19}. Figure \ref{fig_geometry} illustrates the environment represented by the system space $\mathbb{R}_{\text{sys}}^2$, which is defined on a rectangular coordinate with the width and length of $D$ m. Therein, two other RATs are defined: a Wi-Fi and a C-V2X operating based on the IEEE 802.11ac and the 3rd Generation Partnership Project (3GPP) Release 14, respectfully.

\subsubsection{A Homogeneous PPP for Each RAT}
In a $\mathbb{R}_{\text{sys}}^2$, $\lambda_{\mathsf{D}}\left(>0\right)$ vehicles are distributed as a PPP, denoted by $\Phi_{\mathsf{D}}$. The position of the $i$th vehicle is denoted by $\mathtt{x}_{i}=\left(x_{i},y_{i}\right) \in \mathbb{R}_{\text{sys}}^2$. The distributions of Wi-Fi and C-V2X nodes follow a PPP as well, which are denoted by $\Phi_{\mathsf{W}}$ and $\Phi_{\mathsf{C}}$, respectively. Note also that the PPPs discussed in this paper are \textit{stationary point processes} where the density $\lambda$ remains constant according to different points.

\subsubsection{Uniformity Property for Each PPP}
It is important to note that based on the modeling with PPP, the uniformity property of a homogeneous point process can be held \cite{daley}. That is, if a homogeneous point process is defined on a real linear space, then it has the characteristic that the positions of these occurrences on the real line are uniformly distributed. Therefore, we can assume that the DSRC vehicles are uniformly randomly scattered on the road with different values of intensities and CW values, which will be provided in Section \ref{sec_results}.

\subsubsection{Four Types of Packet Transmission Result}
There are four possible results of a packet transmission including successful delivery (SUC) and two types of collision: synchronized transmission (SYNC) and hidden-node collision (HN) \cite{vnc10}:
\begin{itemize}
\item \textit{A SUC does not undergo contention nor collision. Also, with a SUC, we assume that every Rx vehicle in the Tx's transmission range successfully receives the packet.}
\item \textit{A SYNC refers to a situation where more than one Tx's start transmission at the same time due to the same value of backoff in carrier-sense multiple access/collision avoidance (CSMA/CA).}
\item \textit{A HN occurs in relation to carrier-sense threshold.}
\end{itemize}

\subsubsection{Consideration of BSM Broadcast in CCH}
Although non-safety applications may require very high transmission rates surpassing 100 Mbps \cite{bennis_jsac17}, the safety-critical applications are still planned on use of BSMs that are periodically broadcasted \cite{gaurang17}.

The analysis framework and result that will be presented throughout this paper are based on assumption of using the control channel (CCH) only--\textit{i.e.}, channel 178 in the 5.9 GHz band. It means that the result has a room for improvement if the network's channel selection is expanded among the other shared channels (SCHs). As such, the results that will be demonstrated in Section \ref{sec_results} can be regarded as the worst-case, most conservative ones.

\section{Spatiotemporal DSRC Performance Analysis}\label{sec_analysis}
This section presents the proposed spatiotemporal analysis framework, based on the system model characterized in Section \ref{sec_model} where $\mathsf{vT}$, the tagged vehicle, belongs to DSRC while Wi-Fi and C-V2X become the external interference-generating RATs.

\begin{figure}
\vspace{-0.3 in}
\centering
\begin{subfigure}[b]{0.45\linewidth}
\centering
\includegraphics[width = \linewidth]{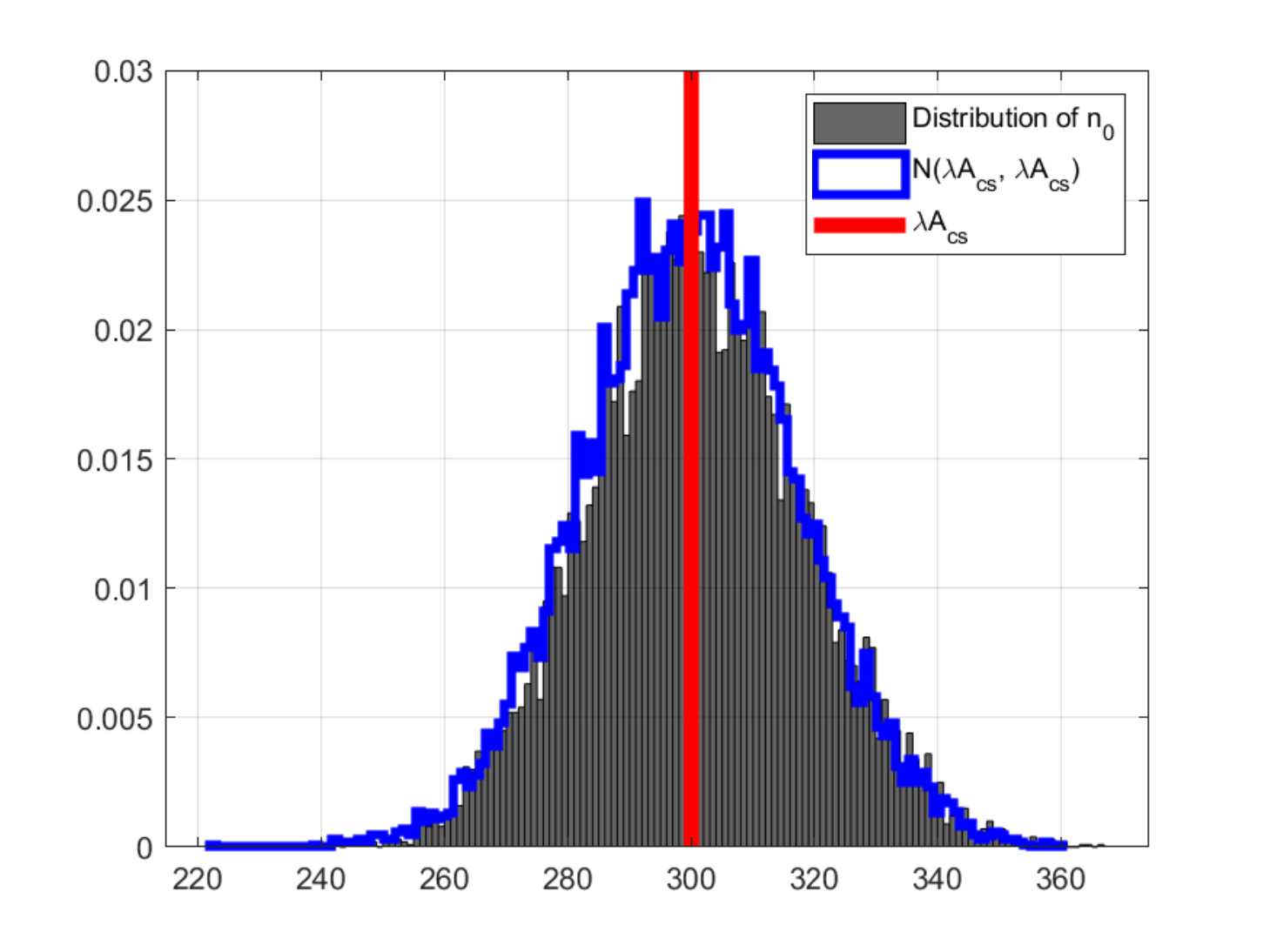}
\caption{PMF}
\label{fig_n0_pmf}
\end{subfigure}\hfill
\begin{subfigure}[b]{0.45\linewidth}
\centering
\includegraphics[width = \linewidth]{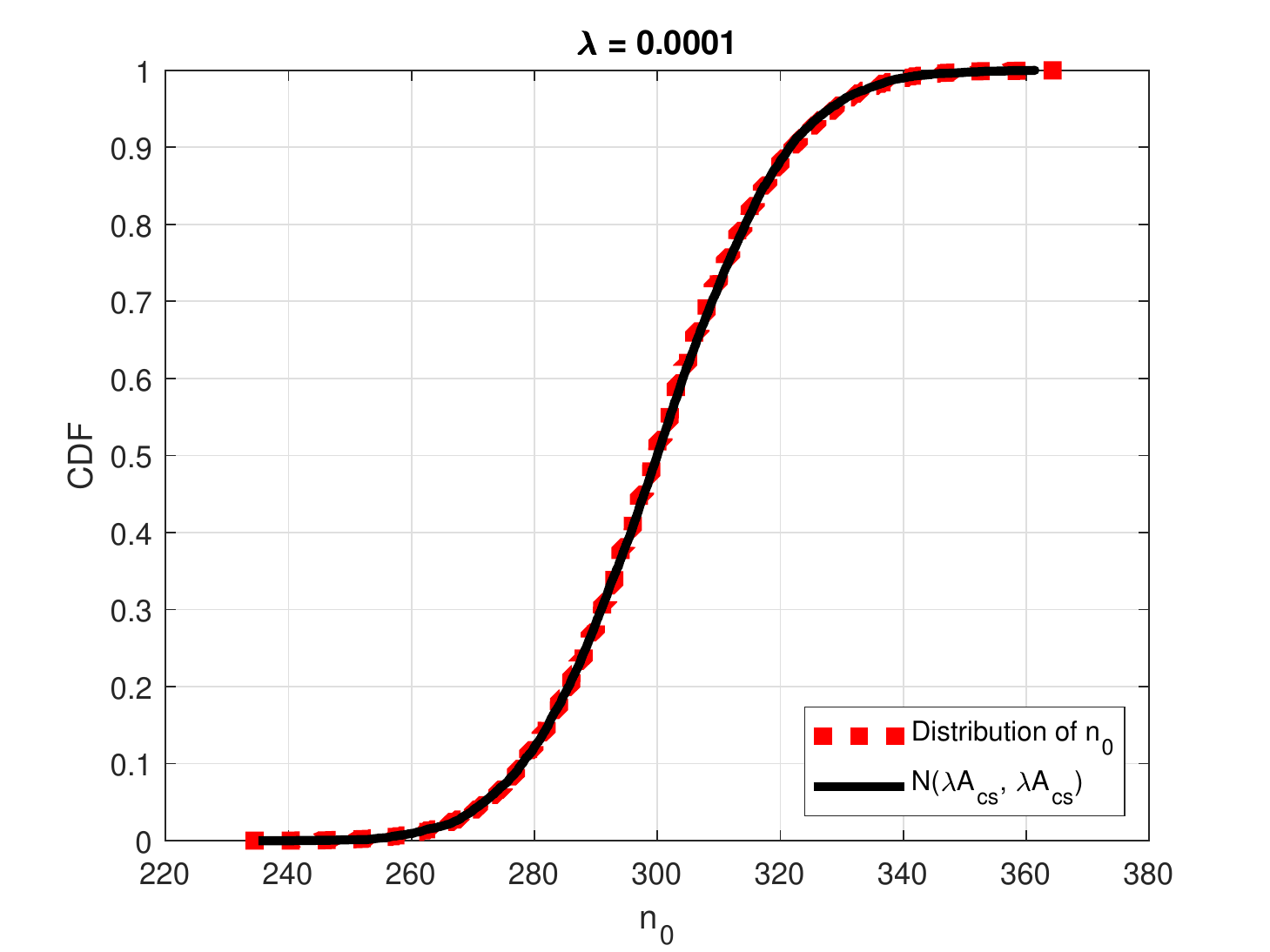}
\caption{CDF}
\label{fig_n0_cdf}
\end{subfigure}
\caption{Model validation: Distribution of $n_{cs}$ (when $\lambda = 300$ (m$^2$))}
\label{fig_n0}
\vspace{-0.3 in}
\end{figure}

\vspace{-0.1 in}
\subsection{Temporal Part of the Analysis: Probabilities of EXP, SYNC, and HN}\label{sec_analysis_temporal}
The key discussion in this section is the fact that the two types of packet collision--SYNC and HN--result in (i) not necessarily a lost message for all the Rx vehicles in a Tx vehicle's $r_{cs}$, and (ii) different performance of broadcast reception from geometric perspective.

This subsection provides a framework for detailed packet behavior, with and without the external interference from the Wi-Fi system. More specifically, it derives closed-form expressions for probabilities of packet results--namely, $\mathsf{P}_{exp}$, $\mathsf{P}_{\text{sync}}$, and $\mathsf{P}_{\text{hn}}$.

\begin{remark}\label{remark_clt}
(Central limit theorem for the number of competing nodes). \textit{Let $n_{cs}$ denote the number of nodes distributed in the area formed by a node's carrier-sense range, $\mathsf{A}_{cs}$. Notice that these $n_{cs}$ nodes compete for the medium with a given node. Then, it is noteworthy that $n_{cs}$ follows the Poisson distribution \cite{haenggi05} with a given value of $\mathsf{A}_{cs} = \mathsf{a}$, which is approximated to the normal distribution due to being sufficiently large (\textit{i.e.}, $n_{cs} > 1000$). Therefore, throughout the rest of this paper, we approximate $n_{cs}$ as a normal random variable, following $\mathcal{N}\left( \lambda \mathsf{A}_{cs}, \lambda \mathsf{A}_{cs} \right)$. This remark is illustrated in Figure \ref{fig_n0}.}
\end{remark}

\subsubsection{Probability of Expiration (EXP)}\label{sec_analysis_exp}
An EXP can occur since a beacon is not always able to start within a beaconing period of $L_{bcn}$. It mainly is attributed to the IEEE 802.11p distributed coordination function (DCF) where a random backoff value is allocated, which is decremented only when a slot is found idle \cite{ieee80211p}. As such, if a vehicle is not able to find an idle slot within a beaconing period, it is supposed to ``drop'' the current BSM and start the clock for the next BSM. This packet drop due to inability of transmitting within a beaconing period is defined as an EXP.

\begin{lemma}\label{lemma_Pb}
(Probability of a busy slot). \textit{The probability that a certain slot is found busy can be written as}
\begin{align}\label{eq_Pb}
&\mathsf{P}_{b}\left(\text{CW}\right)\nonumber\\
&= 1 - \displaystyle \sum_{n_{cs}=0}^{\mathbb{N}\left[\Phi_{cs}\right]} \left( 1 - \left( \text{CW} \displaystyle \sum_{k = 0}^{\text{CW}-1} \frac{1}{\left(1 - \mathsf{P}_{b}\right)^{k}} \prod_{i=1}^{k} \left( \displaystyle \sum_{m=0}^{\min\left(n_{cs}, L_{bcn}-l_{bcn}-k\right)} \left(\mathsf{P}_{b}\right)^{m} \right)^{-1} \right)^{-1} \right)^{n} \mathbb{P}\left[ n_{cs} = n \right]
\end{align}
\textit{where $\mathbb{N}\left[\Phi_{cs}\right]$ denotes the number of points in, $\Phi_{cs}$, a PPP formed in the area of the tagged vehicle's carrier-sense range, $r_{cs}$. It is important to notice that $\Phi_{cs}$ can be equivalent to (i) $\Phi_{\mathsf{D}}$ in case that DSRC is the only enabled system or (ii) a superposition among multiple PPPs, \textit{i.e.}, $\Phi_{\mathsf{D}} + \Phi_{\mathsf{W}} + \Phi_{\mathsf{C}}$ if Wi-Fi and C-V2X are enabled and hence generate external interference into DSRC. Importantly, recall that $\mathbb{P}\left[ n_{cs} = n \right]$ is given by the normal PMF as already discussed in Remark \ref{remark_clt}.}
\end{lemma}

\textit{Proof:} See Appendix \ref{appendix_Pb}. \hfill$\blacksquare$

\begin{figure}
\vspace{-0.3 in}
\hspace{-0.2 in}
\centering
\begin{minipage}{0.45\textwidth}
\centering
\includegraphics[width = \linewidth]{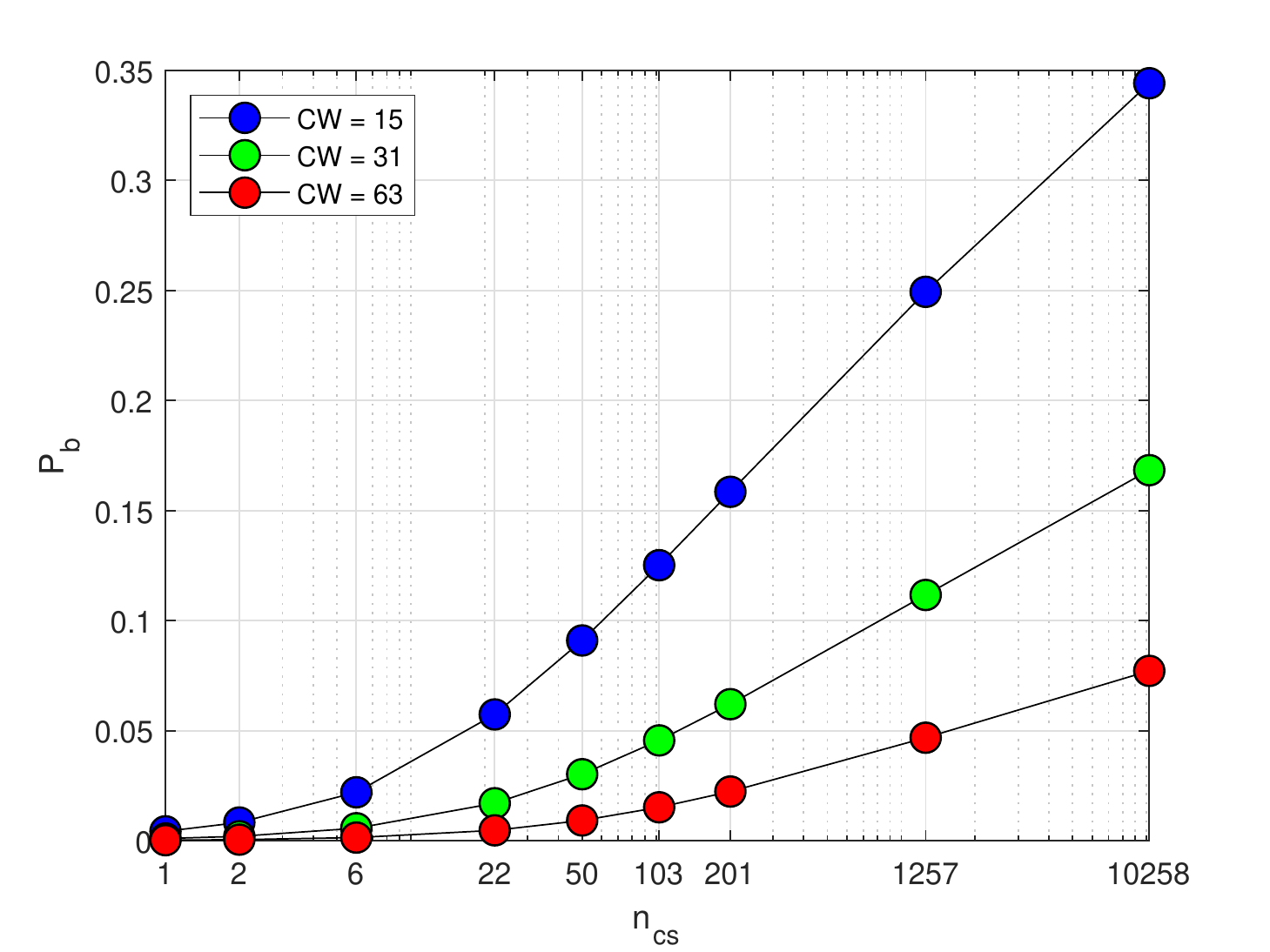}
\caption{$\mathsf{P}_{b}$ versus $n_{cs}$}
\label{fig_Pb}
\end{minipage}\hfill
\begin{minipage}{0.55\textwidth}
\centering
\includegraphics[width = 1.07\linewidth]{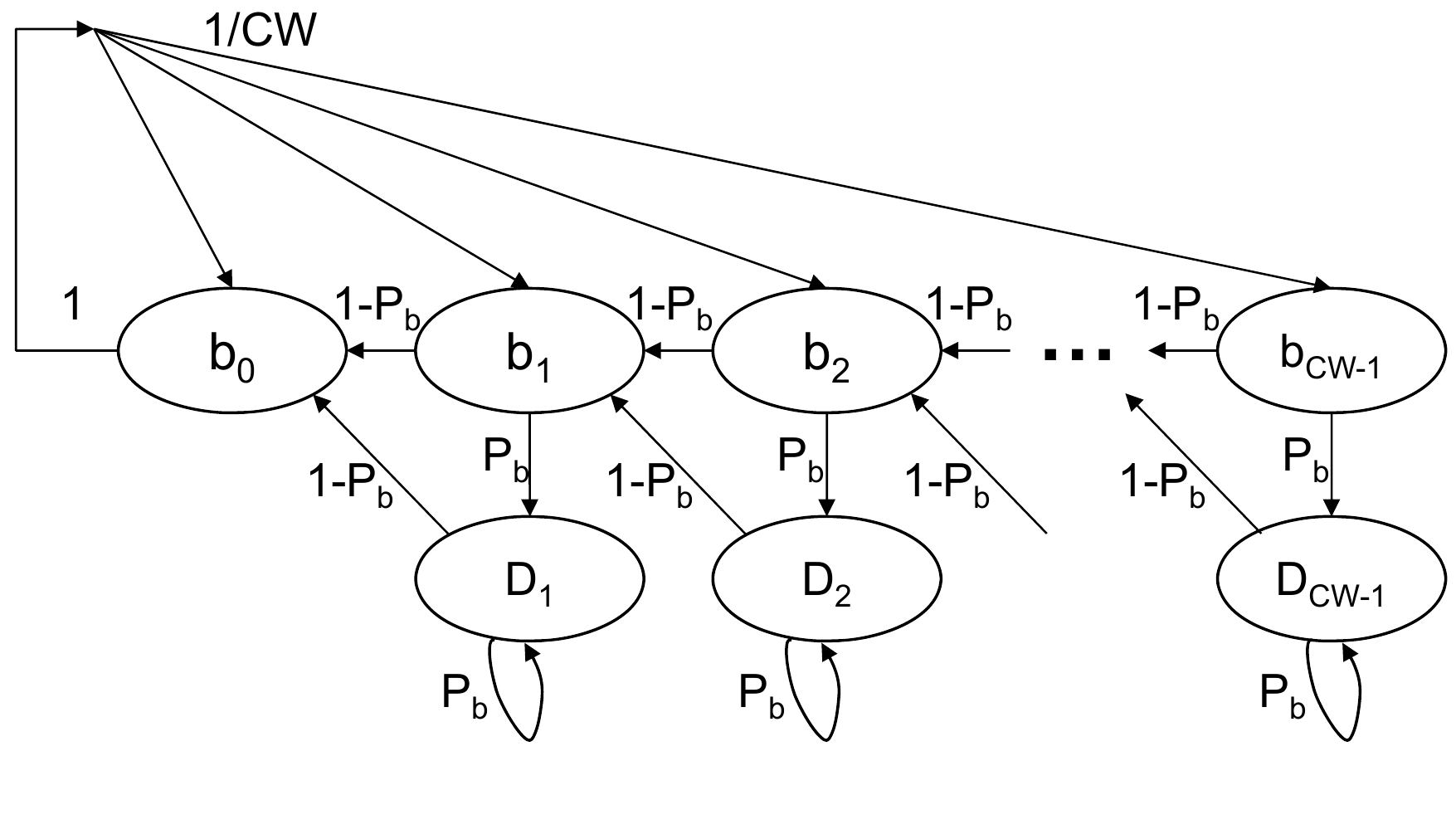}
\caption{Markov process for backoff in DSRC}
\label{fig_markov}
\end{minipage}
\vspace{-0.3 in}
\end{figure}

It is noteworthy that we solve (\ref{eq_Pb}) numerically due to high complexity for solving in a closed form. In the numerical computation, the value of $\mathsf{P}_b$ was incremented by $10^{-5}$, and one yielding the minimal error, $\epsilon$, was chosen as a solution. The resulting $\epsilon$ is kept in the range of several $10^{-6}$'s.

In Figure \ref{fig_markov}, notice also that two key modifications are applied on a prior Markov process model \cite{bianchi}, reflecting the key characteristics in DSRC:
\begin{enumerate}
\item In DSRC, there is no acknowledgement (ACK) for broadcast of a BSM, and thus the CW is never increased and kept constant \cite{eurasip19}.
\item A decrement of a backoff counter occurs only when the channel is sensed idle \cite{vnc10}. This requires change of the transition probability from state $k+1$ to $k$: we modify the probability from 1 \cite{bianchi} to $1-\mathsf{P}_b$.
\end{enumerate}

\begin{remark}\label{remark_Pb_vs_CW}
(Tendency of $\mathsf{P}_b$ versus CW). \textit{Looking at Figure \ref{fig_markov}, one can observe that a lower $\mathsf{P}_{b}$ is resulted as with a greater CW. The reason is that a greater CW makes it more difficult for a packet to make it through state $b_{0}$ due to the need for more $1 - \mathsf{P}_{b}$'s. Accordingly, the results of $\mathsf{P}_{b}$ are presented in Figure \ref{fig_Pb}. The results match one's intuitions: (i) $\mathsf{P}_{b}$ is increased with a greater $n_{cs}$; and (ii) $\mathsf{P}_{b}$ is decreased with a greater CW. The latter makes practical sense in particular, considering that CW is supposed to be doubled every time a packet collision occurs \cite{bianchi}. Although DSRC does not support CW doubling, it would be a reasonable suggestion that a higher CW can resolve packet congestions in a network due to a too large number of nodes competing.}
\end{remark}

\begin{lemma}\label{lemma_Pstart}
(Probability of a BSM transmission within a beaconing period). \textit{The probability that an arbitrary node is able to transmit in any of $L_{bcn}$ slots within a beaconing period can be formulated as}\\
\vspace{-0.3 in}
\begin{align}\label{eq_Pstart}
\mathsf{P}_{\text{start}}\left(\lambda, \text{CW} \right) = \frac{1}{\text{CW}} \displaystyle \sum_{b = 0}^{\text{CW}-1} \displaystyle \sum_{n_{cs}=0}^{\mathbb{N}\left[\Phi_{cs}\right]} \frac{1}{\sigma\left(k\right)} \sum_{k = 0}^{\min\left(n, L_{bcn}-l_{bcn}-k\right)} \left(\begin{array}{c}b+k-1\\k\end{array}\right) \mathsf{P}_{b}^{k} \left( 1 - \mathsf{P}_{b} \right)^{b} \mathbb{P}\left[n_{cs} = n\right]
\end{align}
\textit{where $\sigma\left(k\right)$ denotes the significance of each $k$'s occurrence, which is given by}
\begin{align}
\sigma\left(k\right) = \left(\begin{array}{c}\min\left(n, L_{bcn}-l_{bcn}-k\right)\\k\end{array}\right) \mathsf{P}_{b}^{k} \left( 1 - \mathsf{P}_{b} \right)^{\min\left(n, L_{bcn}-l_{bcn}-k\right) - k}.
\end{align}
\end{lemma}

\textit{Proof:} The $\mathsf{P}_{\text{start}}\left(\lambda, \text{CW} \right)$ is derived as
\begin{align}
&\mathsf{P}_{\text{start}}\left(\lambda, \text{CW} \right)\nonumber\\
&\stackrel{(a)}{=} \mathbb{E}_{\text{CW}} \left[ \mathbb{E}_{n_{cs}} \left[ \mathbb{E}_{k} \left[ \mathbb{P}\left[\text{Transmission in a slot} \right] \right] \right] \right]\nonumber\\
&\stackrel{(b)}{=} \mathbb{E}_{\text{CW}} \left[ \mathbb{E}_{n_{cs}} \left[ \mathbb{E}_{k} \left[ \left(\begin{array}{c}b+k-1\\k\end{array}\right) \mathsf{P}_{b}^{k} \left( 1 - \mathsf{P}_{b} \right)^{b} \right] \right] \right]\nonumber\\
&= \frac{1}{\text{CW}} \displaystyle \sum_{b = 0}^{\text{CW}-1} \displaystyle \sum_{n_{cs}=0}^{\mathbb{N}\left[\Phi_{cs}\right]} \frac{1}{\sigma\left(k\right)} \sum_{k = 0}^{\min\left(n, L_{bcn}-l_{bcn}-k\right)} \left(\begin{array}{c}b+k-1\\k\end{array}\right) \mathsf{P}_{b}^{k} \left( 1 - \mathsf{P}_{b} \right)^{b} \mathbb{P}\left[n_{cs} = n\right]
\end{align}
In (a), index $k$ denotes each of all the possible numbers of busy slots. The maximum of this number is $L_{bcn}-l_{bcn}-k$, subtracting the length of a BSM, $l_{bcn}$, and a backoff counter, $b$, from the total number of slots within a beaconing period, $L_{bcn}$. However, if there are a smaller number of competing nodes than the quantity--\textit{i.e.}, $n < L_{bcn}-l_{bcn}-k$, the maximum number of busy slots will become $n$. As such, the range of $k$ is found as $[0, \min\left(n, L_{bcn}-l_{bcn}-k\right)]$.

Also, in (b), the probability of a transmission in a general slot is given by ${b+k-1 \choose k} \mathsf{P}_{b}^{k} \left( 1 - \mathsf{P}_{b} \right)^{b}$, reflecting the mechanism shown in Figure \ref{fig_markov}. For instance, suppose that the backoff counter is allocated to be 1 and there are 2 busy slots--\textit{i.e.}, $b=1$ and $k=2$. With the probability of $1-\mathsf{P}_b$, the node spends a slot decrementing the backoff counter; and with the probability of $\mathsf{P}_b$, it spends a slot without decrementing the backoff counter, due to seeing a slot busy. This proves the reason of formulating the probability of transmission for each case of $k$ as ${b+k-1 \choose k} \mathsf{P}_{b}^{k} \left( 1 - \mathsf{P}_{b} \right)^{b}$.
\hfill$\blacksquare$

\begin{remark}\label{remark_Pb_vs_CW}
(Tendency of $\mathsf{P}_{\text{start}}$ versus CW). \textit{Referring to (\ref{eq_Pstart}), it is noteworthy that $\mathsf{P}_{\text{start}}$ is decreased as with a greater CW, with the same value for $\mathsf{P}_{b}$. The reason is that the significance $1/\sigma\left(k\right)$ of a smaller backoff value gets smaller as CW increases.}
\end{remark}

\subsubsection{Synchronized Transmission (SYNC)}
Now, we calculate the ``probability of a SYNC'' that corrupts a packet reception at vehicle $\mathsf{vR}$ from $\mathsf{vT}$, which is denoted by $\mathsf{P}_{\text{sync}}$. As illustrated in Figure \ref{fig_geometry_Acol_sync}, a SYNC can be caused by any other vehicle located within the $\mathsf{vT}$'s carrier-sense range. We formally write it as $\mathsf{A}_{\text{sync}} = \pi r_{cs}^2$.

However, it should be noted that a SYNC does not occur for every packet. Rather, it takes a probability that is formulated as a function of $\lambda$ and CW.

\begin{lemma}\label{lemma_Psync}
(Probability of a SYNC). \textit{The probability that a SYNC occurs is given by}
\begin{align}\label{eq_Psync}
\mathsf{P}_{\text{sync}}\left(\lambda, \text{CW}\right) = \Big( 1 - e^{-\lambda \pi r_{cs}^2} \Big) \Big( 1 - \displaystyle \sum_{n_{cs}=0}^{\mathbb{N}\left[\Phi_{cs}\right]} \left(1 - \mathsf{P}_{\text{start}} \right)^{n} \mathbb{P}\left[n_{cs} = n\right] \Big) \tau
\end{align}
\textit{where $\mathsf{P}_{\textnormal{start}}$ is used as defined in (\ref{eq_Pstart}).}
\end{lemma}

\textit{Proof:} See Appendix \ref{appendix_Psync}. \hfill$\blacksquare$

\subsubsection{Hidden-Node Collision (HN)}
By definition \cite{globecom18}, a HN-causing Tx is located outside of $\mathsf{vR}$`s $r_{cs}$ but outside of $\mathsf{vT}$'s $r_{cs}$, as illustrated in Figure \ref{fig_geometry_Acol_hn}. Notice from the figure that the collision area for a HN, which is formally written as $\mathsf{A}_{\text{hn}} = 4\pi r_{cs}^2 - \pi r_{cs}^2 = 3 \pi r_{cs}^2$, is much larger than that for a SYNC.

\begin{remark}\label{remark_difference_HN_SYNC}
(Difference of HN from SYNC in Temporal Perspective). \textit{There is one very important difference between the SYNC and HN, in terms of packet behavior in time slots. That is, while a SYNC only occurs when two colliding transmissions coincide at a certain time slot, a HN can occur even when the starting points of the colliding transmissions are not lined up at the same time instant. Specifically, vehicle $\mathsf{vR}$ is defined to experience a HN not only (i) when the colliding vehicle $\mathsf{vC}$ starts its transmission at the same time slot with $\mathsf{vR}$, but also (ii) when a transmission that $\mathsf{vC}$ started before still remains in effect upon the time of $\mathsf{vR}$'s transmission. This is possible because $\mathsf{vC}$ cannot be sensed by $\mathsf{vR}$ due to being located outside of $\mathsf{vR}$'s $r_{cs}$.}
\end{remark}

\begin{remark}\label{remark_timing_HN}
(Timing of occurrence of a HN). \textit{A HN occurs at the Rx vehicle $\mathsf{vR}$ when any of $n_{cs}$ ($\ge 1$) nodes in $\Phi_{\text{tot}} = \Phi_{\mathsf{D}} \cup \Phi_{\mathsf{W}} \cup \Phi_{\mathsf{C}}$ (which shall be discussed in Proposition \ref{proposition_superposition}) either (i) starts to transmit or (ii) is already in a transmission. In other words, the timing can be identified as (i) any of the $l_{bcn}$ slots, or (ii) any of the preceding $l_{bcn}-1$ slots that are occupied by vehicle $\mathsf{vT}$, respectively.}
\end{remark}

\begin{lemma}\label{lemma_Phn}
(Probability of a HN). \textit{The probability that a HN occurs in any slot within a beaconing period can be written as}
\begin{align}\label{eq_Phn}
\mathsf{P}_{\text{hn}}\left(\text{CW}\right) = \left( 1 - e^{-3\lambda \pi r_{cs}^2} \right) \left( 1 - \left( L_{bcn} - l_{bcn} + 1\right) \displaystyle \sum_{n_{cs}=0}^{\mathbb{N}\left[\Phi_{cs}\right]} \frac{\left( L_{bcn} - l_{bcn} \right)^{n}}{\left(L_{bcn}\right)^{n+1}} \mathbb{P}\left[n_{cs} = n\right] \right) \mathsf{P}_{\text{start}}.
\end{align}
\end{lemma}

\textit{Proof:} See Appendix \ref{appendix_Phn}. \hfill$\blacksquare$

\begin{assumption}\label{assumption_saturated_analysis}
(Saturated analysis). \textit{It is important to note that all the probabilities obtained in this Section via temporal analysis are assumed to characterize a ``saturated' situation of a network. In other words, this temporal model does not take into account the procedures taken in the initiation stages of a network--\textit{e.g.}, a new node's joining to a network. This assumption does not undermine the generality since the BSM exchange in a DSRC network does not necessitate such an initiation process.}
\end{assumption}

\begin{figure}
\vspace{-0.2 in}
\centering
\begin{subfigure}[b]{0.3\linewidth}
\centering
\includegraphics[width = \linewidth]{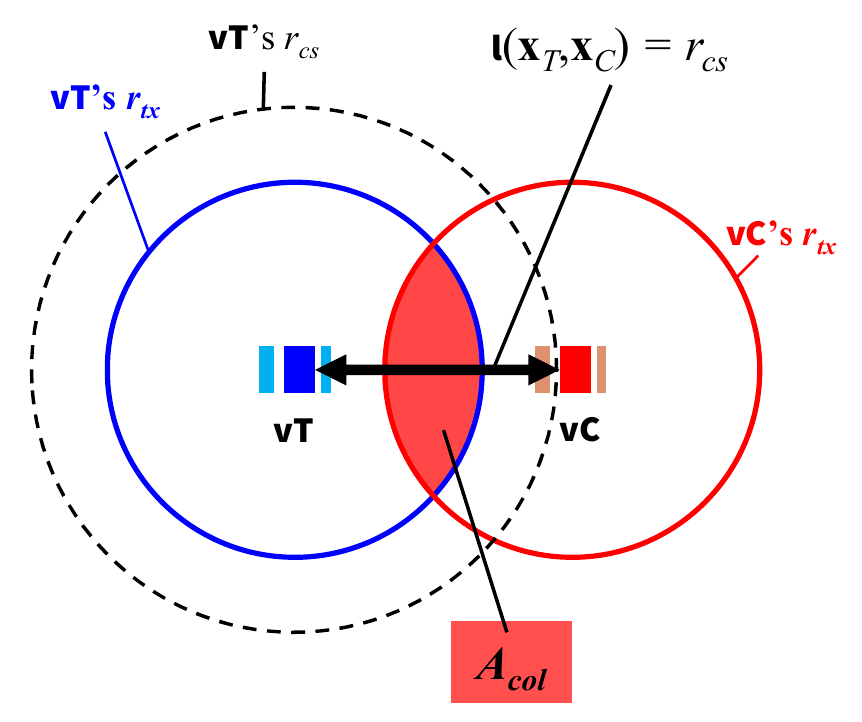}
\caption{Minimum collision area}
\label{fig_geometry_sync_min}
\end{subfigure}\hspace{1.5 in}
\begin{subfigure}[b]{0.3\linewidth}
\includegraphics[width = 0.9\linewidth]{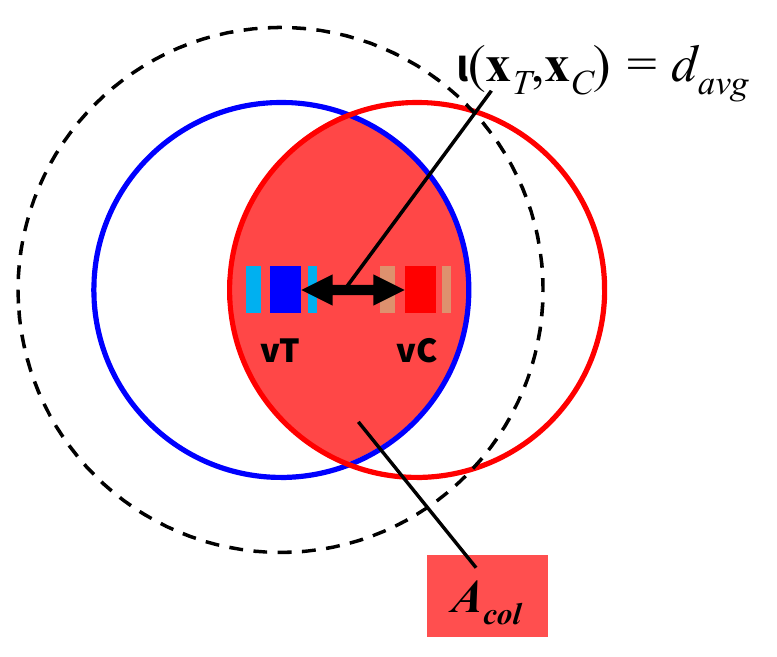}
\caption{Maximum collision area}
\label{fig_geometry_sync_max}
\end{subfigure}
\vspace{-0.2 in}
\caption{Area of region affected by a SYNC}
\label{fig_geometry_sync}
\vspace{0.1 in}
\centering
\begin{subfigure}[b]{0.3\linewidth}
\centering
\includegraphics[width = \linewidth]{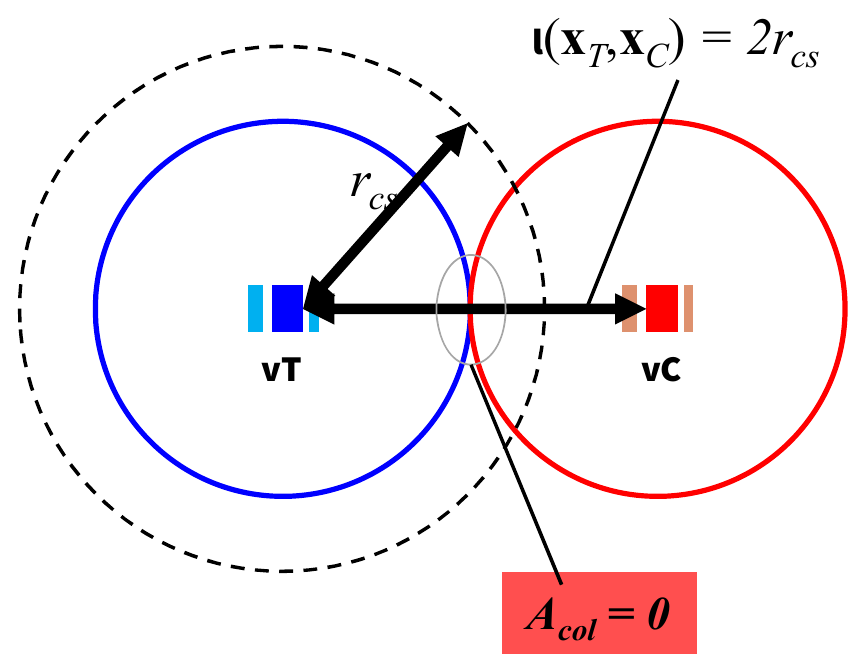}
\caption{Minimum collision area}
\label{fig_geometry_hn_min}
\end{subfigure}\hspace{1.5 in}
\begin{subfigure}[b]{0.3\linewidth}
\centering
\includegraphics[width = 0.95\linewidth]{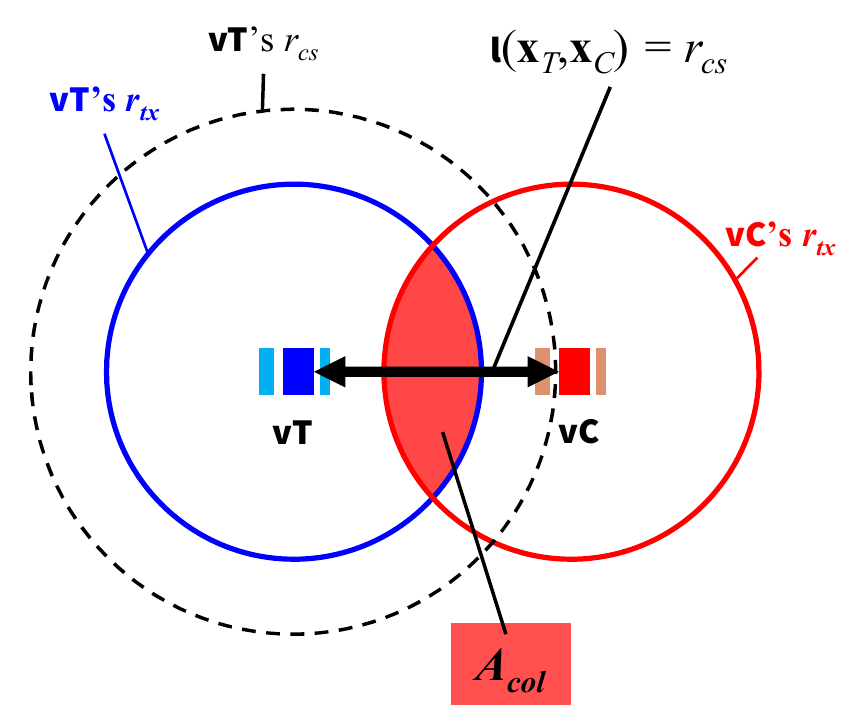}
\caption{Maximum collision area}
\label{fig_geometry_hn_max}
\end{subfigure}
\vspace{-0.2 in}
\caption{Area of region affected by a HN}
\label{fig_geometry_hn}
\vspace{-0.3 in}
\end{figure}

\vspace{-0.3 in}
\subsection{Spatial Part of the Analysis: Rate of Affected Area}\label{sec_analysis_spatial}
The spatial analysis starts from formulating the area that is formed by $\mathsf{vT}$ and $\mathsf{vC}$, as illustrated in Figures \ref{fig_geometry_sync} and \ref{fig_geometry_hn}.

\begin{lemma}\label{lemma_f_L}
(Distribution of Euclidean distance). \textit{The probability density function (PDF) and cumulative distribution function (CDF) of the Euclidean distance between $\mathsf{vT}$ located at the origin and an arbitrary point belonging to $\Phi_{j}$, representing the distance between vehicles $\mathsf{vT}$ and $\mathsf{vR}$, $\mathsf{l} \left( \mathbf{x}_{\mathsf{vT}}, \mathbf{x}_{\mathsf{vR}} \right)$, are given by}
\begin{align}\label{eq_f_L}
f_{\mathsf{L}} \left(\mathsf{l}\right) = \begin{cases}\displaystyle \frac{\pi \mathsf{l}}{2D^2}, {\rm{~~}} 0 \le \mathsf{l} < D\\
\displaystyle \frac{\mathsf{l}}{D^2} \left( \frac{\pi}{2} - 2\arccos \left(\frac{D}{\mathsf{l}}\right) \right), {\rm{~~}} D \le \mathsf{l} \le \sqrt{2}D.\end{cases}
\end{align}
\textit{and}
\begin{align}\label{eq_F_L}
F_{\mathsf{L}} \left(\mathsf{l}\right) = \begin{cases}\displaystyle \frac{\pi \mathsf{l}^2}{4D^2}, {\rm{~~}} 0 \le \mathsf{l} < D\\
\displaystyle \frac{\pi \mathsf{l}^2}{4D^2} - \displaystyle \frac{1}{D^2} \left[ \mathsf{l}^2 \cos^{-1} \left(\frac{D}{\mathsf{l}}\right) - D\mathsf{l} \sqrt{1 - \left(\frac{D}{\mathsf{l}}\right)^2} \right], {\rm{~~}} D \le \mathsf{l} \le \sqrt{2}D,\end{cases}
\end{align}
\textit{where $D$ denotes the boundaries of X and Y axes in $\mathbb{R}^2$, as illustrated in Figure \ref{fig_rgb_definition}. Validation of the distribution is demonstrated in Figure \ref{fig_l_cdf}.}
\end{lemma}

\vspace{-0.1 in}
\textit{Proof:} See Appendix \ref{appendix_f_L}. \hfill$\blacksquare$

\begin{figure}
\vspace{-0.4 in}
\centering
\includegraphics[width = 0.45\linewidth]{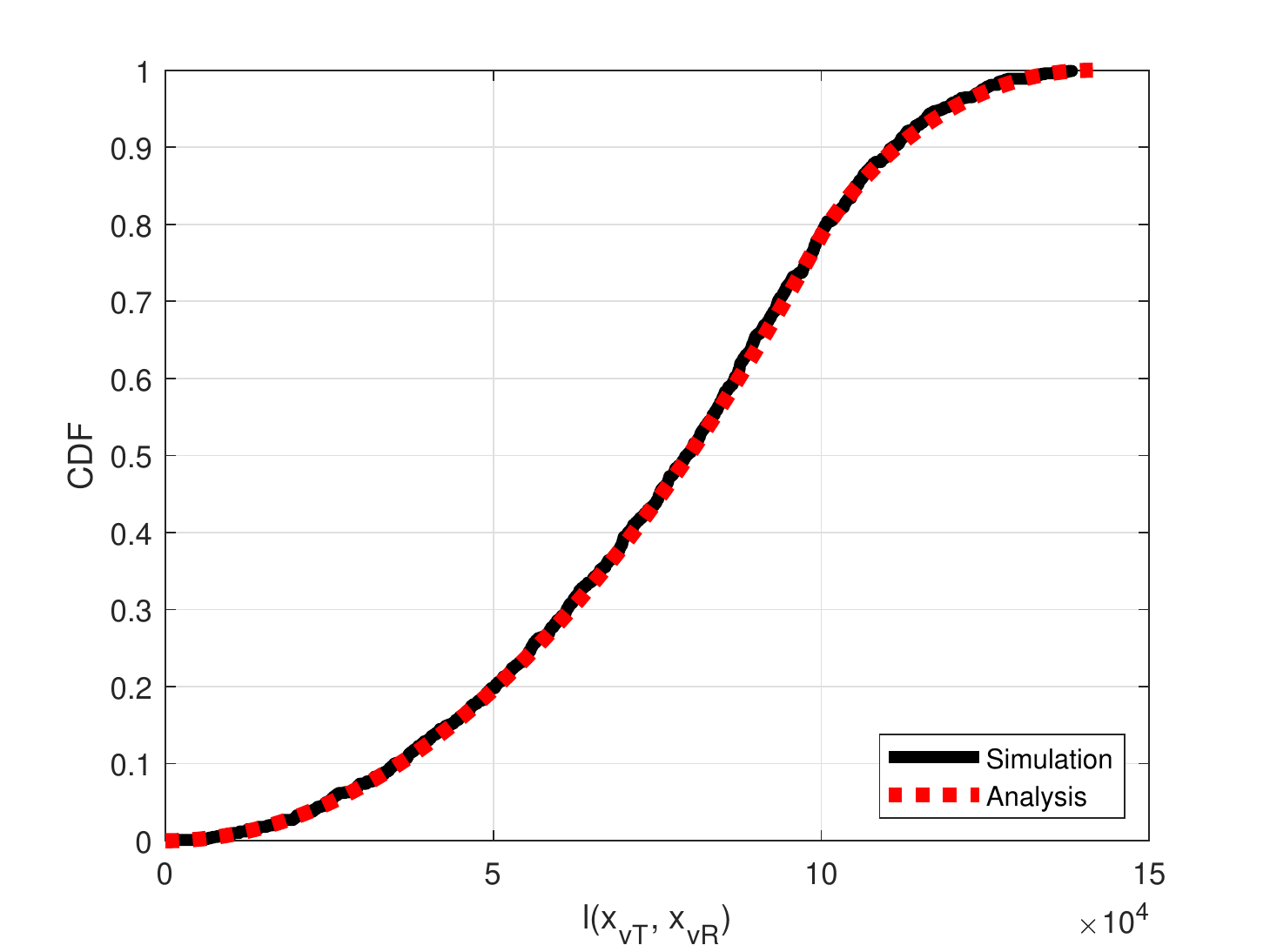}
\caption{Model validation: Distribution of $\mathsf{l}\left( \mathbf{x}_{\mathsf{vT}}, \mathbf{x}_{\mathsf{vR}} \right)$}
\label{fig_l_cdf}
\vspace{-0.3 in}
\end{figure}

\vspace{-0.1 in}
\begin{definition}\label{definition_Acol}
(Area of a collision affected region). \textit{As illustrated in Figure \ref{fig_Acol_vs_l}, the area $\mathsf{A}_{\text{col}}$ is derived as a function of $\mathsf{l}\left( \mathbf{x}_{\mathsf{vT}}, \mathbf{x}_{\mathsf{vR}} \right)$, which is formally written as \cite{globecom18}:}
\begin{align}\label{eq_A_col}
\mathsf{A}_{\text{col}} = 2r_{cs}^2 \cos^{-1} \left(\frac{\mathsf{l}\left( \mathbf{x}_{\mathsf{vT}}, \mathbf{x}_{\mathsf{vR}} \right)}{2r_{cs}}\right) - \frac{\mathsf{l}\left( \mathbf{x}_{\mathsf{vT}}, \mathbf{x}_{\mathsf{vR}} \right)}{2}\sqrt{4r_{cs}^2 - \mathsf{l}^2\left( \mathbf{x}_{\mathsf{vT}}, \mathbf{x}_{\mathsf{vR}} \right)}
\end{align}
\textit{with $0 \le \mathsf{l}\left( \mathbf{x}_{\mathsf{vT}}, \mathbf{x}_{\mathsf{vR}} \right) \le r_{\text{tx}}$. Note that $\mathsf{A}_{\textnormal{col}}$ denotes the area of the intersection of $r_{cs}$'s of vehicles $\mathsf{vT}$ and $\mathsf{vR}$, and $\mathtt{x}_{\mathsf{R}}$ denotes position of $\mathsf{vR}$.}
\end{definition}

\vspace{-0.2 in}
\begin{proposition}\label{proposition_Acol}
(Affected area for SYNC and HN). \textit{Since the range of $\mathsf{l}$ is different between SYNC and HN, the affected area can be formulated differently as}
\begin{align}\label{eq_affected_area}
\begin{split}
\mathsf{A}_{\text{sync}} &= \mathsf{A}_{\text{col}}\left(\mathsf{l}_{\text{sync}}\right), {\rm{~~}} \mathsf{l}_{\text{sync}} = \left\{ \mathsf{l} {\rm{~}} \big| {\rm{~}} 0 \le \mathsf{l} \le r_{cs}\right\}\\
\mathsf{A}_{\text{hn}} &= \mathsf{A}_{\text{col}}\left(\mathsf{l}_{\text{hn}}\right), {\rm{~~}} \mathsf{l}_{\text{hn}} = \left\{ \mathsf{l} {\rm{~}} \big| {\rm{~}} r_{cs} < \mathsf{l} \le2 r_{cs}\right\},
\end{split}
\end{align}
\textit{which are illustrated in Figures \ref{fig_geometry_sync} and \ref{fig_geometry_hn}.}
\end{proposition}

\begin{lemma}\label{lemma_f_A}
(Distribution of $\mathsf{A}_{\text{col}}$). \textit{Distribution of the collision-affected area, $\mathsf{A}_{\text{col}}$, can be formally identified based on the following PDF and CDF:}
\begin{align}\label{eq_f_A}
f_{\mathsf{A}_{\text{col}}}(\mathsf{a}) &= f_{\mathsf{L}} \left( p_{1}\mathsf{a}^2 + p_{2}\mathsf{a} + p_{3} \right) \left( 2 p_{1}\mathsf{a} + p_{2} \right)\\
F_{\mathsf{A}_{\text{col}}}(\mathsf{a}) &= {\begin{cases} \vspace{0.1 in} \displaystyle \frac{\pi}{4r^2} \left(g^{-1}\left(\mathsf{a}\right)\right)^2, {\rm{~~~~}} 0 \le g^{-1}\left(\mathsf{a}\right) \le r\\
\displaystyle \frac{\pi}{4r^2}\left(g^{-1}\left(\mathsf{a}\right)\right)^2 - \frac{1}{r^2}\left(g^{-1}\left(\mathsf{a}\right)\right)^2\cos^{-1}\left(\frac{r}{g^{-1}\left(\mathsf{a}\right)}\right)\nonumber\\
\displaystyle {\rm{~~~}} + \frac{1}{r} g^{-1}\left(\mathsf{a}\right) \sqrt{1 - \frac{r^2}{\left(g^{-1}\left(\mathsf{a}\right)\right)^2}}, {\rm{~~~~}} r \le g^{-1}\left(\mathsf{a}\right) \le \sqrt{2}r
\end{cases}}
\end{align}
\textit{where $p_{1}$, $p_{2}$, and $p_{3}$ denote the coefficients for the quadratic fitting, which can be found in Table \ref{table_g_coefficients} according to different values of $r$. Validation of the distribution is shown in Figure \ref{fig_Acol_cdf}.}
\end{lemma}

\textit{Proof:} See Appendix \ref{appendix_f_A}. \hfill $\blacksquare$

\begin{figure}[t]
\minipage{0.45\textwidth}
\centering
\includegraphics[width = \linewidth]{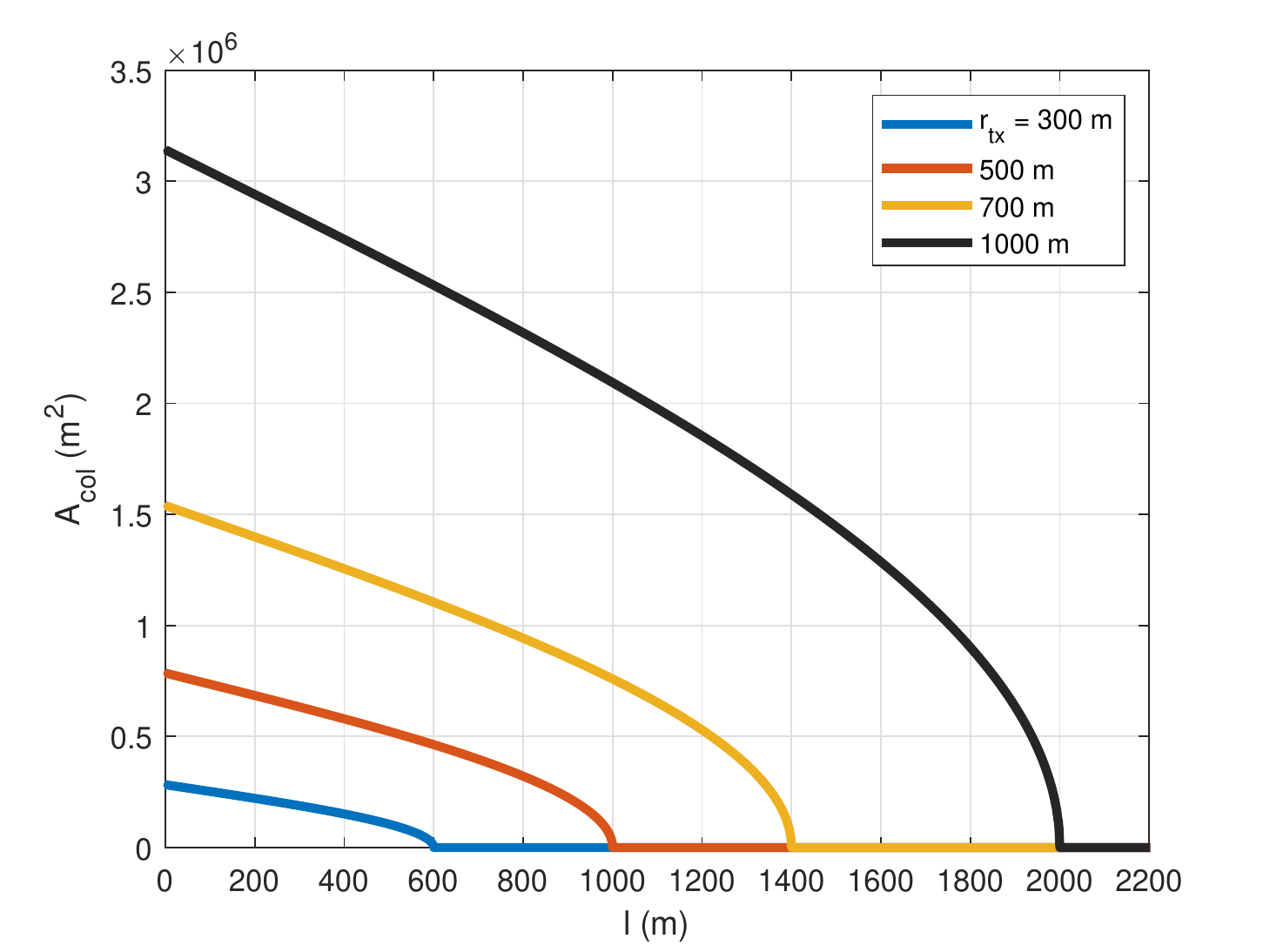}
\caption{$\mathsf{A}_{\text{col}}$ as a function of $\mathsf{l}$}
\label{fig_Acol_vs_l}
\centering
\includegraphics[width = \linewidth]{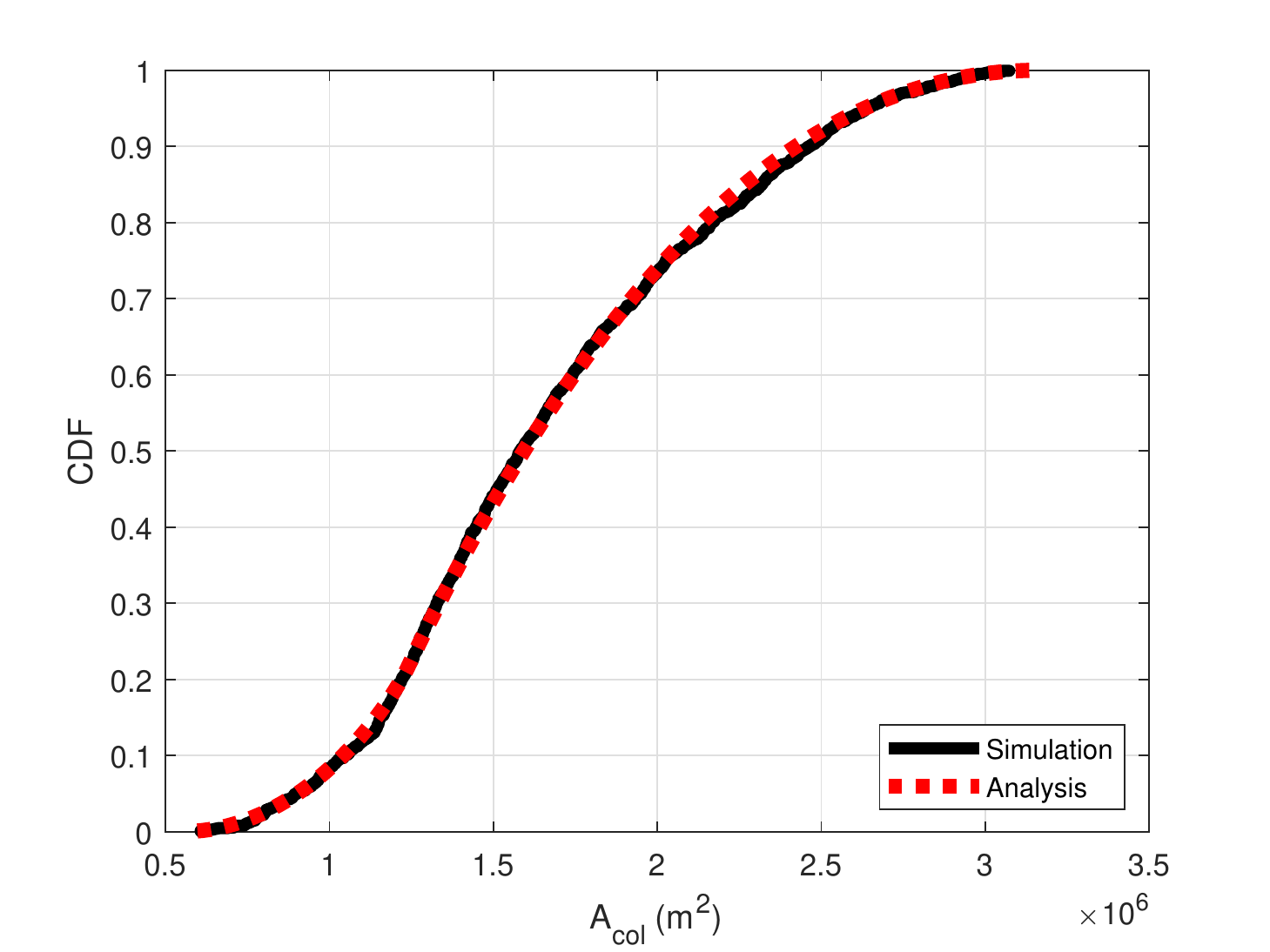}
\caption{Model validation: Distribution of $\mathsf{A}_{\text{col}}$ (when $r_{\text{tx}} = 10^{3}$ m)}
\label{fig_Acol_cdf}
\endminipage\hfill
\minipage{0.45\textwidth}
\centering
\captionsetup{type=table}
\scriptsize
\caption{Coefficients for quadratic fitting function $\mathsf{l} = \widehat{g}^{-1}\left(\mathsf{a}\right) \approx p_{1}\mathsf{a}^2 + p_{2}\mathsf{a} + p_{3}$ for $\mathsf{A}_{\text{col}}$}
\centering
\begin{tabular}{| c || c | c | c || c |}
\hline
\cellcolor{gray!20}$r$ (m) & \cellcolor{gray!20}$p_{1}$ & \cellcolor{gray!20}$p_{2}$ & \cellcolor{gray!20}$p_{3}$ & \cellcolor{gray!20}$\epsilon$\\
\hline\hline
100 & 3.123e-8 & -0.0065709 & 176.45 & 0.0219\\
\hline
200 & 3.9014e-9 & -0.0032847 & 353.04 & 0.0204\\
\hline
300 & 1.1606e-9 & -0.0021914 & 529.7 & 0.0199\\
\hline
400 & 4.8907e-10 & -0.0016432 & 706.2 & 0.0195\\
\hline
500 & 2.5086e-10 & -0.001315 & 882.85 & 0.0194\\
\hline
600 & 1.4505e-10 & -0.0010956 & 1059.4 & 0.0194\\
\hline
700 & 9.1284e-11 & -0.00093898 & 1235.8 & 0.0191\\
\hline
800 & 6.1221e-11 & -0.00082179 & 1412.5 & 0.0191\\
\hline
900 & 4.2974e-11 & -0.0007304 & 1589 & 0.0191\\
\hline
1000 & 3.1354e-11 & -0.00065746 & 1765.7 & 0.0191\\
\hline
1100 & 2.3546e-11 & -0.00059764 & 1942.2 & 0.0190\\
\hline
1200 & 1.8148e-11 & -0.0005479 & 2118.8 & 0.0190\\
\hline
1300 & 1.4268e-11 & -0.00050572 & 2295.3 & 0.0190\\
\hline
1400 & 1.142e-11 & -0.00046956 & 2471.8 & 0.0189\\
\hline
1500 & 9.2897e-12 & -0.0004383 & 2648.5 & 0.0190\\
\hline
1600 & 7.652e-12 & -0.00041088 & 2825 & 0.0189\\
\hline
1700 & 6.3825e-12 & -0.00038675 & 3001.6 & 0.0189\\
\hline
1800 & 5.3751e-12 & -0.00036524 & 3178.1 & 0.0189\\
\hline
1900 & 4.5722e-12 & -0.00034604 & 3354.8 & 0.0189\\
\hline
2000 & 3.919e-12 & -0.00032872 & 3531.3 & 0.0189\\
\hline
\end{tabular}
\label{table_g_coefficients}
\endminipage
\end{figure}

\begin{definition}\label{definition_error}
(Mean absolute fitting error). \textit{Notice that the fitting error, $\epsilon$, is defined as}
\begin{align}\label{eq_error}
\epsilon = \frac{\left| g^{-1}\left(\mathsf{a}\right) - \widehat{g}^{-1}\left(\mathsf{a}\right)\right|}{g^{-1}\left(\mathsf{a}\right)}
\end{align}
\textit{where $\widehat{\cdot}$ denotes the quadratic fitting function. We also note from Table \ref{table_g_coefficients} that the fitting error remains no higher than 2\% for all considered variants of $r$.}
\end{definition}

\begin{definition}\label{definition_RGB_SYNC_HN}
($\mathsf{RGB}$ for SYNC and HN). \textit{Referring to Figure \ref{fig_rgb_definition}, the geometric reception ratio for a SYNC or a HN can be formally written as}
\begin{align}
\mathsf{RGB}_{\text{sync or hn}}\left(\mathsf{l}\right) &=\frac{\mathsf{A}_{\text{sync or hn}}\left(\mathsf{l}\right) }{\pi r_{\text{tx}}^2}
\end{align}
\textit{with $\mathsf{A}_{\textnormal{sync}}$ and $\mathsf{A}_{\textnormal{hn}}$ as defined in (\ref{eq_affected_area}).}
\end{definition}

\subsection{Successful Reception (SUC)}\label{sec_analysis_wo_Psuc}
Now we characterize the geometric rate of a successful packet reception. Once a packet is ``transmitted''--that is, not expired at a $\mathsf{vT}$, it is assumed that all the $\mathsf{vR}$s not experiencing any of SYNC and HN are able to successfully receive the packet and decode the message. We define this behavior as a SUC.

\begin{lemma}\label{lemma_PDR}
(Packet delivery ratio ($\mathsf{PDR}$)). \textit{Combining all the relative quantities from spatial and temporal analyses through this section, the rate of successful packet reception area can be formulated as}
\begin{align}\label{eq_PDR}
\mathsf{PDR} = \mathsf{P}_{\text{start}} \left( 1 - \mathsf{P}_{\text{sync}} - \mathsf{P}_{\text{hn}} \right)
\end{align}
\end{lemma}

\textit{Proof:} 
\begin{align}\label{eq_PDR_proof}
\mathsf{PDR} &= \mathbb{P}\left[ \mathsf{vT} \text{ transmits within a beaconing period} \right] \mathbb{P}\left[ \text{No collision} \right]\nonumber\\
&= \mathbb{P}\left[ \mathsf{vT} \text{ transmits within a beaconing period} \right] \mathbb{P}\left[ \text{No SYNC nor HN} \right]\nonumber\\
&= \mathsf{P}_{\text{start}} \left( 1 - \mathsf{P}_{\text{sync}} - \mathsf{P}_{\text{hn}} \right)
\end{align}
\hfill$\blacksquare$

\begin{definition}\label{definition_STPDR}
(Spatiotemporal $\mathsf{PDR}$ ($\mathsf{STPDR}$)). \textit{Combining all the relative quantities from spatial and temporal analyses through this section, the rate of successful packet reception area can be formulated as}
\begin{align}\label{eq_STPDR}
&\mathsf{STPDR} \left(\lambda, \text{CW}, \mathsf{A}\right)\nonumber\\
&= \mathbb{P}\left[ \mathsf{vT} \text{ transmits within a beaconing period} \right]\nonumber\\
&{\rm{~~~~~~~~~~~~~~~~~~~~}}\cdot \frac{\text{Area of successful BSM reception without collision}}{\text{Area of an entire communication range}}\nonumber\\
&= \mathsf{P}_{\text{start}} \mathsf{RGB}\left[ \text{Reception without collision} \right]\nonumber\\
&= \mathsf{P}_{\text{start}} \Big( \mathsf{RGB}\left[ \text{Reception without SYNC} \right] + \mathsf{RGB}\left[ \text{Reception without HN} \right] \Big)\nonumber\\
&= \mathsf{P}_{\text{start}} \left(\lambda, \text{CW}\right) \bigg[ 1 - \mathsf{P}_{\text{sync}} \left(\lambda, \text{CW}\right) \mathsf{RGB}_{\text{sync}} \left(\mathsf{A}\right) - \mathsf{P}_{\text{hn}} \left(\lambda, \text{CW}\right) \mathsf{RGB}_{\text{hn}} \left(\mathsf{A}\right) \bigg]
\end{align}
\end{definition}

\begin{lemma}\label{lemma_STPDR}
(Average $\mathsf{STPDR}$). \textit{For precise analysis, the randomness in vehicles' geometry must be averaged, which influences $\mathsf{STPDR}$. Specifically, as shown in Figures \ref{fig_geometry_sync} and \ref{fig_geometry_hn}, area $\mathsf{A}_{\text{col}}$ is differentiated according to the geometry determined by the locations of $\mathsf{vT}$ and $\mathsf{vC}$. Since this paper intends to express $\mathsf{STPDR}$ as a function of $\lambda$, $r_{cs}$, and CW, it is integrated over the other two critical factors: the area of a collision, $\mathsf{A}_{\text{col}}$, and the number of colliding nodes, $n_{\text{col}}$, which is given by}\\\vspace{-0.4 in}
\begin{align}
&\overline{\mathsf{STPDR}}\left(\lambda, \text{CW}\right)\nonumber\\
&{\rm{~~~~~~~~~~~~}}= \mathsf{P}_{\text{start}} \left( 1 - \frac{1}{\pi r_{\text{tx}}^2} \Big( \mathsf{P}_{\text{sync}} \int_{\floor{\mathsf{A}}}^{\ceil{\mathsf{A}}} \mathsf{a}_{\text{sync}} f_{\mathsf{A}_{\text{col}}}(\mathsf{a}) \text{d}\mathsf{a}_{\text{sync}} - \mathsf{P}_{\text{hn}} \int_{\floor{\mathsf{A}}}^{\ceil{\mathsf{A}}} \mathsf{a}_{\text{hn}} f_{\mathsf{A}_{\text{col}}}(\mathsf{a}) \text{d}\mathsf{a}_{\text{hn}} \Big) \right).
\end{align}
\textit{Referring to Figures \ref{fig_geometry_sync}, \ref{fig_geometry_hn}, and \ref{fig_Acol_vs_l}, $\ceil{\mathsf{A}}$ and $\floor{\mathsf{A}}$ in the integral range are defined as}
\begin{itemize}
\item For SYNC:
\begin{align}\label{eq_range_SYNC}
\begin{aligned}\ceil{\mathsf{A}} &= \max \mathsf{A}_{\text{sync}} = \mathsf{A}_{\text{col}}\left(\min \mathsf{l}_{\text{sync}}\right) = \mathsf{A}_{\text{col}}\left( 0 \right)\\
\floor{\mathsf{A}} &= \min \mathsf{A}_{\text{sync}} = \mathsf{A}_{\text{col}}\left(\max \mathsf{l}_{\text{sync}} \right) = \mathsf{A}_{\text{col}}\left(r_{\text{tx}}\right)\end{aligned}
\end{align}
\item For HN:
\begin{align}\label{eq_range_HN}
\begin{aligned}\ceil{\mathsf{A}} &= \max \mathsf{A}_{\text{hn}} = \mathsf{A}_{\text{col}}\left(\min \mathsf{l}_{\text{sync}}\right) = \mathsf{A}_{\text{col}}\left( r_{\text{tx}} \right)\\
\floor{\mathsf{A}} &= \min \mathsf{A}_{\text{hn}} = \mathsf{A}_{\text{col}}\left(\max \mathsf{l}_{\text{sync}} \right) = \mathsf{A}_{\text{col}}\left(2r_{\text{tx}}\right)\end{aligned}
\end{align}
\end{itemize}
\textit{where $\mathsf{A}_{\text{col}}$ has been defined in (\ref{eq_affected_area}).}
\end{lemma}

\textit{Proof:}
Taking the average only to the related variables yields
\begin{align}
&\overline{\mathsf{STPDR}}\left(\lambda, \text{CW}\right)\nonumber\\
&{\rm{~~~~~~~~~~~}}= \mathbb{E}_{\mathsf{A}_{\text{col}}} \Big[ \mathsf{P}_{\text{start}} \left( 1 - \mathsf{P}_{\text{sync}} \mathsf{RGB}_{\text{sync}} - \mathsf{P}_{\text{hn}} \mathsf{RGB}_{\text{hn}} \right) \Big]\nonumber\\
&{\rm{~~~~~~~~~~~}}= \mathsf{P}_{\text{start}} \left( 1 - \mathbb{E}_{\mathsf{A}_{\text{col}}} \left[ \mathsf{P}_{\text{sync}} \mathsf{RGB}_{\text{sync}} \right] - \mathbb{E}_{\mathsf{A}_{\text{col}}} \left[ \mathsf{P}_{\text{hn}} \mathsf{RGB}_{\text{hn}} \right] \right)\nonumber\\
&{\rm{~~~~~~~~~~~}}= \mathsf{P}_{\text{start}} \left( 1 - \frac{1}{\pi r_{\text{tx}}^2} \Big( \mathsf{P}_{\text{sync}} \mathbb{E}_{\mathsf{A}_{\text{col}}} \left[ \mathsf{a}_{\text{sync}} \right] - \mathsf{P}_{\text{hn}} \mathbb{E}_{\mathsf{A}_{\text{col}}} \left[ \mathsf{a}_{\text{hn}} \right] \Big) \right)\nonumber\\
&{\rm{~~~~~~~~~~~}}= \mathsf{P}_{\text{start}} \left( 1 - \frac{1}{\pi r_{\text{tx}}^2} \Big( \mathsf{P}_{\text{sync}} \int_{\floor{\mathsf{A}}}^{\ceil{\mathsf{A}}} \mathsf{a}_{\text{sync}} f_{\mathsf{A}_{\text{col}}}(\mathsf{a}) \text{d}\mathsf{a}_{\text{sync}} - \mathsf{P}_{\text{hn}} \int_{\floor{\mathsf{A}}}^{\ceil{\mathsf{A}}} \mathsf{a}_{\text{hn}} f_{\mathsf{A}_{\text{col}}}(\mathsf{a}) \text{d}\mathsf{a}_{\text{hn}} \Big) \right)
\end{align}
where the integral ranges have already been defined in (\ref{eq_range_SYNC}) and (\ref{eq_range_HN}), and the PDF for the packet collision area, $f_{\mathsf{A}_{\text{col}}}(\mathsf{a})$, in (\ref{eq_f_A}).
\hfill$\blacksquare$

\section{External Interference}
Now we analyze the impact of the external interference. Reflecting the FCC's current consideration on scenarios of operating the 5.9 GHz band \cite{pai19}, this paper identifies the 3GPP C-V2X \cite{tr36300} and IEEE 802.11ac \cite{wifi6ghz} as the coexisting wireless technologies in the band. Recall that the focus of this paper is assess the performance of a DSRC system only; as such, the networking behaviors of Wi-Fi and C-V2X are analyzed only for the purpose of measuring the interference into DSRC.

The external interference can be formulated as a superposition additional PPPs to that for DSRC, $\Phi_{\mathsf{D}}$.

\begin{proposition}\label{proposition_superposition}
(PPP superposition for external interference). \textit{The external interference by Wi-Fi and C-V2X can be formally written as a superposition of three distinct PPPs, which is given by}
\begin{align}\label{eq_Phi_superposition}
\Phi_{\text{tot}} = \Phi_{\mathsf{D}} \cup \Phi_{\mathsf{W}} \cup \Phi_{\mathsf{C}}
\end{align}
\textit{with the intensity of}
\begin{align}\label{eq_lambda_superposition}
\lambda_{\text{tot}} = \lambda_{\mathsf{D}} + \lambda_{\mathsf{W}} + \lambda_{\mathsf{C}}.
\end{align}
\textit{Now one can understand that this proposition incurs larger value of $\lambda$ in the spatiotemporal analyses shown in Section \ref{sec_analysis}, which as a consequence degrades $\mathsf{PDR}$ and $\mathsf{RGB}$.}
\end{proposition}

\begin{figure}
\vspace{-0.2 in}
\centering
\begin{subfigure}[b]{0.49\linewidth}
\centering
\includegraphics[width = \linewidth]{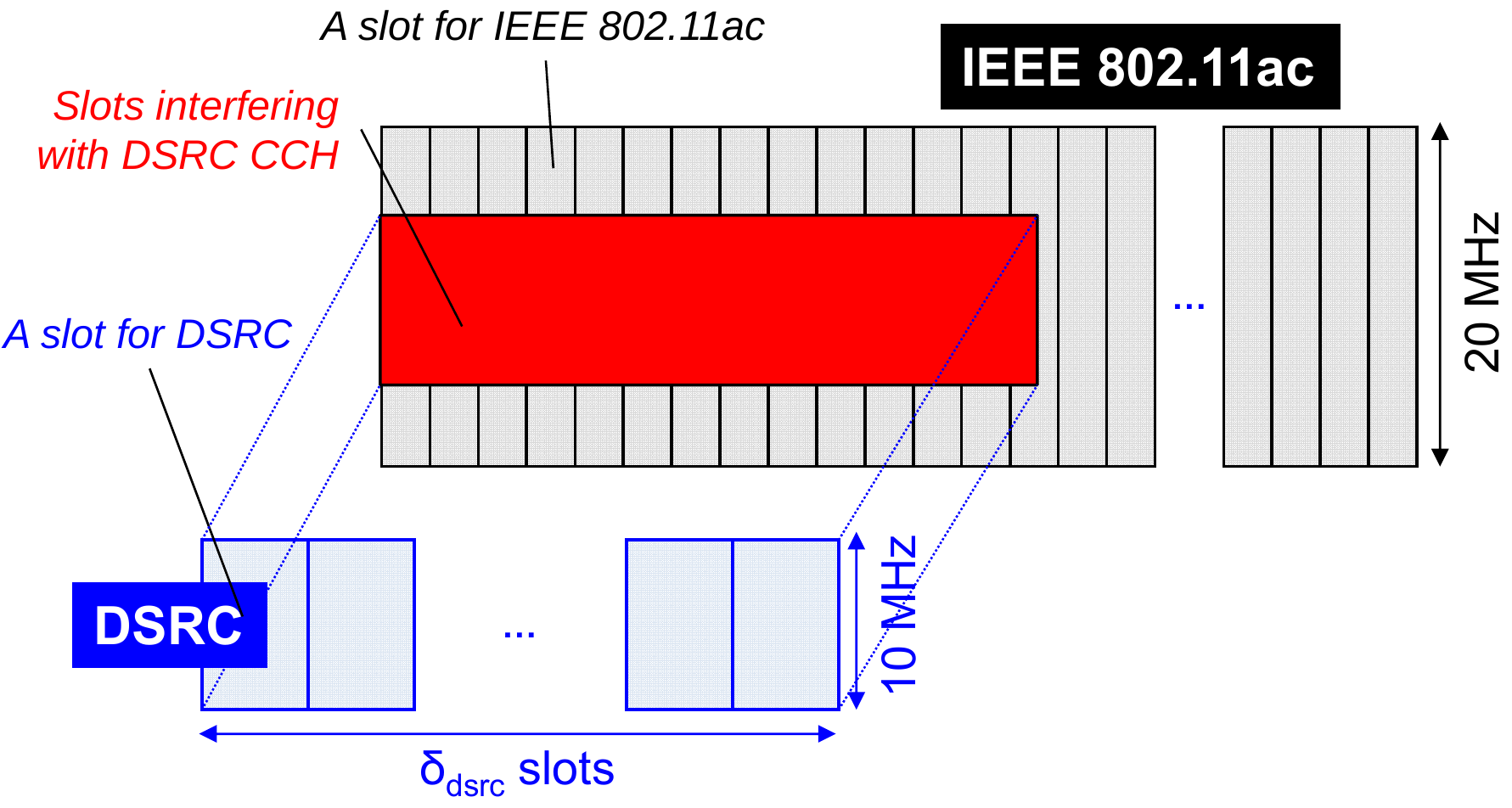}
\caption{With Wi-Fi}
\label{fig_external_interference_wifi}
\end{subfigure}\hfill
\begin{subfigure}[b]{0.49\linewidth}
\centering
\includegraphics[width = \linewidth]{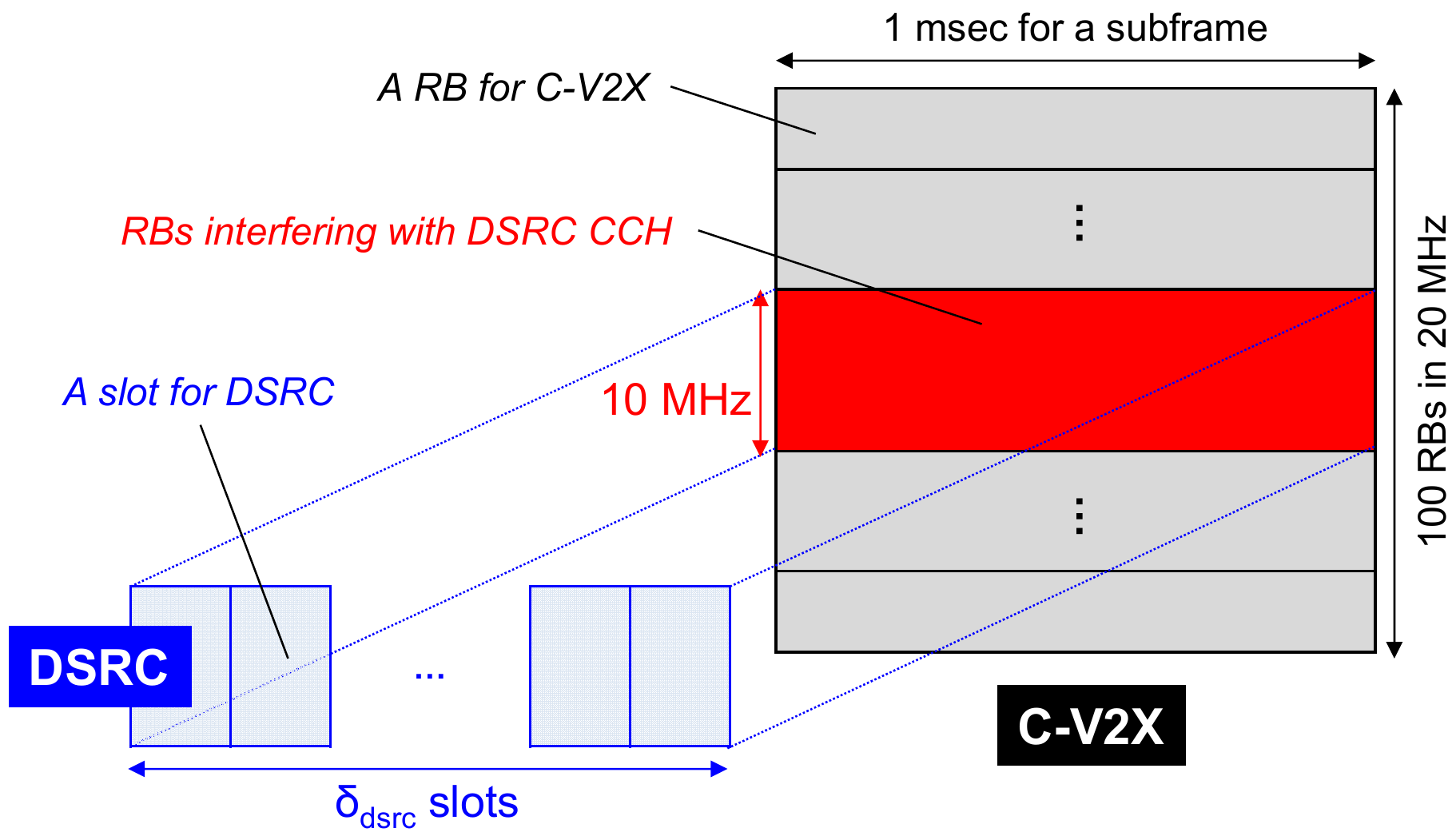}
\caption{With C-V2X}
\label{fig_external_interference_cv2x}
\end{subfigure}
\caption{Interference from C-V2X and IEEE 802.ac to DSRC in time and frequency}
\label{fig_external_interference}
\vspace{-0.2 in}
\end{figure}

\subsection{From Wi-Fi}
Same with DSRC, the nodes belonging to a Wi-Fi system access to the medium based on the IEEE 802.11 CSMA-CA. However, due to difference in (i) the backoff process and (ii) networking parameters (such as interframe space, slot time, etc), the procedure of calculating $\mathsf{P}_{\text{start}}$ \cite{icc05} is different from that for DSRC.

Regarding the intensity of PPP, this paper assumes $\lambda_{\text{dsrc}} > \lambda_{\text{wifi}}$, which is reasonable considering that the Wi-Fi is adopted by general users (not for V2X).

\begin{remark}\label{remark_conflict_wifi}
(Time- and frequency-conflict of a DSRC beaconing period with a Wi-Fi frame). \textit{Figure \ref{fig_external_interference_wifi} shows how a DSRC's beaconing period (1 msec and 10 MHz) conflicts with an IEEE 802.11ac frame. From the temporal perspective, a slot time for 802.11ac is 9 $\mu$sec \cite{mag14}, which yields that the $\ceil{\delta_{\text{dsrc}}/9}$th slot of an 802.11ac Tx causes the 2nd slot of DSRC corrupted. In frequency, assuming the bandwidth of IEEE 802.11ac to be 20 MHz, an IEEE 802.11ac packet can cause a complete collision with the DSRC's CCH being 10 MHz wide.}
\end{remark}

\subsection{From C-V2X}
Compared to Wi-Fi, operation principles of the C-V2X have greater difference from the DSRC. The key difference is that a C-V2X system is ``synchronized'' while a DSRC is not.

A C-V2X system can operate in two different modes: mode 3 with support from the infrastructure and mode 4 without the infrastructure. This paper assumes mode 4 since mode 3 has no specified resource management algorithm available by 3GPP. Also, mode 4 can be assumed to have less delicate mechanisms to avoid inter-RAT interference with DSRC than mode 3 that is scheduled by base stations. Also, we assume 10 packets per second (10 pps), although the 3GPP defines as frequent as 1 pps. At 10 pps, a C-V2X subframe occupies 100 msec, matching to one beaconing period of DSRC: notice that this paper assumes a 100\% duty cycle for the CCH \cite{ieee1609_4}.

\begin{remark}\label{remark_conflict_cv2x}
(Time- and frequency-conflict of a DSRC beaconing period with a C-V2X subframe). \textit{Figure \ref{fig_external_interference_cv2x} shows how a DSRC's beaconing period (1 msec and 10 MHz) conflicts with a C-V2X subframe. A C-V2X network allocates the resource both in terms of time and frequency: the smallest scheduling unit is a resource block (RB), which is 180 kHz each and thus narrower than the DSRC's CCH. This makes the calculation of a conflict more complicated than in the external interference from Wi-Fi. Assuming 20 MHz for C-V2X, 100 RBs are allocated per subframe. The DSRC CCH (10 MHz in 5.885-5.895 GHz) matches to 55.56 RBs in C-V2X. Since a vehicle takes 12 RBs, 4.63 nodes collide with a DSRC's BSM. As illustrated in Figure \ref{fig_external_interference_cv2x}, combined in time and frequency, $1 \text{ msec} / \delta_{\text{dsrc}} \times 0.5556$ slots are interfered by one C-V2X subframe.}
\end{remark}

\begin{remark}\label{remark_Pb_cv2x}
(Probability of a busy RB). \textit{As shown in Section \ref{sec_analysis_temporal}, formulation of a networking behavior started from the probability that a slot is busy, $\mathsf{P}_{b}$. Similarly, one needs to know the probability of a busy RB in order to characterize a C-V2X system's scheduling of RB. Although it has recently been modeled as a probability distribution \cite{etri18}, this paper does not adopt it for the following reasons: (i) the model was not provided in a closed form; (ii) the base station-scheduled RBs may not be applied to the mode 4's behavior. Therefore, this paper parameterizes $P_b$ of a C-V2X system from 0 to 1.}
\end{remark}

\begin{table}[t]
\caption{Key system parameters and values}
\centering
\begin{tabular}{| c || l || l |}
\hline
\multicolumn{2}{|c||}{\cellcolor{gray!30}{\textbf{Parameter}}} & \multicolumn{1}{|c|}{\cellcolor{gray!30} \textbf{Value}}\\ \hline\hline
{\cellcolor{gray!20}{Common}} & $D$ & 2 km \\ \hline
{\cellcolor{gray!20}{}} & Slot time & 66.7 $\mu$s\\
{\cellcolor{gray!20}{}} & Broadcast interval of BSM & 100 msec\\
{\cellcolor{gray!20}{}} & $r_{cs}, r_{\text{tx}}$ & 500 m\\
\multirow{-4}{*}{\cellcolor{gray!20}{DSRC}} & $\lambda_{d}$ & \{3, 5, 6, 9, 13, 20, 35, 160, 641, 1257, 2718\} nodes per $\mathsf{A}_{cs} (= \pi r_{cs}^2)$\\ \hline
{\cellcolor{gray!20}{}} & System & IEEE 802.11ac\\
{\cellcolor{gray!20}{}} & Slot time & 9 $\mu$s\\
\multirow{-3}{*}{{\cellcolor{gray!20}{Wi-Fi}}} & $\lambda_{w}$ & 500 m$^{-2}$\\ \hline
{\cellcolor{gray!20}{}} & System & 3GPP Release 14\\
{\cellcolor{gray!20}{}} & Subframe & 1 ms\\
\multirow{-3}{*}{{\cellcolor{gray!20}{C-V2X}}} & $\lambda_{c}$ & 300 m$^{-2}$\\ \hline
\end{tabular}
\label{table_parameters}
\end{table}

\section{Results and Discussions}\label{sec_results}
This section verifies the accuracy of the analysis presented in Section \ref{sec_analysis}, by comparing to Monte Carlo simulations of the network defined in Section \ref{sec_model}. Table \ref{table_parameters} summarizes the key parameters and their values that were used for production of the results.

\subsection{Temporal Analysis--$\mathsf{PDR}$}\label{sec_results_temporal}
Figures \ref{fig_PDR} through \ref{fig_Phn} demonstrate the results of the the temporal analyses provided in Section \ref{sec_analysis_temporal}. Notice, as indicated in Table \ref{table_parameters}, that the three figures commonly have the intensity of DSRC system, $\lambda_{d}$, on the horizontal axis ranged in one node in \{500, 400, 350, 300, 250, 200, 150, 70, 35, 25, 17\}$^{2}$ m$^{2}$, which is translated to \{3, 5, 6, 9, 13, 20, 35, 160, 641, 1257, 2718\} nodes per $\mathsf{A}_{cs} (= \pi r_{cs}^2)$. A wide range of the intensity is selected in order to see the tendency of all the probabilities accurately.

\subsubsection{Probability of a SYNC}
The results of $\mathsf{P}_{\text{sync}}$, are not shown in this section. The reason is that the value is very close to 0 regardless of the parameters because of its definition as given in (\ref{eq_Psync}): the probability that more than one nodes start a transmission in an exactly same slot out of $L_{bcn} = 1500$ slots. DSRC is already designed such that this type of collision can be avoided: the probability that a node transmits in a particular slot, $\tau$, is very small referring to Figure \ref{fig_markov}; whereas the probability that it transmits in any of the $L_{bcn}$ slots is far higher referring to $\mathsf{P}_{\text{start}}$.

Due to ignoring the impact of SYNC as such, referring to its definition as given in (\ref{eq_PDR}), the key type of collision degrading $\mathsf{PDR}$ in a DSRC system is HN.

\begin{figure}
\vspace{-0.2 in}
\centering
\begin{subfigure}[b]{0.45\linewidth}
\centering
\includegraphics[width = \linewidth]{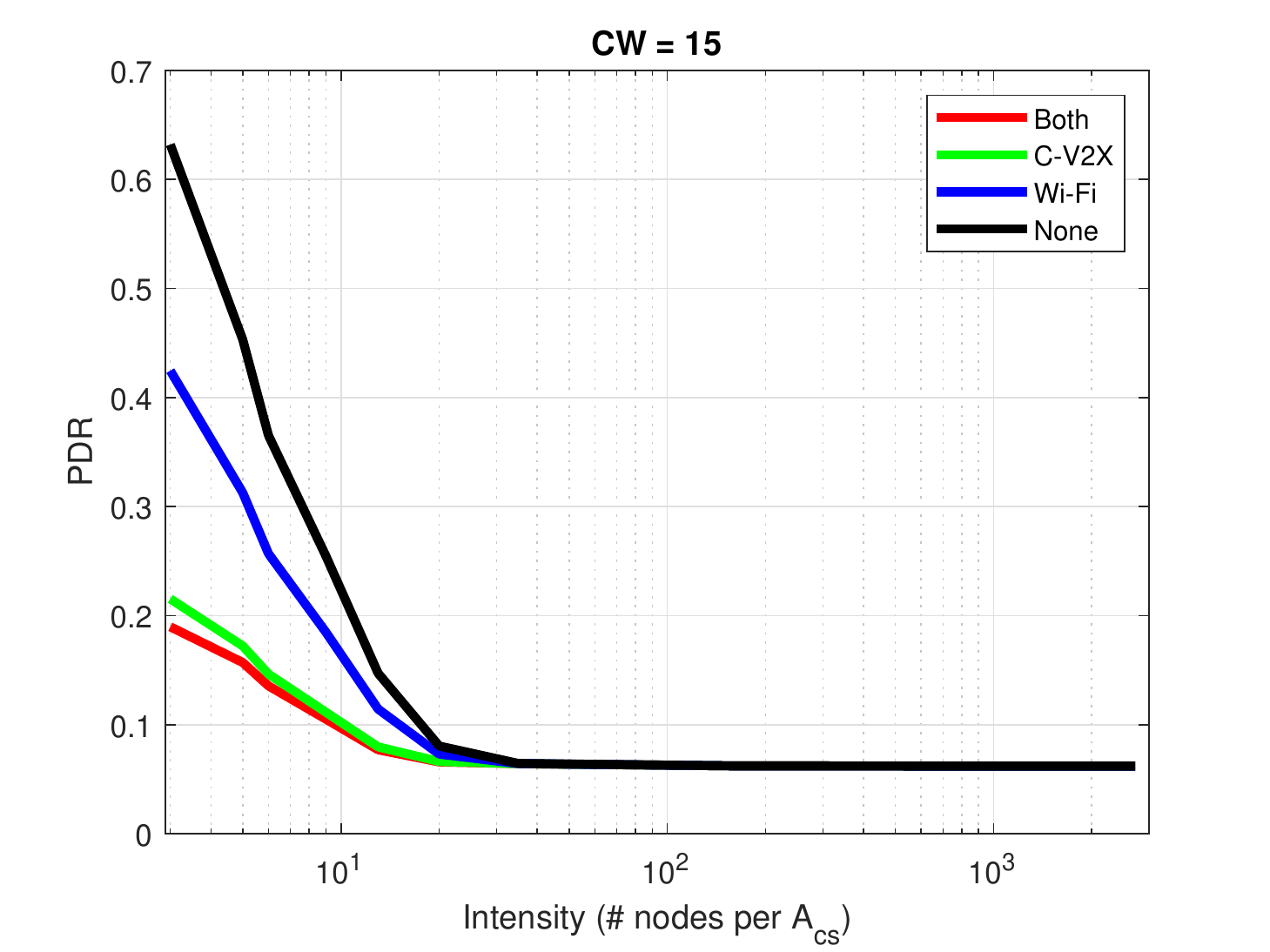}
\caption{With CW = 15}
\label{fig_PDR_CW15}
\end{subfigure}\hfill
\begin{subfigure}[b]{0.45\linewidth}
\centering
\includegraphics[width = \linewidth]{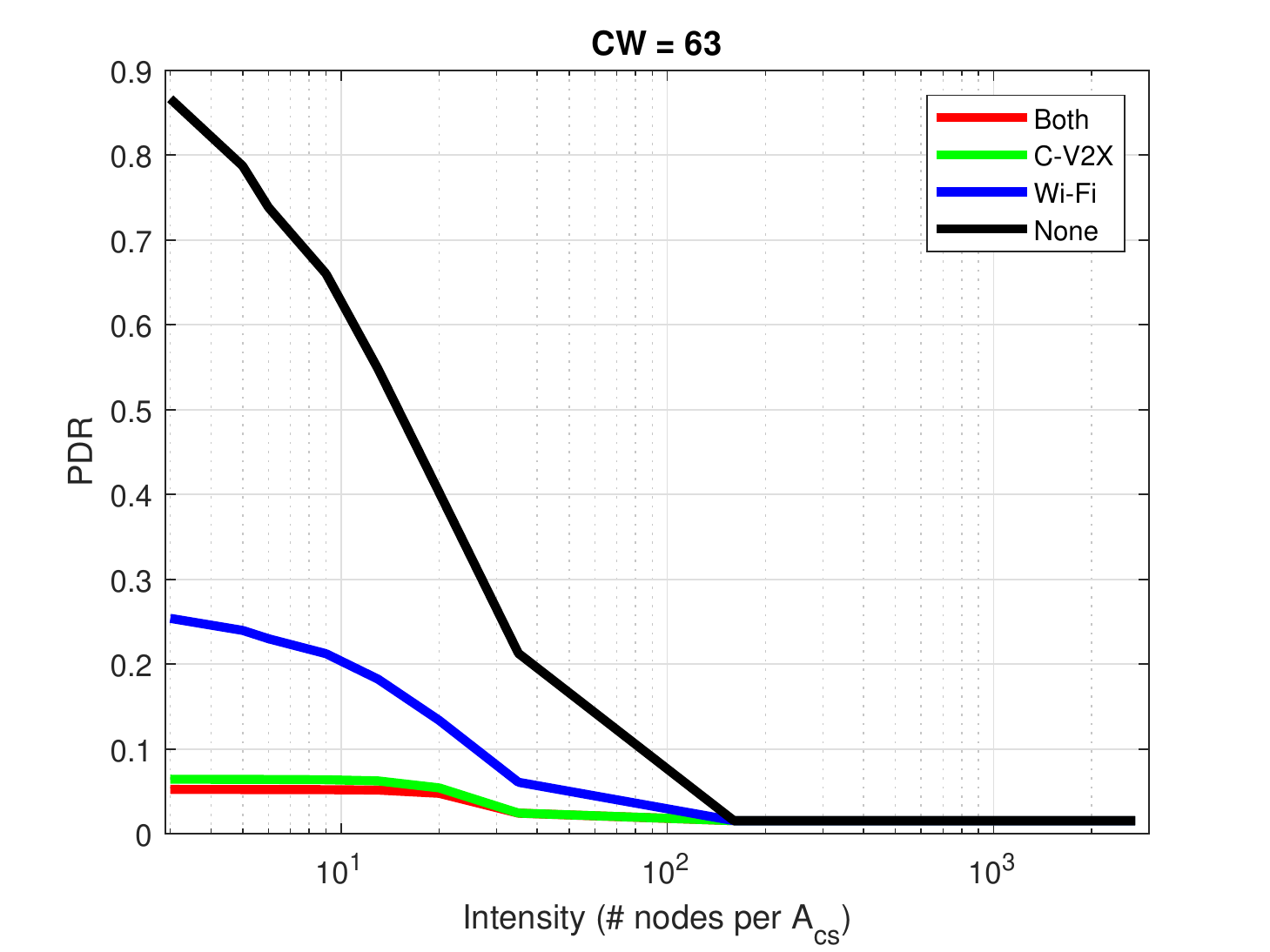}
\caption{With CW = 63}
\label{fig_PDR_CW63}
\end{subfigure}
\begin{subfigure}[b]{0.45\linewidth}
\centering
\includegraphics[width = \linewidth]{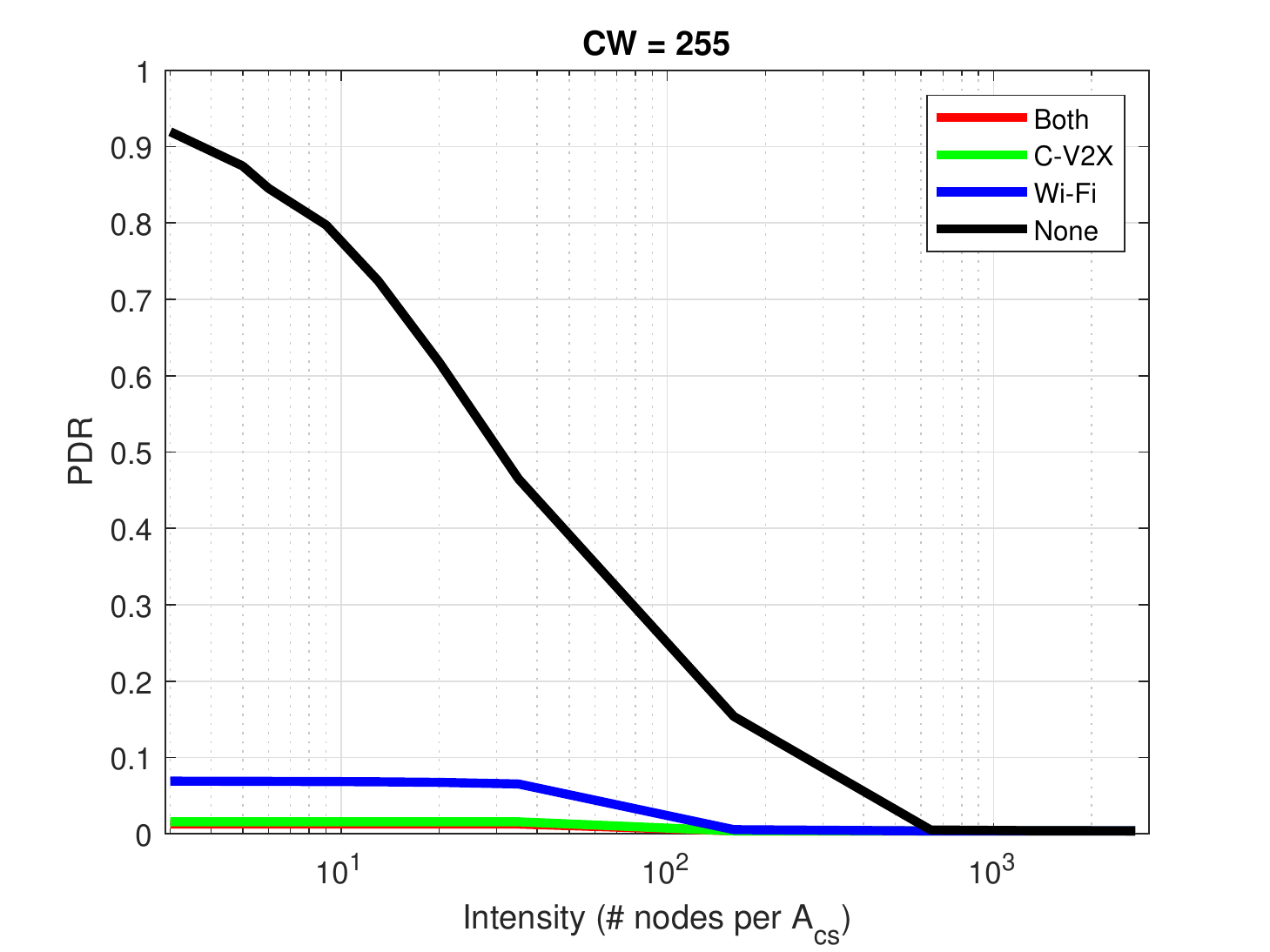}
\caption{With CW = 255}
\label{fig_PDR_CW255}
\end{subfigure}\hfill
\begin{subfigure}[b]{0.45\linewidth}
\centering
\includegraphics[width = \linewidth]{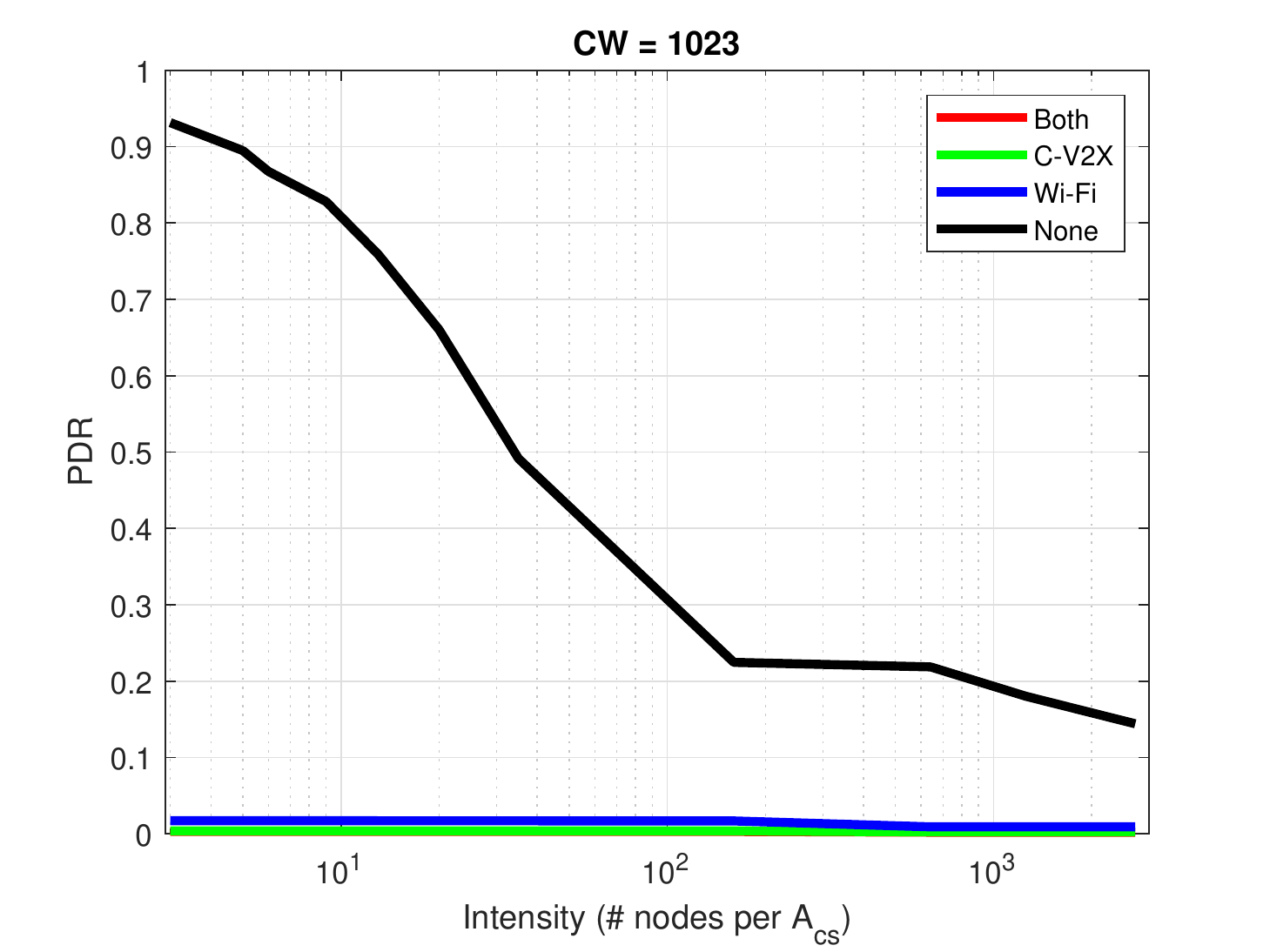}
\caption{With CW = 1023}
\label{fig_PDR_CW1023}
\end{subfigure}
\caption{$\mathsf{PDR}$ versus $\lambda$ according to the type of external interference}
\label{fig_PDR}
\vspace{-0.2 in}
\end{figure}

\begin{figure}
\vspace{-0.2 in}
\centering
\begin{subfigure}[b]{0.45\linewidth}
\centering
\includegraphics[width = \linewidth]{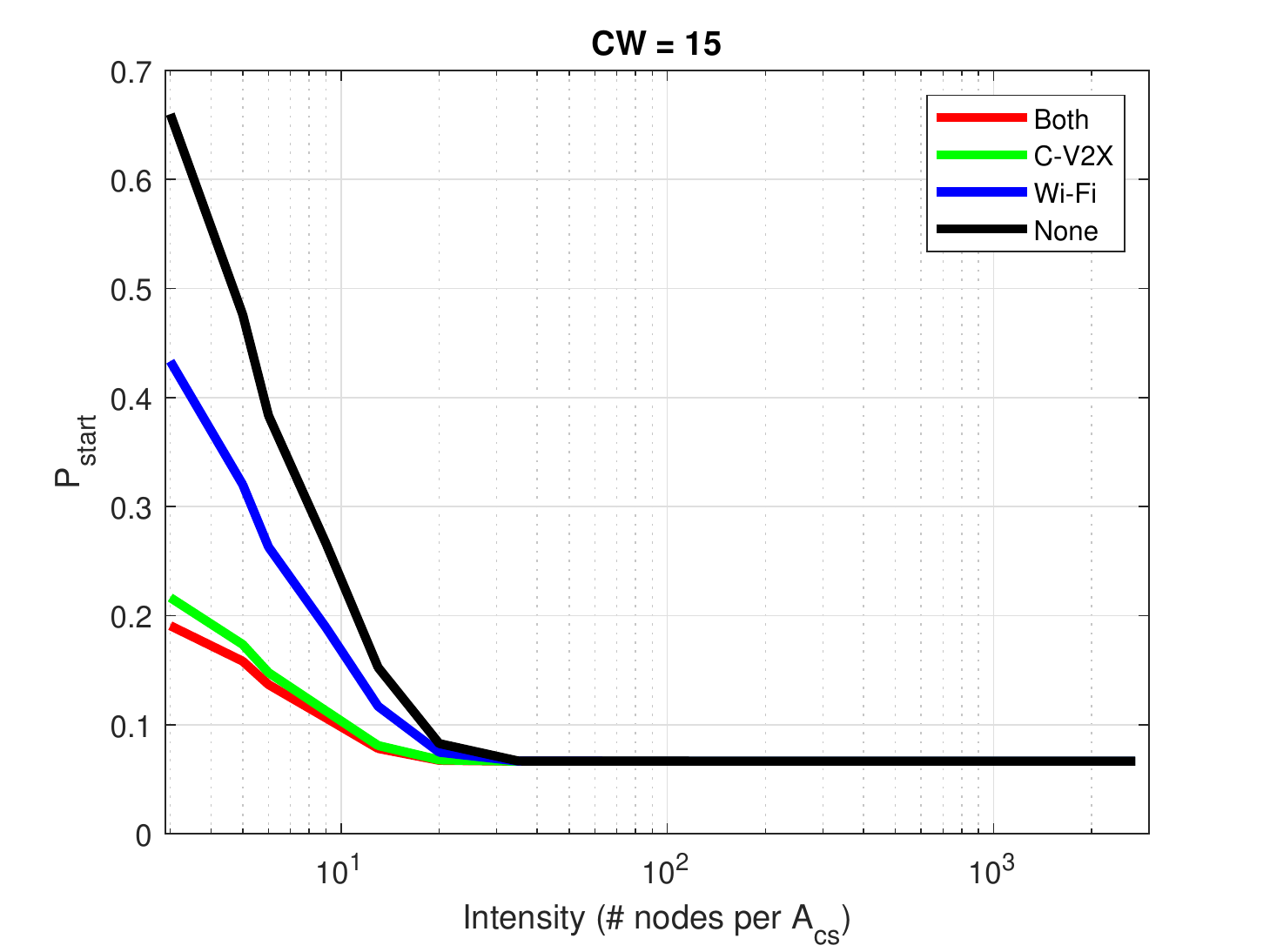}
\caption{With CW = 15}
\label{fig_Pstart_CW15}
\end{subfigure}\hfill
\begin{subfigure}[b]{0.45\linewidth}
\centering
\includegraphics[width = \linewidth]{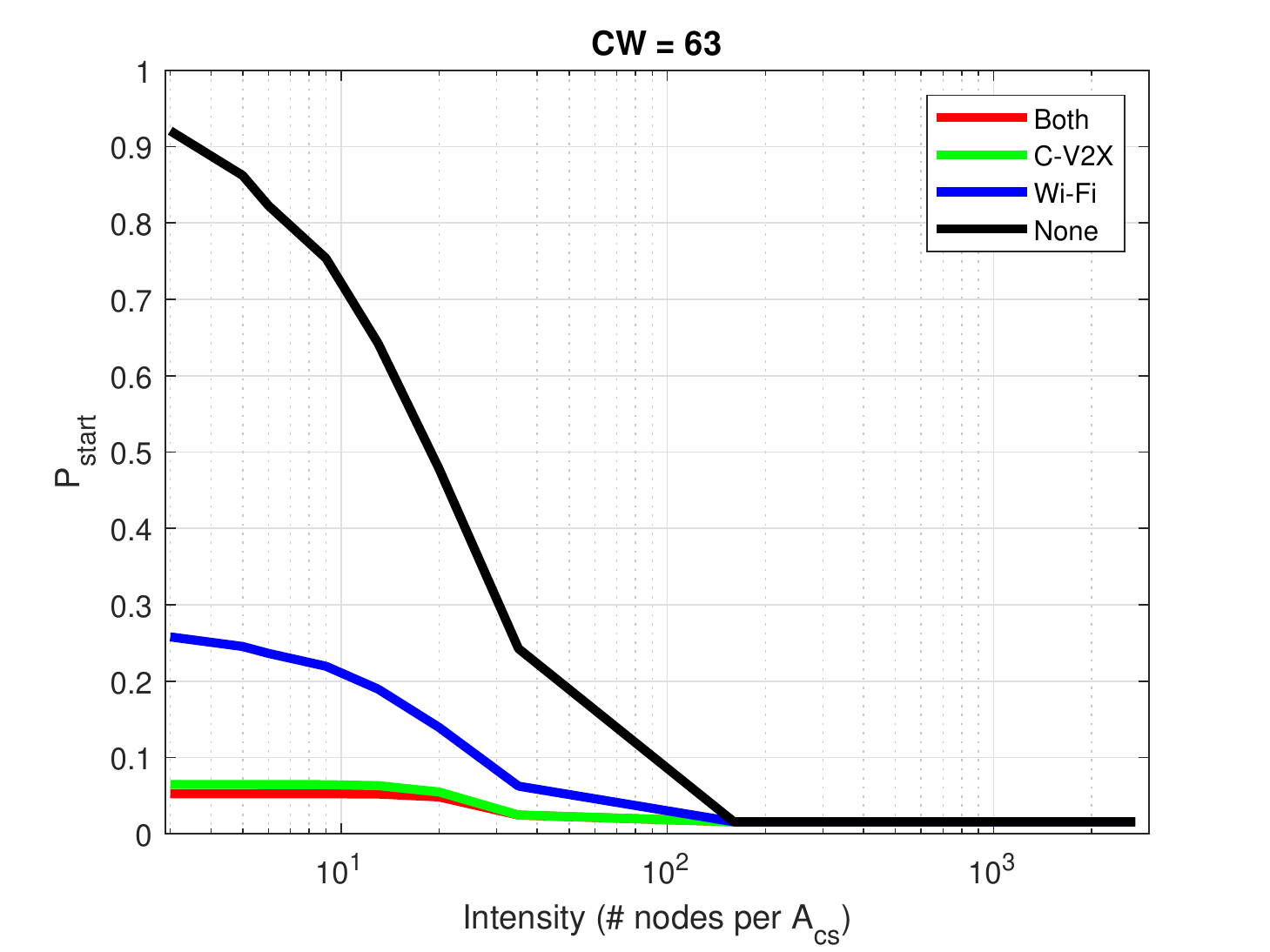}
\caption{With CW = 63}
\label{fig_Pstart_CW63}
\end{subfigure}
\begin{subfigure}[b]{0.45\linewidth}
\centering
\includegraphics[width = \linewidth]{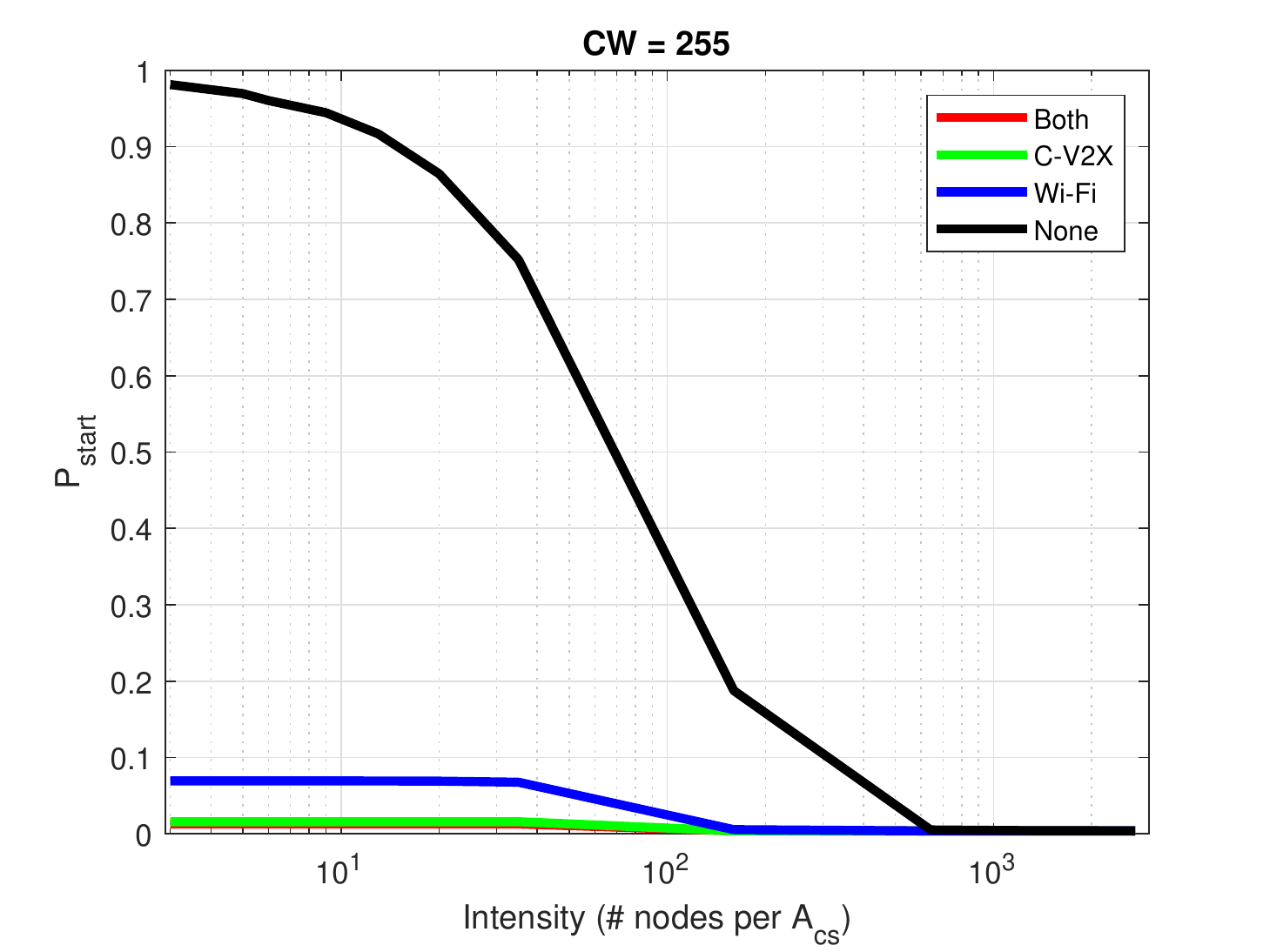}
\caption{With CW = 255}
\label{fig_Pstart_CW255}
\end{subfigure}\hfill
\begin{subfigure}[b]{0.45\linewidth}
\centering
\includegraphics[width = \linewidth]{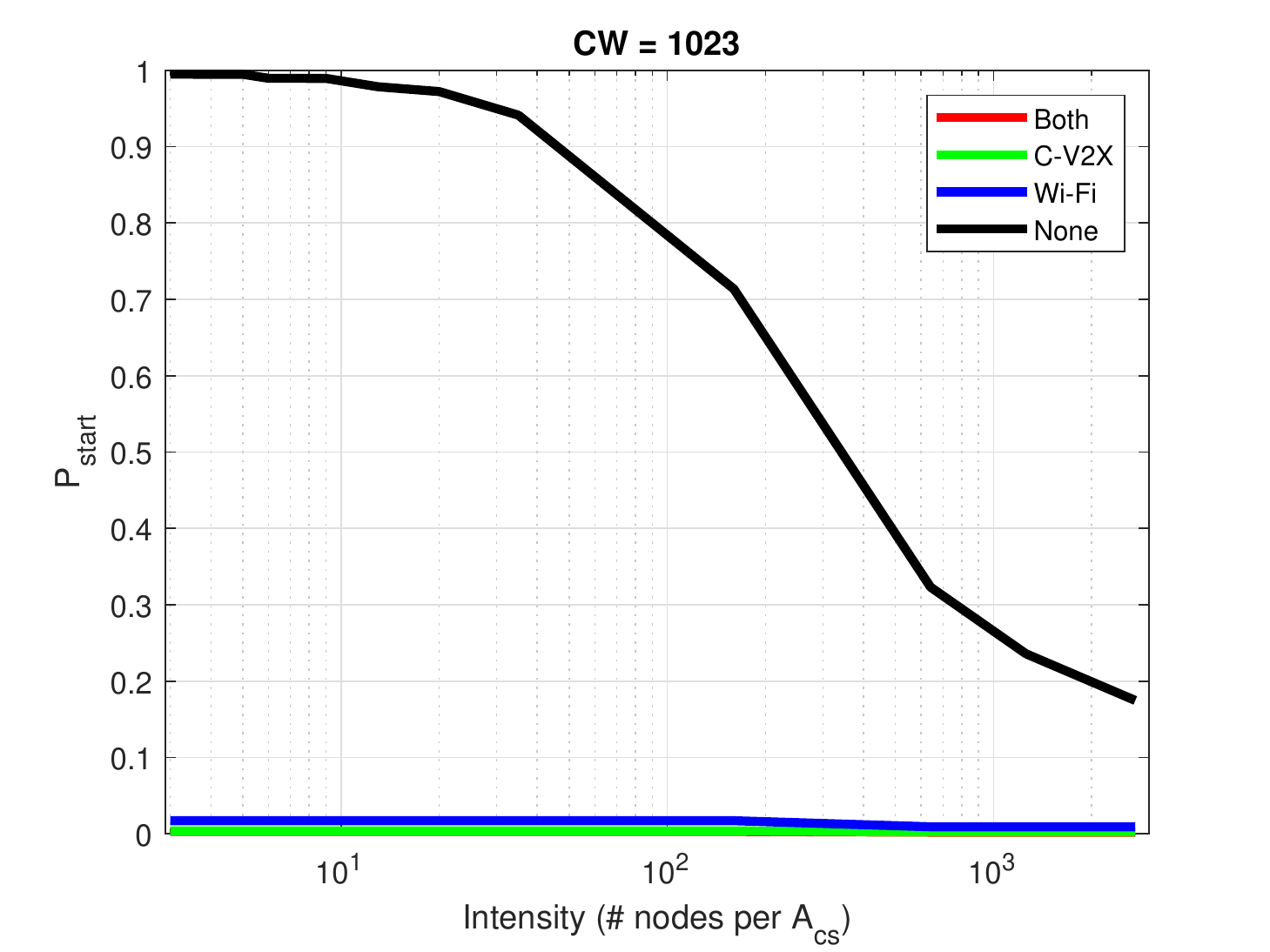}
\caption{With CW = 1023}
\label{fig_Pstart_CW1023}
\end{subfigure}
\caption{$\mathsf{P}_{\text{start}}$ versus $\lambda$ according to the type of external interference}
\label{fig_Pstart}
\vspace{-0.2 in}
\end{figure}

\begin{figure}
\vspace{-0.2 in}
\centering
\begin{subfigure}[b]{0.45\linewidth}
\centering
\includegraphics[width = \linewidth]{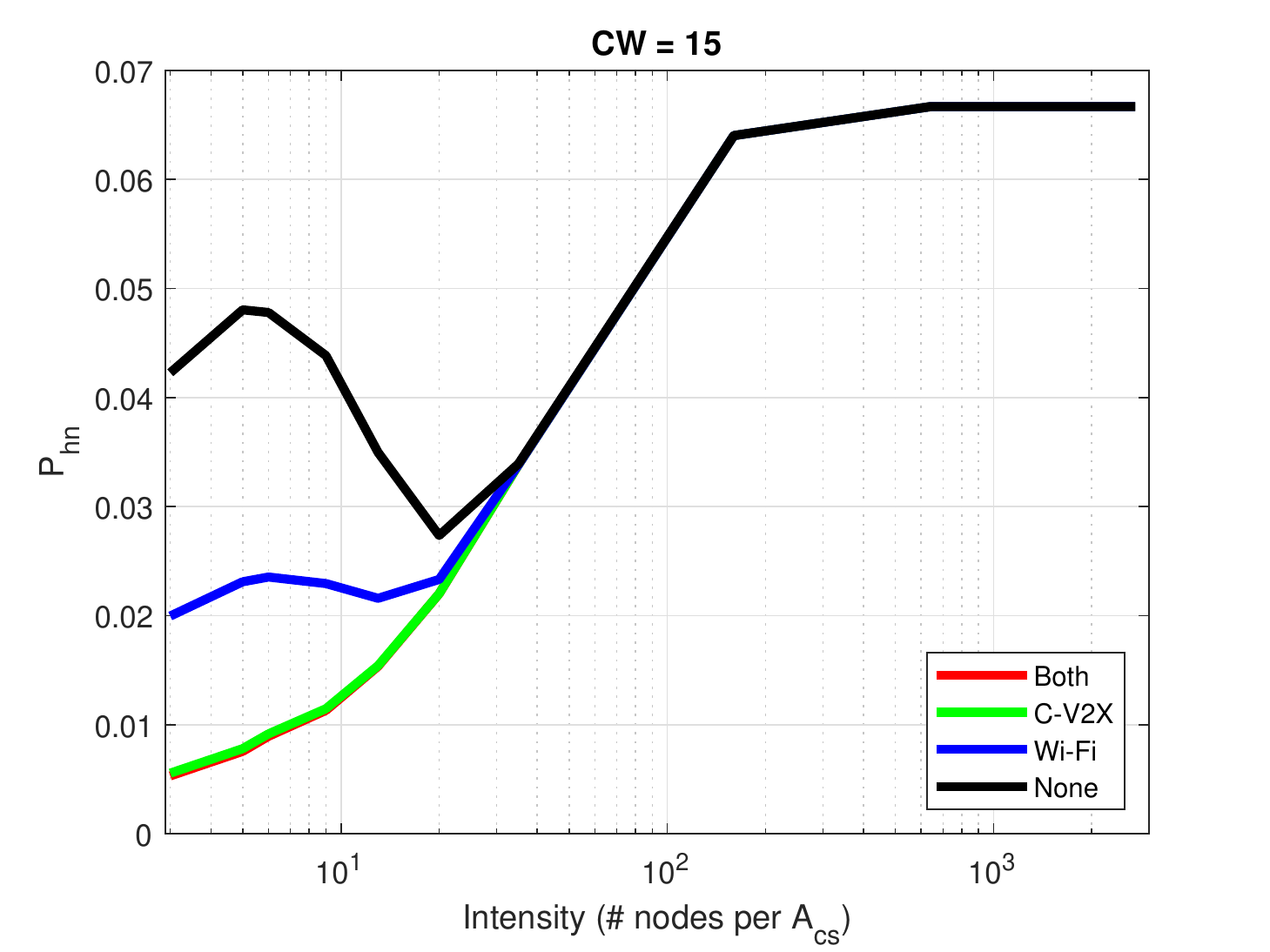}
\caption{With CW = 15}
\label{fig_Phn_CW15}
\end{subfigure}\hfill
\begin{subfigure}[b]{0.45\linewidth}
\centering
\includegraphics[width = \linewidth]{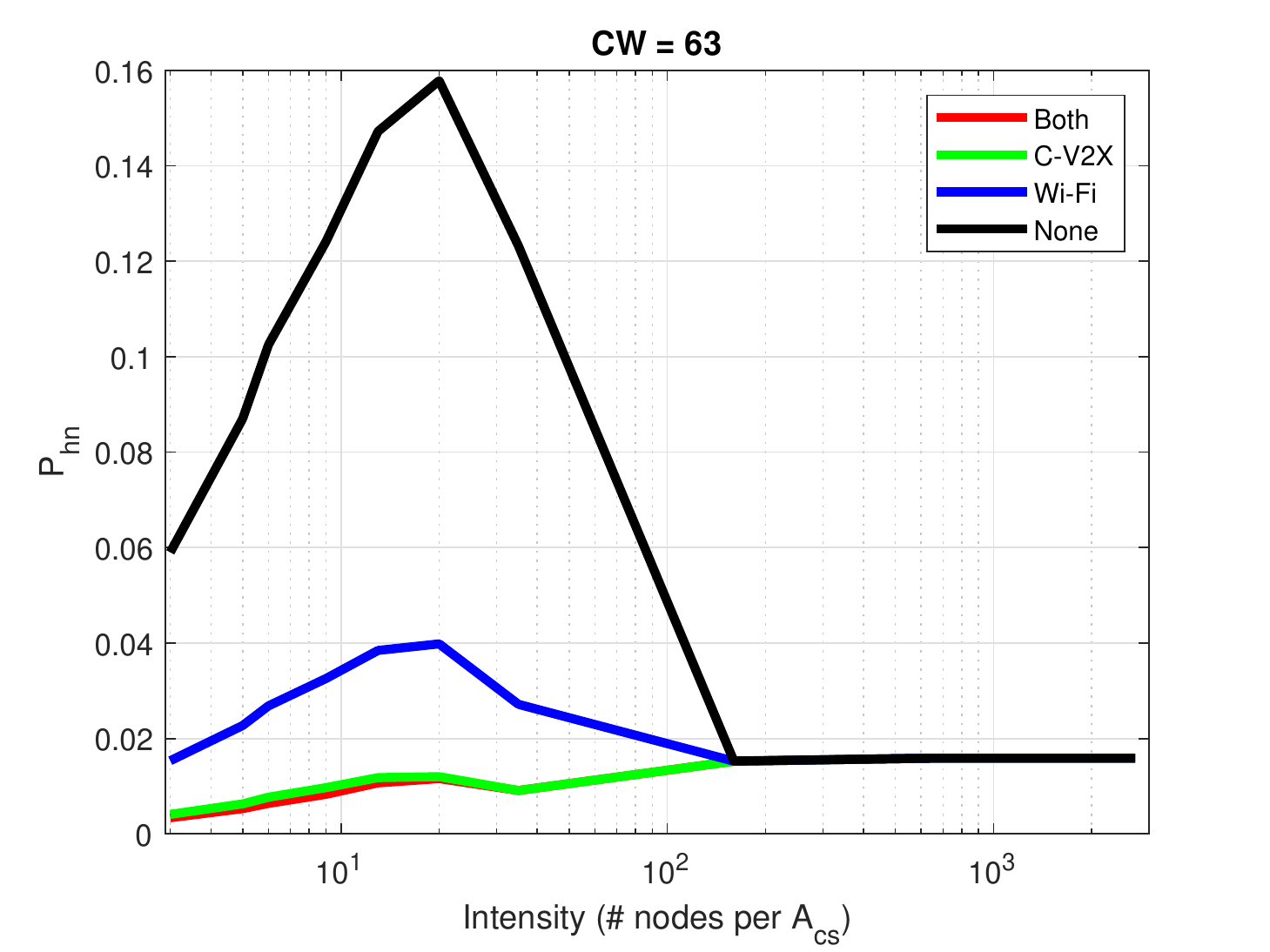}
\caption{With CW = 63}
\label{fig_Phn_CW63}
\end{subfigure}
\begin{subfigure}[b]{0.45\linewidth}
\centering
\includegraphics[width = \linewidth]{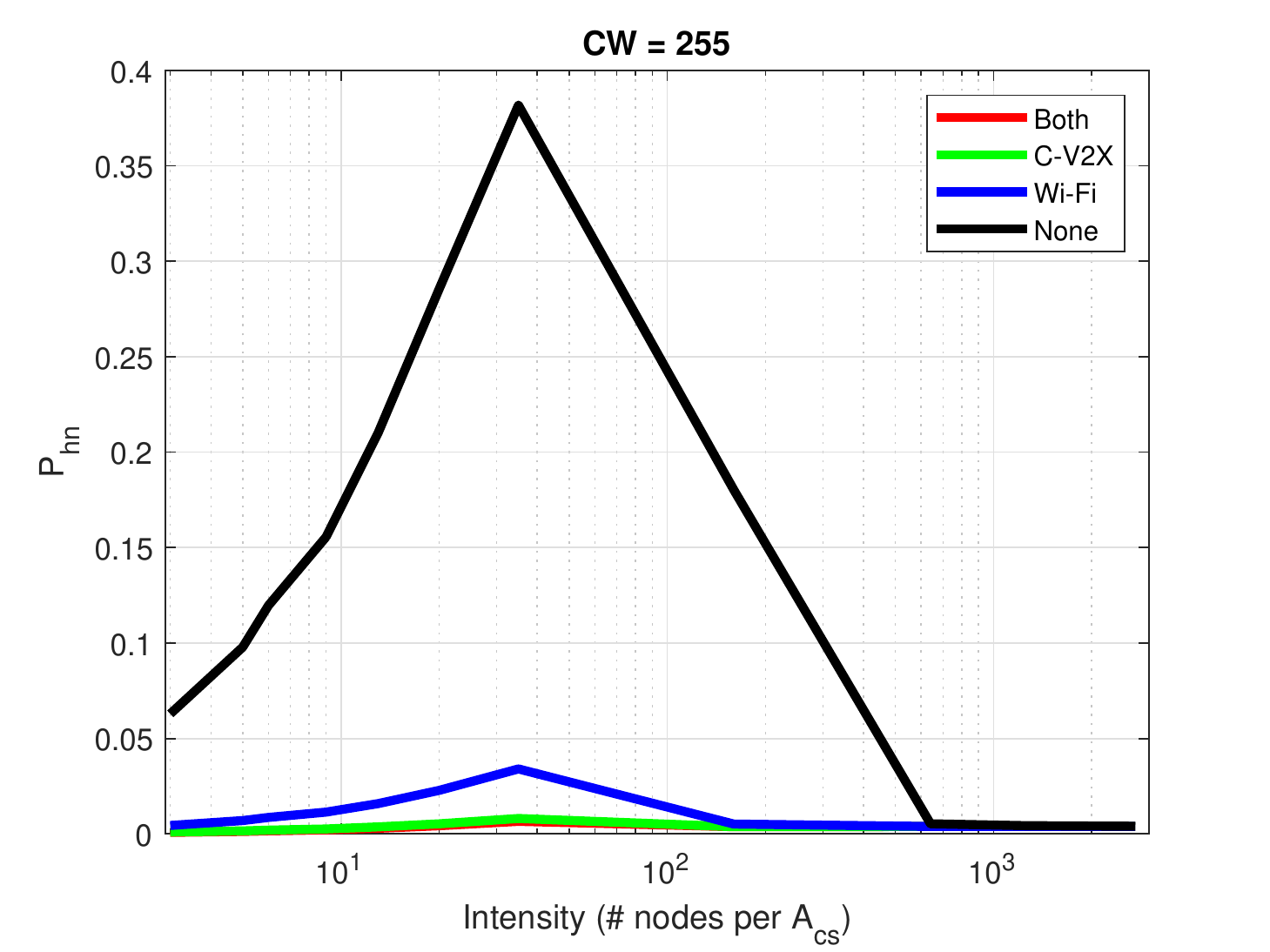}
\caption{With CW = 255}
\label{fig_Phn_CW255}
\end{subfigure}\hfill
\begin{subfigure}[b]{0.45\linewidth}
\centering
\includegraphics[width = \linewidth]{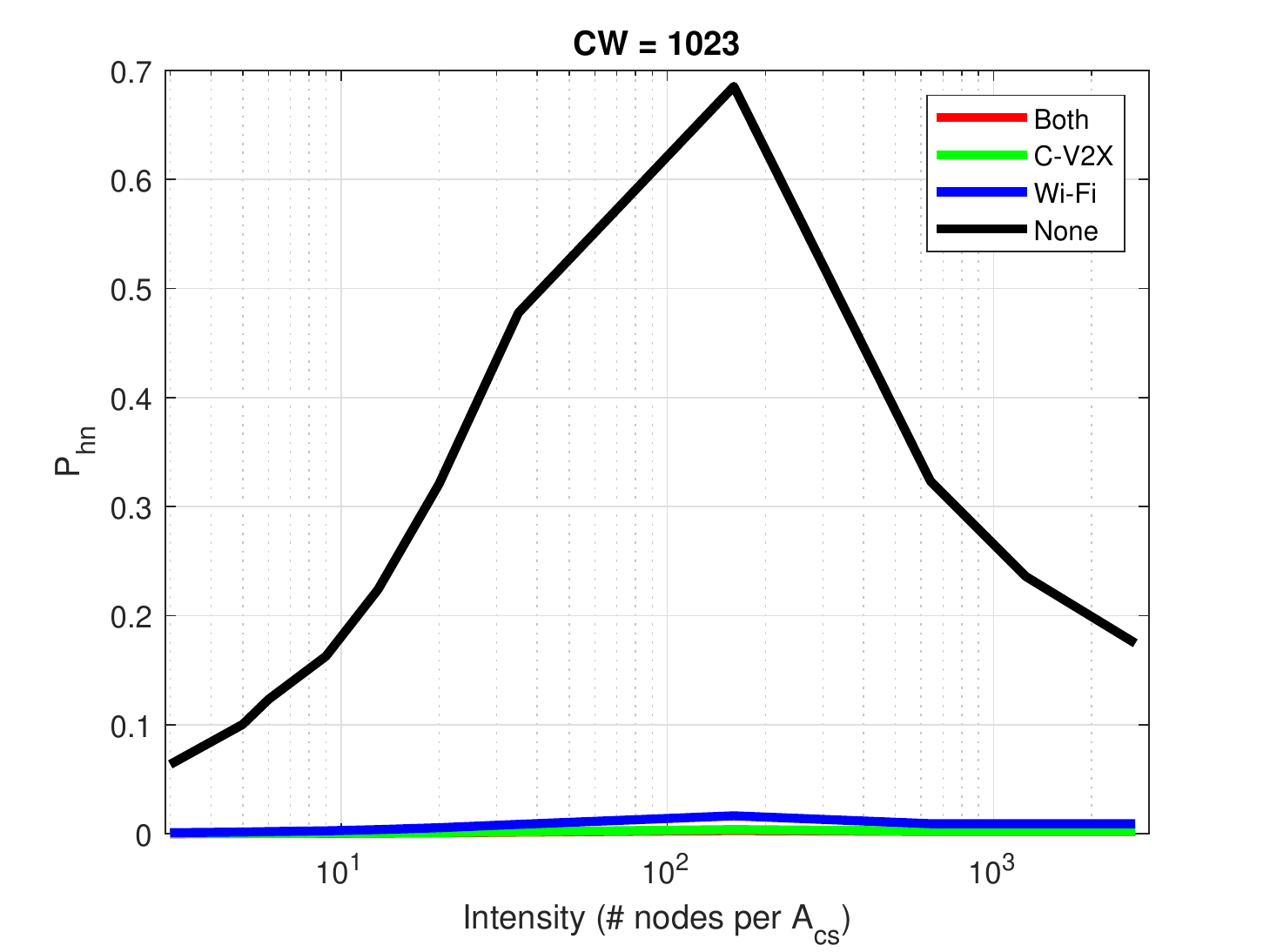}
\caption{With CW = 1023}
\label{fig_Phn_CW1023}
\end{subfigure}
\caption{$\mathsf{P}_{\text{hn}}$ versus $\lambda$ according to the type of external interference}
\label{fig_Phn}
\vspace{-0.2 in}
\end{figure}

\subsubsection{$\mathsf{PDR}$, $\mathsf{P}_{\text{start}}$, and $\mathsf{P}_{\text{hn}}$}
Figure \ref{fig_PDR} plots the $\mathsf{PDR}$, which is given in (\ref{eq_PDR}), versus $\lambda$ according to the type of external interference. Notice that ``none'' indicates no external interference and thus presence of internal interference among DSRC nodes only.

\begin{remark}\label{remark_myopic}
The results provided in Figure \ref{fig_PDR} is the most myopic and zoomed-in view, in order to be the most accurate on (i) the broadcast performance of a DSRC network and (ii) the impact of external interference. Related to it, notice the following remarks:
\begin{itemize}
\item The result regards only one BSM within a beaconing period of 100 msec. It means that over a longer period of time, other BSMs may be successfully received. Also, depending on the applications, exact number of correctly received BSMs should vary.
\end{itemize}
\end{remark}

\begin{remark}\label{remark_latency_requirement}
This analysis framework measuring within every beaconing period becomes more useful, considering the latency requirement according to safety-critical application \cite{daesik17}. The latency requirement is 100 msec at most: any missed delivery of a BSM may incur a serious malfunction of a safety-critical application. Hence, the $\mathsf{PDR}$s provided in Figure \ref{fig_PDR} gets more significant.
\end{remark}

Several key points to discuss are found:
\begin{itemize}
\item A higher $\mathsf{PDR}$ is shown with a higher CW. It is attributed from the tendency shown in Figure \ref{fig_Pstart}: $\mathsf{P}_{\text{start}}$ is increased with a higher CW. It is due to the fact that $\mathsf{P}_{b}$ decreases significantly as CW increases when there is no external interference.
\item However, this tendency versus CW is inverted in presence of external interference. It also is due to the same tendency found in $\mathsf{P}_{\text{start}}$, which is because $\mathsf{P}_{b}$ is kept almost the same as CW increases with no external interference. Referring to (\ref{eq_Pstart}) and the associated Remark , $\mathsf{P}_{\text{start}}$ is decreased as with a greater CW, with a fixed $\mathsf{P}_{b}$.
\item As shown in Figure \ref{fig_PDR_CW15}, the impact of interference from C-V2X is greater than that from Wi-Fi. It is due to the relative impacts as shown in Figure \ref{fig_external_interference}: a subframe of a C-V2X system is far longer compared to a Wi-Fi slot (\textit{i.e.}, 1 msec for a C-V2X subframe versus 9 $\mu$sec for an IEEE 802.11ac time slot \cite{mag14}). Hence, a C-V2X subframe causes a higher interference to DSRC than an IEEE 802.11ac does. Moreover, this also explains the reason that the amount of external interference is not very different between scenarios of (i) only C-V2X causes interference and (ii) both C-V2X and Wi-Fi cause interference.
\item The $\mathsf{PDR}$ is observed to be lower than practical intuition. The reason is congestion in the air interface, which is again attributed to DSRC's relatively long range (at least 300 m) but relatively narrow bandwidth (10 MHz) \cite{rcs17}; in fact, the results in this section are produced with $r_{cs} =$ 500 m.
\end{itemize}

\begin{figure}
\vspace{-0.2 in}
\centering
\begin{subfigure}[b]{0.45\linewidth}
\centering
\includegraphics[width = \linewidth]{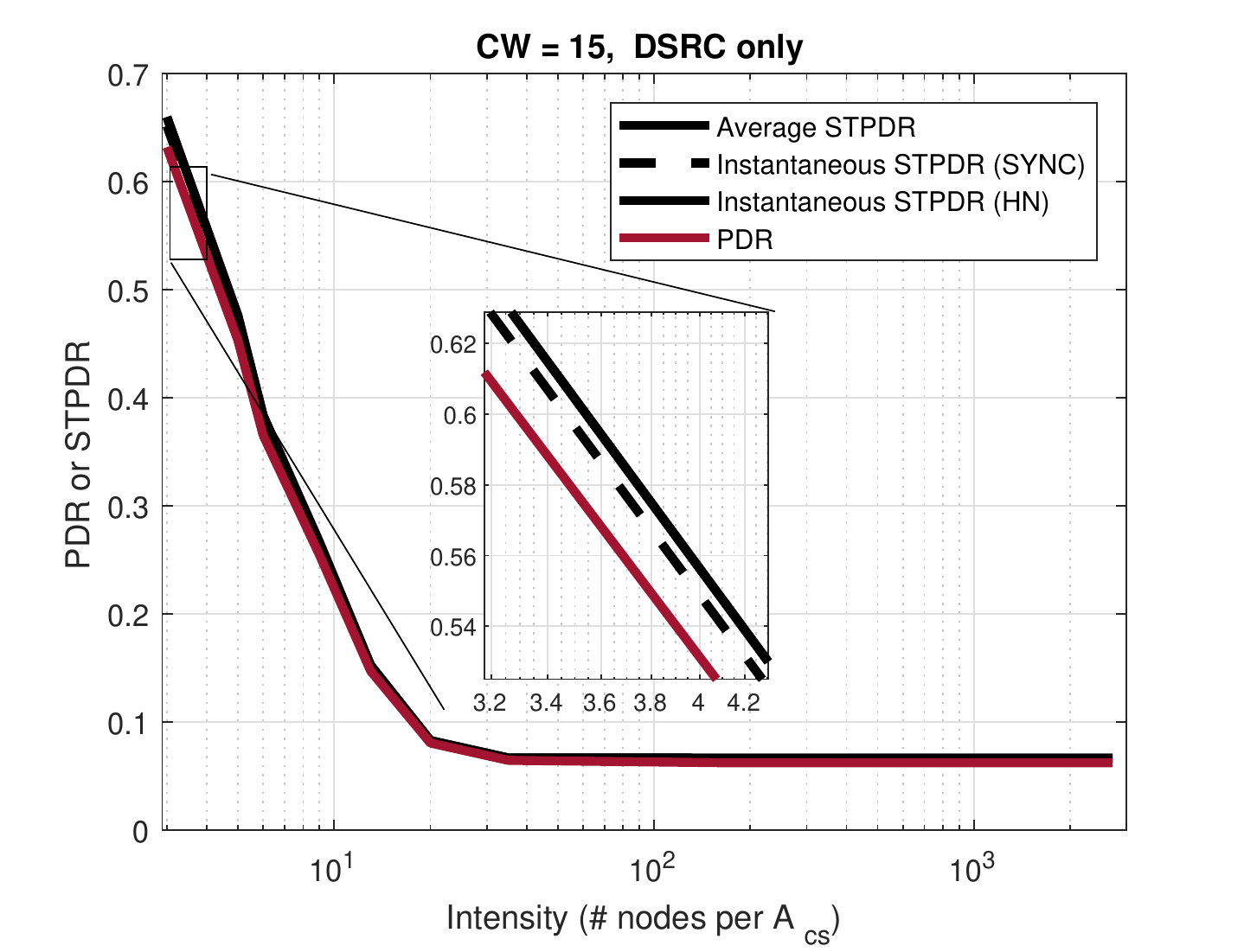}
\caption{With CW = 15}
\label{fig_STPDR_CW15}
\end{subfigure}\hfill
\begin{subfigure}[b]{0.45\linewidth}
\centering
\includegraphics[width = \linewidth]{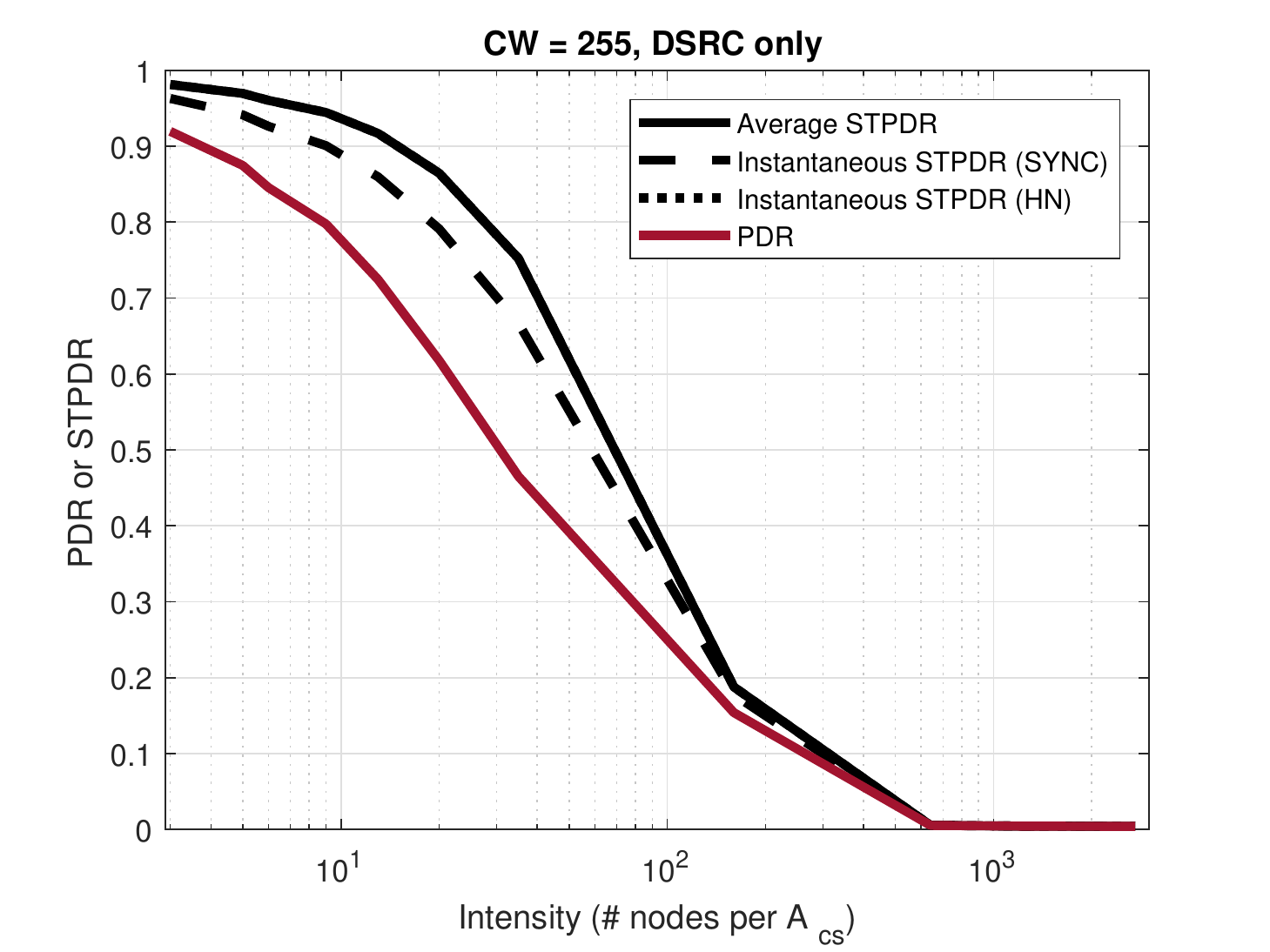}
\caption{With CW = 255}
\label{fig_STPDR_CW255}
\end{subfigure}
\vspace{-0.1 in}
\caption{$\mathsf{STPDR}$}
\label{fig_STPDR}
\end{figure}
\begin{figure}
\centering
\begin{subfigure}[b]{0.45\linewidth}
\centering
\includegraphics[width = \linewidth]{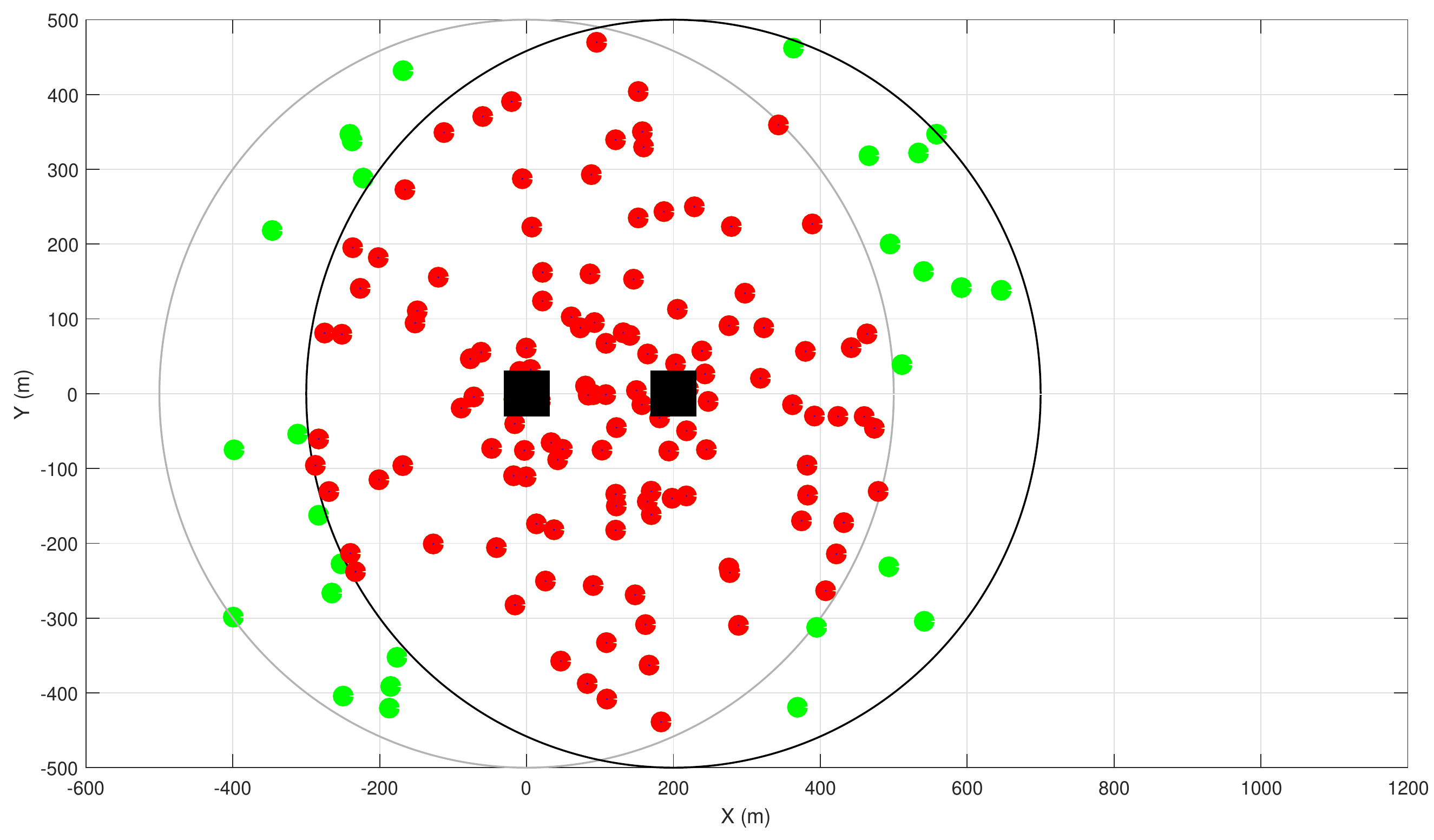}
\caption{SYNC ($\mathsf{vC}$ within $\mathsf{vT}$'s $r_{cs}$)}
\label{fig_STPDR_snapshot_200m_apart}
\end{subfigure}\hfill
\begin{subfigure}[b]{0.45\linewidth}
\centering
\includegraphics[width = \linewidth]{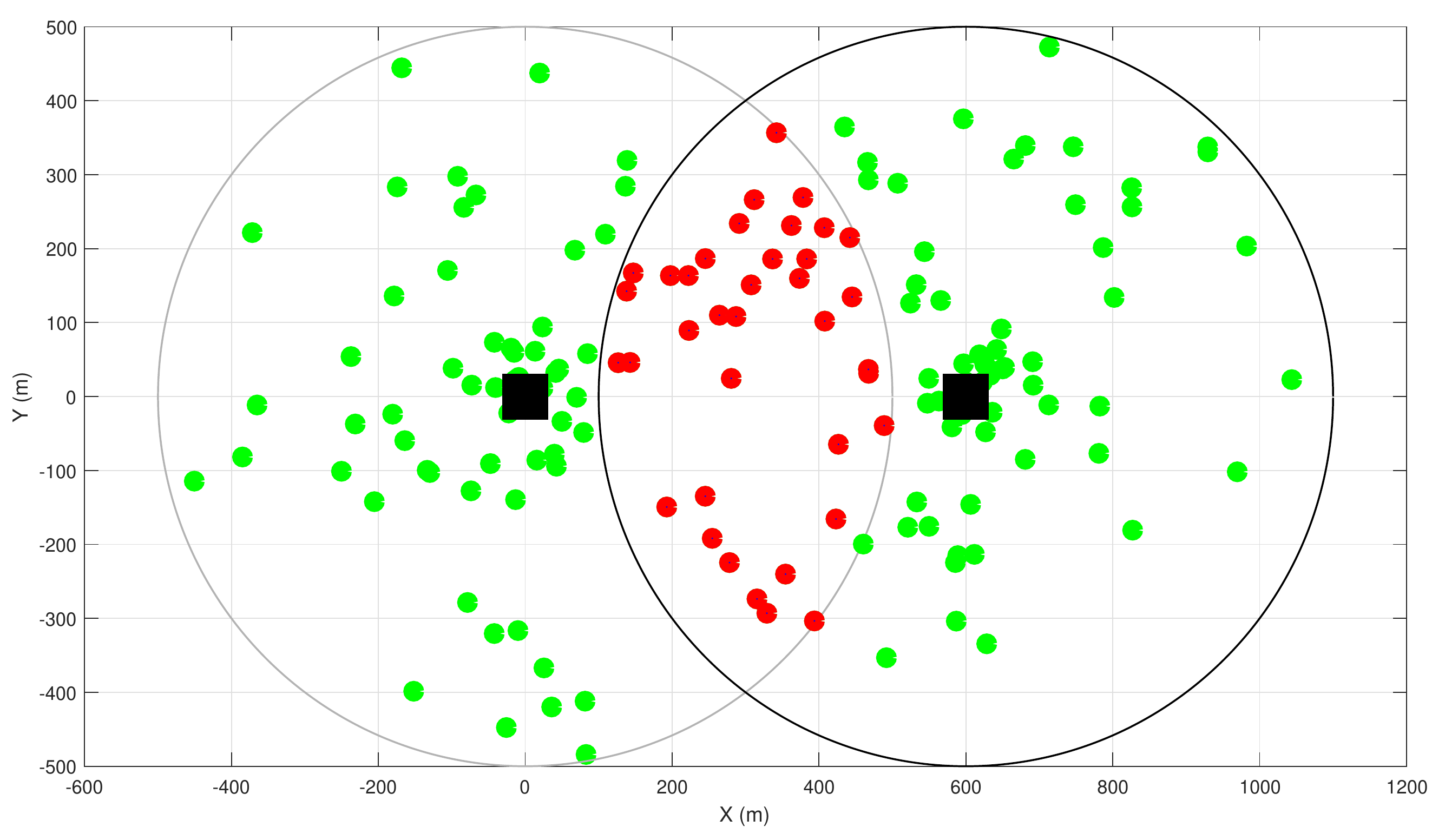}
\caption{HN ($\mathsf{vC}$ out of $\mathsf{vT}$'s $r_{cs}$)}
\label{fig_STPDR_snapshot_600m_apart}
\end{subfigure}
\vspace{-0.1 in}
\caption{A snapshot of $\mathsf{A}_{\text{col}}$ according to distance between $\mathsf{vT}$ and $\mathsf{vC}$}
\label{fig_STPDR_snapshot}
\vspace{-0.2 in}
\end{figure}

\subsection{Spatial Impact--$\mathsf{STPDR}$}\label{sec_results_spatial}
Figure \ref{fig_STPDR} demonstrates the results of $\mathsf{STPDR}$. Notice that not all the cases of interference and CWs since a similar tendency is shown for the $\mathsf{PDR}$ in Figure \ref{fig_PDR}. The similarity is attributed from the fact that they both are functions of $\mathsf{P}_{\text{start}}$. However, the $\mathsf{STPDR}$ shows a critical difference: it is a ``softer'' version of $\mathsf{PDR}$ due to the discounts on $\mathsf{P}_{\text{sync}}$ and $\mathsf{P}_{\text{hn}}$ by $\mathsf{RGB}_{\text{sync}}$ and $\mathsf{RGB}_{\text{hn}}$, respectively. As already discussed in Section \ref{sec_results_temporal}, similar to $\mathsf{PDR}$, a $\mathsf{STPDR}$ is dominantly determined by $\mathsf{P}_{\text{start}}$ and $\mathsf{P}_{\text{hn}}$ since $\mathsf{P}_{\text{sync}} \approx 0$.

As observed from comparison of Figures \ref{fig_STPDR_CW15} and \ref{fig_STPDR_CW255}, the spatial impact gets greater with a larger value of CW. Instantaneous $\mathsf{STPDR}$s from SYNC and HN are produced from $\mathsf{vT}$-$\mathsf{vC}$ separation distance of 200 and 600 m, respectively, whose snapshots are provided in Figures \ref{fig_STPDR_snapshot_200m_apart} and \ref{fig_STPDR_snapshot_600m_apart}. As shown in Figure \ref{fig_STPDR}, $\mathsf{STPDR}$ presents a higher value reflecting the ``softer'' calculation. To remind, the $\mathsf{STPDR}$ considers the possibility of BSM reception by the receiver nodes in the unaffected area, which are expressed as green dots in Figure \ref{fig_STPDR_snapshot}.

Figure \ref{fig_STPDR_snapshot} shows a snapshot of a situation where an $\mathsf{STPDR}$ is calculated. Green dots indicate the nodes that can receive a BSM from $\mathsf{vT}$ without collision, and red dots indicate those cannot receive it due to collision from $\mathsf{vC}$.

\begin{remark}\label{remark_RGB_advantage}
(The key advantage of $\mathsf{STPDR}$). \textit{As highlighted in the snapshot shown in Figures \ref{fig_STPDR_snapshot}, the advantage of $\mathsf{STPDR}$ over $\mathsf{PDR}$ is the ability to capture the spatial ratio of a BSM reception, which the $\mathsf{PDR}$ neglects. The $\mathsf{STPDR}$ takes into account the $\mathsf{RGB}$, which we had coined \cite{globecom18} and illustrated in Figure \ref{fig_rgb_definition}.} The $\mathsf{RGB}_{\text{sync}}$ and $\mathsf{RGB}_{\text{hn}}$ averaged over all the possible values of the link length, $\mathsf{l}$, are found to be 0.0576 and 0.0742, respectively. This suggests that $\mathsf{STPDR}$ does not derail extremely from the $\mathsf{PDR}$, which provides the backward compatibility while being more accurate by comprehending both spatial and temporal factors.
\end{remark}

\section{Conclusions}
This paper has presented an analysis framework that models the intertwined impacts between the temporal and spatial factors. It also proposed a metric that can capture the spatiotemporal impact--namely, $\mathsf{STPDR}$. Based on the metric, the results demonstrated the impacts of the external interference from Wi-Fi and/or C-V2X into DSRC.

Multiple key design insights were drawn from the results: (i) selection of CW has a significant influence on PDR: (i-i) a greater CW increases PDR without presence of external interference; and (i-ii) a smaller CW increases PDR when external interference exists; (ii) C-V2X has far greater influence than Wi-Fi on the performance of DSRC due to its smallest scheduling unit (\textit{i.e.}, subframe) is far longer than a slot of Wi-Fi.

\appendix
\subsection{Proof of Lemma \ref{lemma_Pb}}\label{appendix_Pb}
The probability that a particular slot out of $L_{bcn}$ slots in a beaconing period is occupied by any of the other nodes is defined as \cite{bianchi}
\begin{align}\label{eq_Pb}
\mathsf{P}_{b}\left(\text{CW}\right) &= \mathbb{E}_{n_{cs}} \left[ 1 - \mathbb{P}\left[\text{No other vehicle transmits in the slot}\right] \right]\nonumber\\
&= 1 - \displaystyle \sum_{n_{cs}=0}^{\mathbb{N}\left[\Phi_{cs}\right]} \left[ 1 - \tau\left(\text{CW}\right) \right]^{n} \mathbb{P}\left[ n_{cs} = n \right]
\end{align}
where $\tau$ indicates the probability that a station transmits in a time slot.

The transition matrix for the Markov process given in Figure \ref{fig_markov} can be found as
\begin{align}
\begin{bmatrix}
1/\text{CW} &1/\text{CW} &1/\text{CW} & &\cdots & &1/\text{CW}\\
\mathbb{P}\left(b_{1} \rightarrow b_{0}\right) &0 &0 & &\cdots & &0\\
0 &\mathbb{P}\left(b_{2} \rightarrow b_{1}\right) &0 & &\cdots & &0\\
\vdots &\vdots &\vdots & \ddots &\cdots & &\vdots\\
0 &0 &0 &\cdots & \mathbb{P}\left(b_{k} \rightarrow b_{k-1}\right) & \cdots &0\\
\vdots &\vdots &\vdots & &\cdots & \ddots &\vdots\\
0 &0 &0 & &\cdots & &\mathbb{P}\left(b_{\text{CW}-1} \rightarrow b_{\text{CW}-2}\right)
\end{bmatrix}
\end{align}
where the probability of a one-step transition can be obtained in a general form of
\begin{align}
\mathbb{P}\left(b_{k} \rightarrow b_{k-1}\right) &= \underbrace{\left(1 - \mathsf{P}_{b}\right)}_{\text{Directly } b_{k} \rightarrow b_{k-1}} + \underbrace{\mathsf{P}_{b}\left(1 - \mathsf{P}_{b}\right) + \mathsf{P}_{b}^{2} \left(1 - \mathsf{P}_{b}\right) + \cdots + \mathsf{P}_{b}^{\min\left(n_{cs}, L_{bcn}-l_{bcn}-k\right)} \left(1 - \mathsf{P}_{b}\right)}_{\text{Via } D_{k}}\nonumber\\
&\stackrel{(a)}{=} \left(1 - \mathsf{P}_{b}\right) \displaystyle \sum_{m=0}^{\min\left(n_{cs}, L_{bcn}-l_{bcn}-k\right)} \left(\mathsf{P}_{b}\right)^{m}\nonumber\\
&\stackrel{(b)}{\approx} \left(1 - \mathsf{P}_{b}\right) \frac{1}{1 - \mathsf{P}_{b}} = 1.
\end{align}
In (a), notice that quantity $\min\left(n_{cs}, L_{bcn}-l_{bcn}-k\right)$ has already been discussed in the proof of Lemma \ref{lemma_Pstart}. Also, from (b), we assure that the transition matrix is valid since it yields each row to be a 1.

Also, it is noteworthy that the slot busy probability, $\mathsf{P}_{b}$, is determined by not only other competing nodes in DSRC but those in C-V2X and Wi-Fi as well. This is formally written as
\begin{align}\label{eq_Pb_external}
\mathsf{P}_{b} = \min \big( \mathsf{P}_{b, \text{dsrc}} + \mathsf{P}_{b, \text{cv2x}} + \mathsf{P}_{b, \text{wifi}}, 1 \big).
\end{align}
The formulation follows from the fact that ``either'' of the three RATs can cause interference to a general node's transmission. It is intuitive in the sense that as the external interference gets greater, the value of $\mathsf{P}_b$ is increased; as a direct consequence, the state propagation in the Markov chain becomes less likely, which in turn incurs a smaller $\mathsf{P}_{\text{start}}$. (See Figure \ref{fig_markov}.) A sum being greater than 1 means that the node has no chance to transmit already, which is equivalent to $\mathsf{P}_{b} = 1$.

The following steady-state relationship for the Markov chain can be formulated as
\begin{align}\label{eq_b0}
b_{0} &= b_{k}\frac{1}{\text{CW}} \prod_{i=1}^{k} \mathbb{P}\left(b_{k} \rightarrow b_{k-1}\right)\nonumber\\
&= b_{k}\frac{\left(1 - \mathsf{P}_{b}\right)^{k}}{\text{CW}} \prod_{i=1}^{k} \left( \displaystyle \sum_{m=0}^{\min\left(n_{cs}, L_{bcn}-l_{bcn}-k\right)} \left(\mathsf{P}_{b}\right)^{m} \right).
\end{align}
Then, the normalization condition is formally expressed as
\begin{align}\label{eq_normalization}
1 &= \displaystyle \sum_{k = 0}^{\text{CW}-1} b_{k}\nonumber\\
&= b_{0} \text{CW} \displaystyle \sum_{k = 0}^{\text{CW}-1} \frac{1}{\left(1 - \mathsf{P}_{b}\right)^{k}} \prod_{i=1}^{k} \left( \displaystyle \sum_{m=0}^{\min\left(n_{cs}, L_{bcn}-l_{bcn}-k\right)} \left(\mathsf{P}_{b}\right)^{m} \right)^{-1},
\end{align}
which yields
\begin{align}\label{eq_b0_final}
b_{0} &= \left( \text{CW} \displaystyle \sum_{k = 0}^{\text{CW}-1} \frac{1}{\left(1 - \mathsf{P}_{b}\right)^{k}} \prod_{i=1}^{k} \left( \displaystyle \sum_{m=0}^{\min\left(n_{cs}, L_{bcn}-l_{bcn}-k\right)} \left(\mathsf{P}_{b}\right)^{m} \right)^{-1} \right)^{-1}\nonumber\\
&\stackrel{(a)}{\coloneqq} \tau\left(\text{CW}\right).
\end{align}
Notice that (a) follows from the fact that since $b_{0}$ is the only state that a node is allowed to transmit, it equivalent to the probability that a node transmits in a time slot.

Lastly, we remind that the number of competing nodes, $n_{cs}$, is a normal random variable as shown in Remark \ref{remark_clt} and Figure \ref{fig_n0} in Section \ref{sec_analysis_temporal}.

As a result, the probability that an arbitrary slot is busy can be derived via an algebra plugging (\ref{eq_b0_final}) into (\ref{eq_Pb}) as
\begin{align}\label{eq_Pb_algebra}
&\mathsf{P}_{b}\left(\text{CW}\right)\nonumber\\
&= \mathbb{E}_{n_{cs}} \left[ 1 - \left( 1 - \tau\left(\text{CW}\right) \right)^{n_{cs}} \right]\nonumber\\
&= 1 - \mathbb{E}_{n_{cs}} \left[ \left( 1 - \left( \text{CW} \displaystyle \sum_{k = 0}^{\text{CW}-1} \frac{1}{\left(1 - \mathsf{P}_{b}\right)^{k}} \prod_{i=1}^{k} \left( \displaystyle \sum_{m=0}^{\min\left(n_{cs}, L_{bcn}-l_{bcn}-k\right)} \left(\mathsf{P}_{b}\right)^{m} \right)^{-1} \right)^{-1} \right)^{n_{cs}} \right]\nonumber\\
&= 1 - \displaystyle \sum_{n_{cs}=0}^{\mathbb{N}\left[\Phi_{cs}\right]} \left( 1 - \left( \text{CW} \displaystyle \sum_{k = 0}^{\text{CW}-1} \frac{1}{\left(1 - \mathsf{P}_{b}\right)^{k}} \prod_{i=1}^{k} \left( \displaystyle \sum_{m=0}^{\min\left(n_{cs}, L_{bcn}-l_{bcn}-k\right)} \left(\mathsf{P}_{b}\right)^{m} \right)^{-1} \right)^{-1} \right)^{n} \mathbb{P}\left[ n_{cs} = n \right]
\end{align}
which completes the proof.

\subsection{ Proof of Lemma \ref{lemma_Psync}}\label{appendix_Psync}
The probability that a vehicle $\mathsf{vT}$ experiences a SYNC can be modeled as

\begin{align}\label{eq_Psync_proof}
\mathsf{P}_{\text{sync}}\left(\lambda, \text{CW}\right) &= \mathbb{P} \left( n_{cs} > 0 \right) \nonumber\\
&{\rm{~~~~}}\cdot \mathbb{E}_{n_{cs}} \left[ \mathbb{P} \left( \text{At least one of the } n_{cs} \text{ other nodes transmit} \right) \right]\nonumber\\
&{\rm{~~~~}}\cdot \mathbb{P} \left( \mathsf{vT} \text{ transmits in a given slot} \right).
\end{align}
We can quantify each of the two terms as follows.

One critical condition for a point process to be a PPP is that the number of points falling in a bounded Borel set $\mathcal{A}$ is a Poisson random variable with the parameter of $\lambda |\mathcal{A}|$, which is given by \cite{haenggi05}
\begin{align}\label{eq_p_n_A}
\mathbb{P}(\mathcal{A}) = \frac{\left( \lambda |\mathcal{A}| \right)^{n_{cs}} e^{-\lambda |\mathcal{A}|}}{n_{cs}!}
\end{align}
where $|\mathcal{A}|$ denotes the area of an arbitrary two-dimensional space $\mathcal{A}$. Recall from Figures \ref{fig_geometry_sync} and \ref{fig_geometry_hn} that $\mathsf{A}_{\text{col}}$ forms a circular space in which a point $x$ is located at the origin of the center and another point $y$ is placed $r$ away from the origin. That is, $|\mathsf{A}_{\text{col}}| = \pi r_{cs}^2$. Based on this, we can exploit the CDF of the distance $\mathsf{l}$ between the two arbitrary points for calculation of $1 - \mathbb{P} \left( \text{No other node in } \mathsf{A}_{\text{col}} \right)$, which is written as
\begin{align}\label{eq_first}
\mathbb{P} \left( n_{cs} > 0 \right) &\stackrel{(a)}{=} \mathbb{P}(n > 0, {\rm{~}} \mathsf{A}_{\text{col}})\nonumber\\
&{\rm{~}}= 1 - \mathbb{P}(n_{cs} = 0, \mathsf{A}_{\text{col}})\nonumber\\
&{\rm{~}}= 1 - \frac{\left( \lambda \pi r_{cs}^2 \right)^0 e^{-\lambda \pi r_{cs}^2}}{0!}\nonumber\\
&{\rm{~}}= 1 - e^{-\lambda \pi r_{cs}^2}, {\rm{~~}} r_{cs} \ge 0
\end{align}
where $\mathsf{l}(\cdot,\cdot)$ denotes the distance between two arbitrary points placed in a two-dimensional space that can be expressed as a bounded Borel set. For (a), we assume the existence of two nodes at least: one for $\mathsf{vT}$ and the other as a potential SYNC-causing node.

Next, the second term of (\ref{eq_Psync_proof}) can be modeled as
\begin{align}\label{eq_second}
&\mathbb{E}_{n_{cs}} \left[ \mathbb{P} \left( \text{At least one of the } n_{cs} \text{ other nodes transmit} \right) \right]\nonumber\\
&= 1 - \mathbb{E}_{n_{cs}} \left[ \left(1 - \mathsf{P}_{\text{start}} \right)^{n_{cs}} \right]\nonumber\\
&= 1 - \displaystyle \sum_{n_{cs} \in \lambda \mathsf{A_{cs}}} \left(1 - \mathsf{P}_{\text{start}} \right)^{n} \mathbb{P}\left[n_{cs} = n\right]
\end{align}
where $\mathsf{P}_{\text{start}}$ has been defined in Lemma \ref{lemma_Pstart}, and $n_{cs}>1$ as already assumed in (\ref{eq_first}).

Similarly, the third term of (\ref{eq_Psync_proof}) can be found as $\mathsf{P}_{\text{start}}$. It expresses that a SYNC never occurs until $\mathsf{vT}$, the vehicle of interest, actually transmits.

As a result, combining the three terms, the probability of a SYNC can be found as
\begin{align}\label{eq_Psync_proof_final}
\mathsf{P}_{\text{sync}}\left(\lambda, \text{CW}\right) = \Big( 1 - e^{-\lambda \pi r_{cs}^2} \Big) \Big( 1 - \displaystyle \sum_{n_{cs} \in \lambda \mathsf{A_{cs}}} \left(1 - \mathsf{P}_{\text{start}} \right)^{n} \mathbb{P}\left[n_{cs} = n\right] \Big) \tau
\end{align}
where $r_{cs} \ge 0$.

\begin{figure}
\vspace{-0.2 in}
\centering
\begin{subfigure}[b]{0.4\linewidth}
\includegraphics[width = \linewidth]{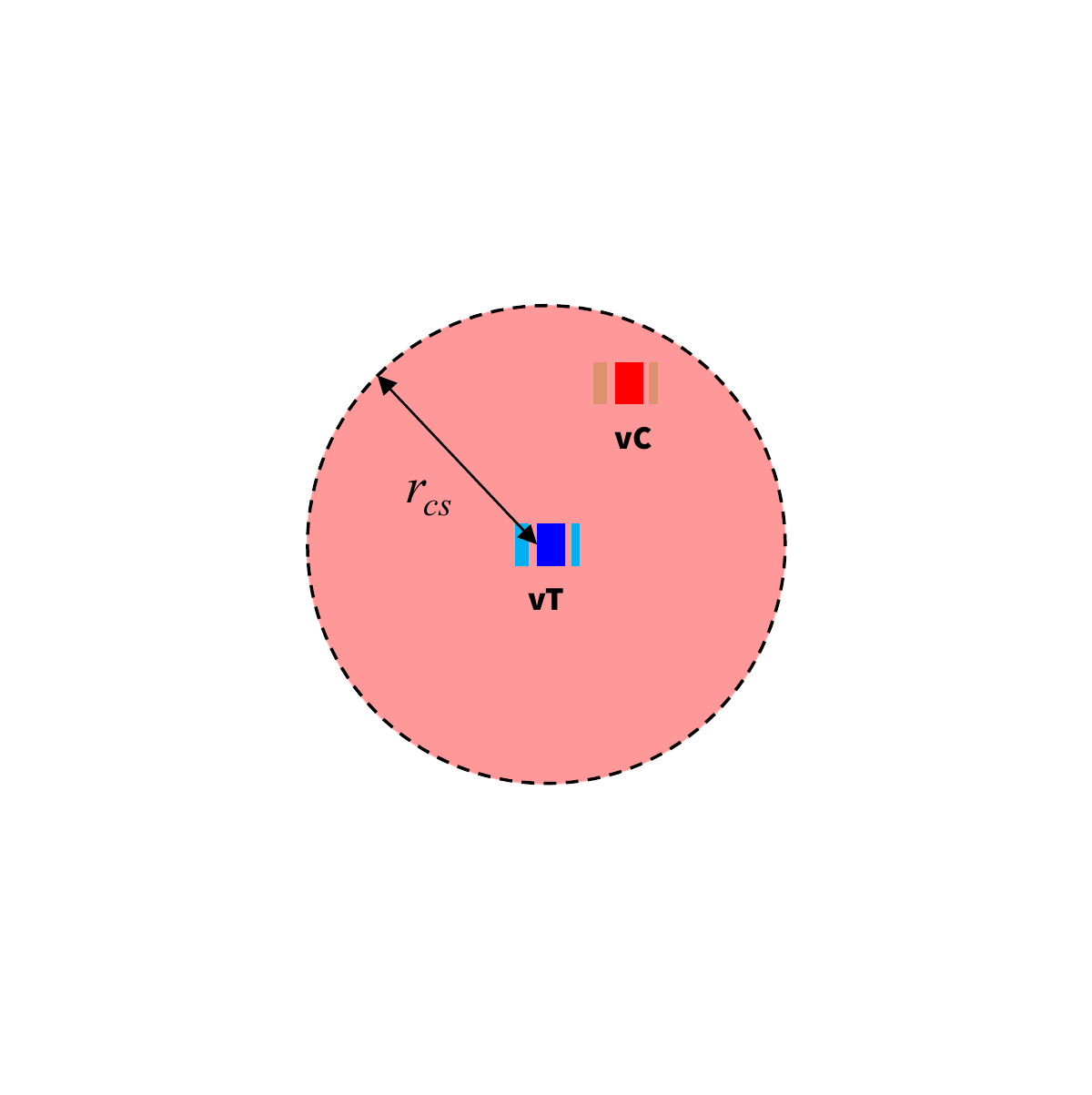}
\caption{SYNC}
\label{fig_geometry_Acol_sync}
\end{subfigure}
\begin{subfigure}[b]{0.4\linewidth}
\includegraphics[width = \linewidth]{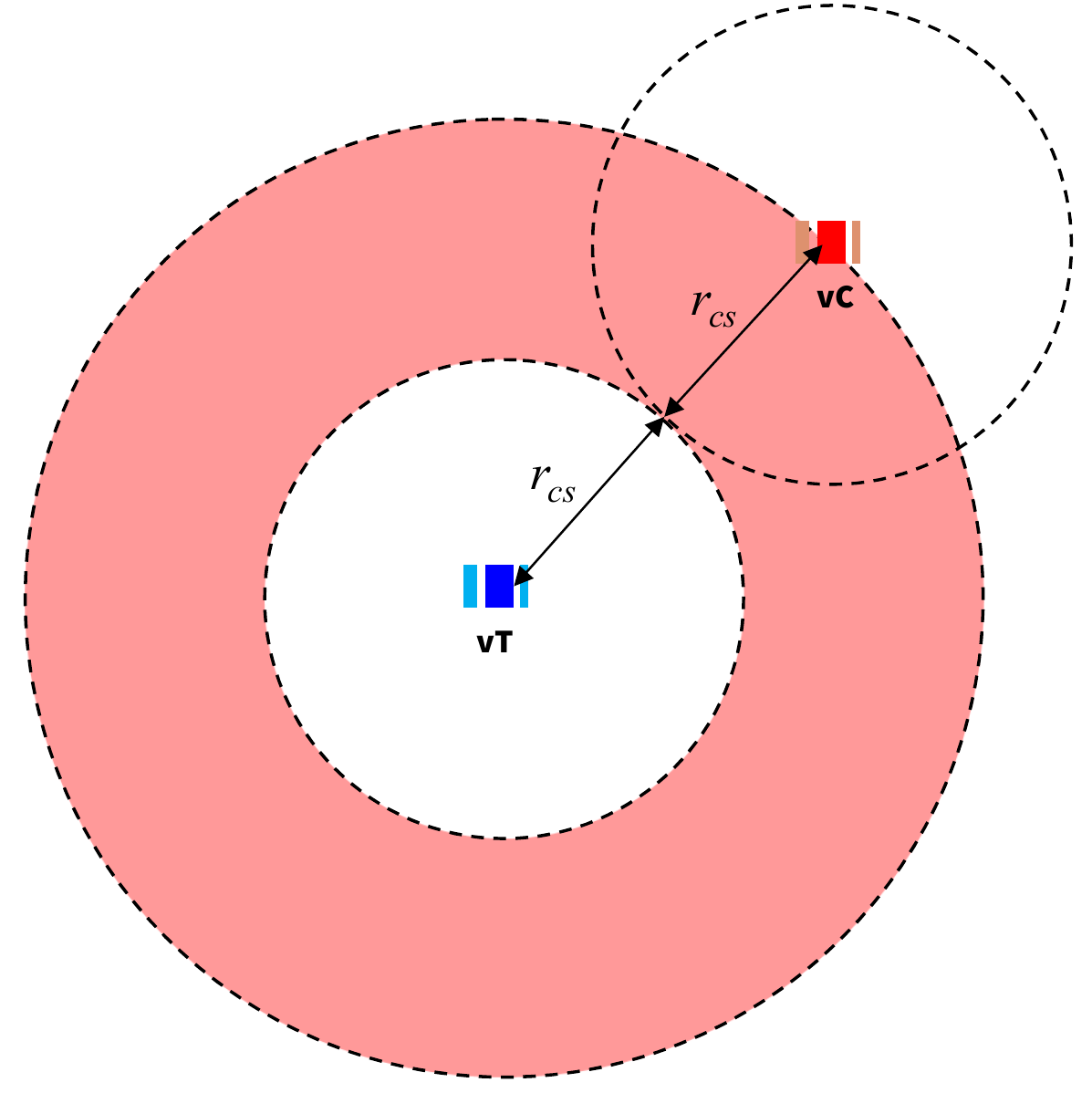}
\caption{HN}
\label{fig_geometry_Acol_hn}
\end{subfigure}
\caption{Area causing a collision--\textit{i.e.}, SYNC and HN}
\label{fig_geometry_Acol}
\vspace{-0.2 in}
\end{figure}

\subsection{Proof of Lemma \ref{lemma_Phn}}\label{appendix_Phn}
Similarly to (\ref{eq_Psync}), the probability of a HN can be modeled as
\begin{align}\label{eq_Phn_proof}
\mathsf{P}_{\text{hn}}\left(\lambda, \text{CW}\right) &= \mathbb{P} \left( n_{cs} > 0 \right)\nonumber\\
&{\rm{~~~}} \cdot \mathbb{E}_{n_{cs}} \left[ \mathbb{P} \left(\text{At least one HN interruption during a } \mathsf{vT} \text{`s BSM} \right) \right]\nonumber\\
&{\rm{~~~}} \cdot \mathbb{P} \left( \mathsf{vT} \text{ transmits in any of the } L_{bcn} \text{ slots} \right).
\end{align}

The first term can be obtained in a similar manner with (\ref{eq_p_n_A}) but with a different area of collision, which is formally written as
\begin{align}
\mathbb{P}\left(n_{cs} > 0\right) &{\rm{~}}= 1 - e^{-\lambda \|\mathsf{A}_{\text{col}}\|}\nonumber\\
&\stackrel{(a)}{=} 1 - e^{-3\lambda \pi r_{cs}^2}.
\end{align}
Note that $\mathsf{A}_{\text{col}}$ in (a) follows from Figure \ref{fig_geometry_hn} that $\|\mathsf{A}_{\text{col}}\| = \pi\left(4r_{cs}^2 - r_{cs}^2\right) = 3 \pi r_{cs}^2$.

\begin{figure}[t]
\vspace{-0.3 in}
\centering
\includegraphics[width = 0.7\linewidth]{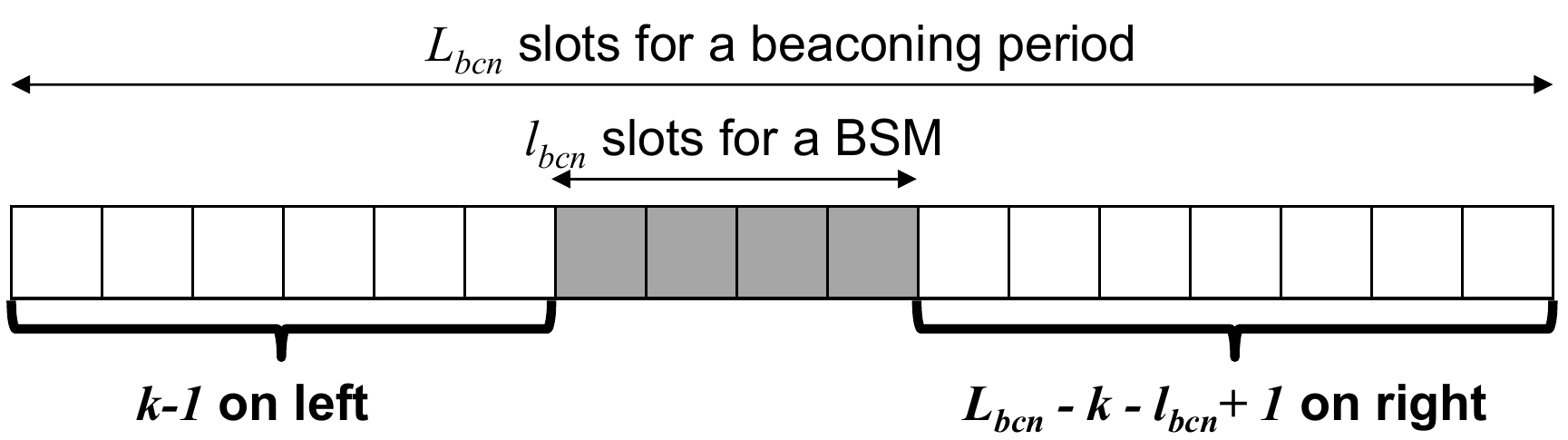}
\caption{Slots distributed among $\mathsf{vT}$ and other competing nodes}
\label{fig_Phn_proof}
\vspace{-0.3 in}
\end{figure}

For a certain value of the number of nodes causing a HN, the second term of (\ref{eq_Phn_proof}) can be derived as
\begin{align}
&\mathbb{E}_{n_{cs}} \left[ \mathbb{P} \left(\text{At least one HN interruption during a } \mathsf{vT} \text{`s BSM} \right) \right]\nonumber\\
&{\rm{~~~~~~~~~~~~~~~~~~~~~~~~~~~~~~~~}}= 1 - \mathbb{E}_{n_{cs}} \left[ \mathbb{P} \left( \text{No interruption during a BSM by } \mathsf{vT} \right) \right]
\end{align}
where
\begin{align}\label{eq_Phn_proof_second}
&\mathbb{E}_{n_{cs}} \left[ \mathbb{P} \left( \text{No interruption during a BSM by } \mathsf{vT} \right) \right]\nonumber\\
&=\mathbb{E}_{n_{cs}} \bigg[ \mathbb{P} \left( \text{``Contiguous'' } l_{bcn} \text{ slots taken by } \mathsf{vT} \right)\nonumber\\
&{\rm{~~~~~~~~~~~~~~}}\cdot \mathbb{P} \big(\text{No overlap with the } n_{cs} \text{ nodes } \big| \text{ ``Contiguous'' } l_{bcn} \text{ slots taken by } \mathsf{vT} \big) \bigg]\nonumber\\
&= \displaystyle \sum_{n_{cs}=0}^{\mathbb{N}\left[\Phi_{cs}\right]} \frac{{L_{bcn} - l_{bcn} + 1 \choose 1} \cdot \left( L_{bcn} - l_{bcn} \right)^{n}}{\left(L_{bcn}\right)^{n+1}} \mathbb{P}\left[n_{cs} = n\right]\nonumber\\
&\stackrel{(a)}{=} \left( L_{bcn} - l_{bcn} + 1\right) \displaystyle \sum_{n_{cs}=0}^{\mathbb{N}\left[\Phi_{cs}\right]} \frac{\left( L_{bcn} - l_{bcn} \right)^{n}}{\left(L_{bcn}\right)^{n+1}} \mathbb{P}\left[n_{cs} = n\right].
\end{align}
Notice that (a) follows from the is derived by
\begin{align}
&\mathbb{N}\left[ \text{\# slots taken by } \mathsf{vT} \right]\mathbb{N}\left[ \text{\# slots taken by other nodes} \right]\nonumber\\
&= \displaystyle \sum_{k = 1}^{L_{bcn} - l_{bcn} + 1} {L_{bcn} - l_{bcn} + 1 \choose 1} \left( \left( k - 1 \right) + \left( L_{bcn} - k - l_{bcn} + 1 \right) \right)^{n_{cs}}\nonumber\\
&= \displaystyle \sum_{k = 1}^{L_{bcn} - l_{bcn} + 1} {L_{bcn} - l_{bcn} + 1 \choose 1} \left( L_{bcn} - l_{bcn} \right)^{n_{cs}}.
\end{align}
Also, notice that the total number of scenarios can be calculated as $\left(L_{bcn}\right)^{n_{cs}+1}$ with the total of $L_{bcn}$ slots and $n_{cs}+1$ nodes competing for a slot, which forms the denominator of (\ref{eq_Phn_proof_second}).

As a result, (\ref{eq_Phn_proof}) can be rewritten as
\begin{align}
&\mathsf{P}_{\text{hn}}\left(\lambda, \text{CW}\right)\nonumber\\
&=\left( 1 - e^{-3\lambda \pi r_{cs}^2} \right) \left( 1 - \left( L_{bcn} - l_{bcn} + 1\right) \displaystyle \sum_{n_{cs}=0}^{\mathbb{N}\left[\Phi_{cs}\right]} \frac{\left( L_{bcn} - l_{bcn} \right)^{n}}{\left(L_{bcn}\right)^{n+1}} \mathbb{P}\left[n_{cs} = n\right] \right) \mathsf{P}_{\text{start}},
\end{align}
which completes the proof.

\subsection{Proof of Lemma \ref{lemma_f_L}}\label{appendix_f_L}
Assume that the transmitting vehicle, $\mathsf{vT}$, is located at the origin of a quadrant and the vehicle transmitting a colliding packet, $\mathsf{vC}$, is located at an arbitrary point $(x,y)$. Since we are deriving the area in which two packets from $\mathsf{vT}$ and $\mathsf{vC}$ collide, the calculation proceeds with respect to a node's transmission range, $r_{\text{tx}}$. Referring to Figure \ref{fig_geometry_hn}, the range for $\mathsf{l}$ can be found as $\left[0, 2r_{\text{tx}}\right]$--\textit{i.e.}, $\left[0, r_{\text{tx}}\right]$ causing a SYNC and $\left[r_{\text{tx}}, 2r_{\text{tx}}\right]$ causing a HN. Any value of $\mathsf{l}$ greater than $2r_{\text{tx}}$ does not affect reception of a packet since the transmission ranges of two nodes causing a packet collision do not overlap.

Looking at the problem from the spatial point of view, it is clear that neither the PDF nor CDF cannot but be defined piecewise, as illustrated in Figure \ref{fig_euclidean}. In other words, it is straightforward that one can derive the PDF from the area of the segment of the circular annulus between $\mathsf{l}$ and $\mathsf{l}+\text{d}\mathsf{l}$ intercepted by the square, divided by the area of the whole square, $D^2$, \textit{i.e.},
\begin{align}\label{eq_l_pdf_proof}
f_{\mathsf{L}} \left(\mathsf{l}\right) &= \begin{cases}\displaystyle \frac{\pi \mathsf{l}}{2D^2}, {\rm{~~}} 0 \le \mathsf{l} < D\\
\displaystyle \frac{\mathsf{l}}{D^2} \left( \frac{\pi}{2} - 2\arccos \left(\frac{D}{\mathsf{l}}\right) \right), {\rm{~~}} D \le \mathsf{l} \le \sqrt{2}D.\end{cases}
\end{align}
Integration of (\ref{eq_l_pdf_proof}) with respect to $\mathsf{l}$ yields the CDF as
\begin{align}
F_{\mathsf{L}} \left(\mathsf{l}\right) = \begin{cases}\displaystyle \frac{\pi \mathsf{l}^2}{4D^2}, {\rm{~~}} 0 \le \mathsf{l} < D\\
\displaystyle \frac{\pi \mathsf{l}^2}{4D^2} - \displaystyle \frac{1}{D^2} \left[ \mathsf{l}^2 \cos^{-1} \left(\frac{D}{\mathsf{l}}\right) - D\mathsf{l} \sqrt{1 - \left(\frac{D}{\mathsf{l}}\right)^2} \right], {\rm{~~}} D \le \mathsf{l} \le \sqrt{2}D,\end{cases}
\end{align}
which complete the proof.

\begin{figure}[t]
\vspace{-0.3 in}
\centering
\includegraphics[width = 0.3\linewidth]{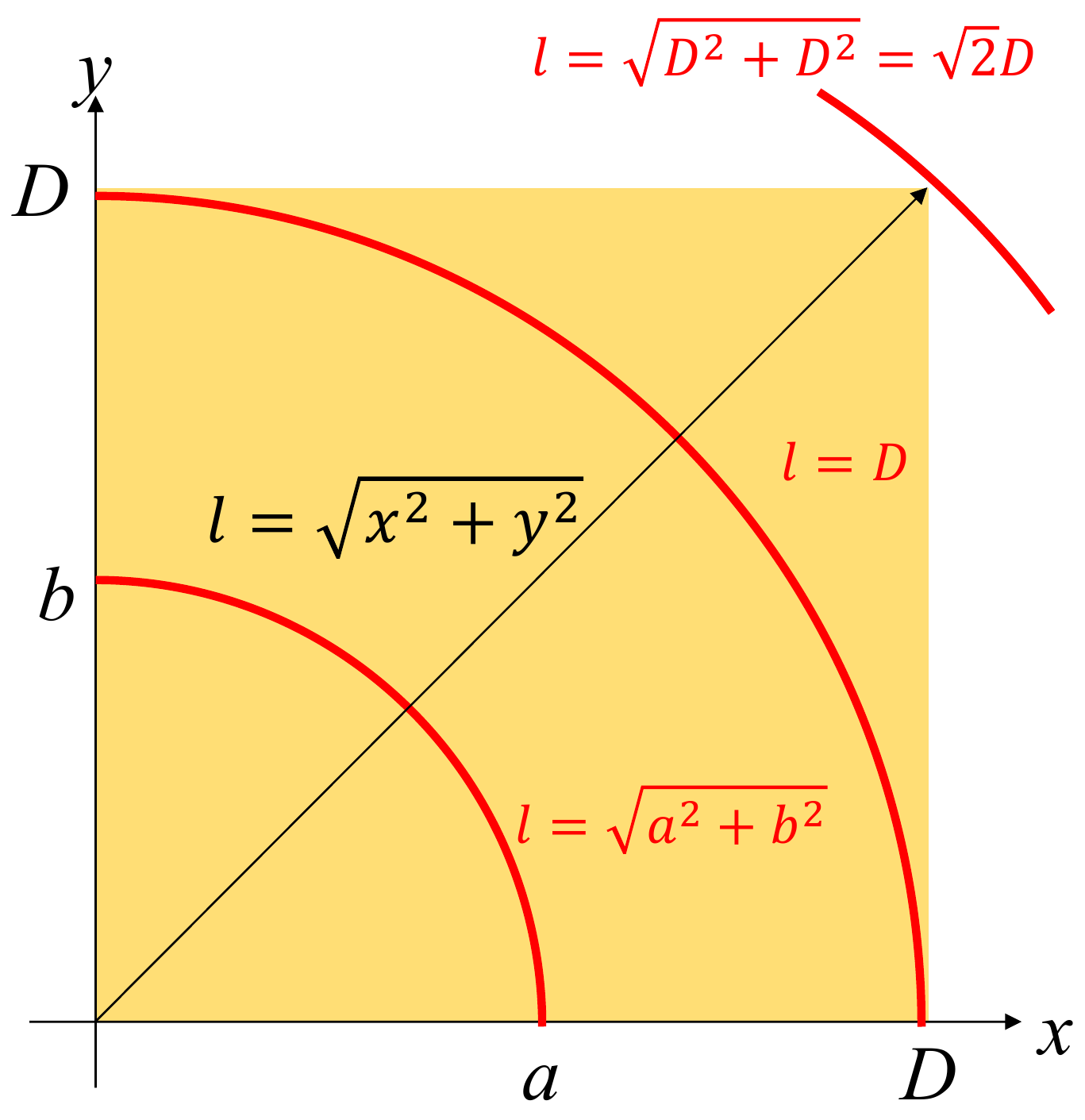}
\caption{Finding the distribution of the length between $\mathsf{vC}$ and $\mathsf{vT}$ (denoted by $\mathsf{l}$)}
\label{fig_euclidean}
\vspace{-0.3 in}
\end{figure}

\subsection{Proof of Lemma \ref{lemma_f_A}}\label{appendix_f_A}
We remind that the fitting table for $\mathsf{l}$ is written as an inverse function of $\mathsf{a}$, which is given by
\begin{align}\label{eq_fit_quadratic}
\mathsf{l} &= g^{-1}\left( \mathsf{a} \right)\nonumber\\
&= p_{1}\mathsf{a}^2 + p_{2}\mathsf{a} + p_{3}.
\end{align}
Also, the coefficients, $p_{1}$, $p_{2}$, and $p_{3}$, have been presented in Table \ref{table_g_coefficients}.

The resulting PDF is formulated as
\begin{align}\label{eq_f_A_proof}
f_{\mathsf{A}_{\text{col}}}(\mathsf{a}) &= f_{\mathsf{L}}\left(\mathsf{l} = g^{-1}\left( \mathsf{a} \right) \right) \left|\frac{\partial}{\partial \mathsf{a}}g^{-1}\left( \mathsf{a} \right)\right|\nonumber\\
&= f_{\mathsf{L}} \left( p_{1}\mathsf{a}^2 + p_{2}\mathsf{a} + p_{3} \right) \left( 2p_{1}\mathsf{a} + p_{2} \right).
\end{align}
Since (\ref{eq_f_A_proof}) is identical to (\ref{eq_f_A}), it completes the proof of the PDF.

Therefore, the CDF is given by
\begin{align}\label{eq_F_A_proof}
F_{\mathsf{A}_{\text{col}}} \left(\mathsf{a}\right) &= \mathbb{P}\Big( \mathsf{A} = g\left(l\right) \le \mathsf{a}\Big)\nonumber\\
&= \mathbb{P}\Big( \mathsf{l} \le g^{-1}\left(\mathsf{a}\right) \Big)
\end{align}
According to (\ref{eq_f_L}), the range of $\mathsf{l} = g^{-1}\left(\mathsf{a}\right)$ should be divided into two separate ranges:
\begin{itemize}
\item When $0 \le g^{-1}\left(\mathsf{a}\right) \le r$,
\end{itemize}
\begin{align}\label{eq_F_A_proof_case_r}
\mathbb{P}\Big( \mathsf{l} \le g^{-1}\left(\mathsf{a}\right) \Big) &= \displaystyle \int_{0}^{g^{-1}\left(\mathsf{a}\right)} f_{\mathsf{L}}(\mathsf{l}) \text{d}\mathsf{l}\nonumber\\
&= \displaystyle \frac{1}{r^2} \int_{0}^{g^{-1}\left(\mathsf{a}\right)} \frac{\pi}{2}\mathsf{l} \text{d}\mathsf{l}\nonumber\\
&= \frac{\pi}{4r^2} \left(g^{-1}\left(\mathsf{a}\right)\right)^2.
\end{align}
\begin{itemize}
\item When $0 \le g^{-1}\left(\mathsf{a}\right) \le r$,
\end{itemize}
\begin{align}\label{eq_F_A_proof_case_sqrt_2r}
&\mathbb{P}\Big( \mathsf{l} \le g^{-1}\left(\mathsf{a}\right) \Big)\nonumber\\
&= \displaystyle \int_{0}^{r} f_{\mathsf{L}}(\mathsf{l}) \text{d}\mathsf{l} + \displaystyle \int_{r}^{g^{-1}\left(\mathsf{a}\right)} f_{\mathsf{L}}(\mathsf{l}) \text{d}\mathsf{l}\nonumber\\
&= \displaystyle \frac{1}{r^2}\int_{0}^{r} \frac{\pi}{2}\mathsf{l} \text{d}\mathsf{l} + \displaystyle \frac{1}{r^2} \int_{r}^{g^{-1}\left(\mathsf{a}\right)} \left( \frac{\pi}{2} - 2\cos^{-1}\left(\frac{r}{\mathsf{l}}\right) \right) \text{d}\mathsf{l}\nonumber\\
&= \frac{\pi}{4} + \frac{\pi}{4r^2} \bigg( \left(g^{-1}\left(\mathsf{a}\right)\right)^2 - r^2 \bigg) - \frac{1}{r^2} \bigg[\mathsf{l}^2\cos^{-1}\left(\frac{r}{\mathsf{l}}\right) - r \mathsf{l} \sqrt{1 - \frac{r^2}{\mathsf{l}^2}}\bigg]_{r}^{g^{-1}\left(\mathsf{a}\right)}\nonumber\\
&= \frac{\pi}{4r^2}\left(g^{-1}\left(\mathsf{a}\right)\right)^2 - \frac{1}{r^2}\left(g^{-1}\left(\mathsf{a}\right)\right)^2\cos^{-1}\left(\frac{r}{g^{-1}\left(\mathsf{a}\right)}\right) + \frac{1}{r} g^{-1}\left(\mathsf{a}\right) \sqrt{1 - \frac{r^2}{\left(g^{-1}\left(\mathsf{a}\right)\right)^2}}.
\end{align}
As a result, the CDF can be formally idenfied as
\begin{eqnarray}\label{eq_F_A_proof_final}
\left.\begin{aligned}
F_{\mathsf{A}_{\text{col}}} \left(\mathsf{a}\right) &= {\begin{cases} \vspace{0.1 in} \displaystyle \frac{\pi}{4r^2} \left(g^{-1}\left(\mathsf{a}\right)\right)^2, {\rm{~~~~}} 0 \le g^{-1}\left(\mathsf{a}\right) \le r\\
\displaystyle \frac{\pi}{4r^2}\left(g^{-1}\left(\mathsf{a}\right)\right)^2 -\frac{1}{r^2}\left(g^{-1}\left(\mathsf{a}\right)\right)^2\cos^{-1} \left(\frac{r}{g^{-1}\left(\mathsf{a}\right)}\right)\nonumber\\
\displaystyle {\rm{~~~~~~~~~~~~~~~~~~}} + \frac{1}{r} g^{-1}\left(\mathsf{a}\right) \sqrt{1 - \frac{r^2}{\left(g^{-1}\left(\mathsf{a}\right)\right)^2}}, {\rm{~~~~}} r \le g^{-1}\left(\mathsf{a}\right) \le \sqrt{2}r
\end{cases}}
\end{aligned}\right.
\end{eqnarray}
which completes the proof.



\begin{thebibliography}{99}\setlength{\parskip}{0.000001 em}
\bibitem{fcc_dsrc} Federal Communications Commission (FCC), \textit{Dedicated short range communications (DSRC) service}, Apr. 2019. [Online]. Available: \url{https://www.fcc.gov/wireless/bureau-divisions/mobility-division/dedicated-short-range-communications-dsrc-service}

\bibitem{fcc1668a1} Federal Communications Commission (FCC), \textit{The commission seeks to update and refresh the record in the ``unlicensed national information infrastructure (U-NII) devices in the 5 GHz band'' Proceeding}, FCC 16-68A1.

\bibitem{5gaa} 5GAA, ``The case for cellular V2X for safety and cooperative driving,'' Nov. 2016. [Online]. Available: \url{http://5gaa.org/wp-content/uploads/2017/10/5GAA-whitepaper-23-Nov-2016.pdf}


\bibitem{europe} J. Van Roy, ``EU parliament finally votes for wifi to connect cars,'' \textit{New Mobility News}, Apr. 2019. [Online]. Available: \url{https://newmobility.news/2019/04/18/eu-parliament-finally-votes-for-wifi-to-connect-cars/}

\bibitem{dot} American Association of State Highway and Transportation Officials (AASHTO), ``State DOTs sign letter supporting preservation of 5.9 GHz spectrum,'' \textit{AASHTO J.}, Aug. 2019. [Online]. Available: \url{https://aashtojournal.org/2019/08/23/state-dots-sign-letter-supporting-preservation-of-5-9-ghz-spectrum/}.

\bibitem{fcc_phase1} FCC, ``Phase I testing of prototype U-NII-4 devices,'' TR 17-1006, Oct. 2018.

\bibitem{aashto} AASHTO, \textit{Re: Docket No. DOT-OST-2018-0210}, Feb. 2019. [Online]. Available: \url{https://policy.transportation.org/wp-content/uploads/sites/59/2019/02/AASHTO-Comments-USDOT-V2X-Communication-RFC-FINAL.pdf}

\bibitem{bennis_lett18} C-F. Liu and M. Bennis, ``Ultra-reliable and low-latency vehicular transmission: an extreme value theory approach,'' \textit{IEEE Commun. Lett.}, vol. 22, iss. 6, Jun. 2018.


\bibitem{infocom_6} X. Ma and X. Chen, ``Delay and broadcast reception rates of highway safety applications in vehicular ad hoc networks,'' in \textit{Proc. IEEE Mobile Netw. Veh. Environ. 2007}.

\bibitem{infocom_7} X. Ma, X. Chen, and H. H. Refai, ``Performance and reliability of DSRC vehicular safety communication: a formal analysis,'' \textit{EURASIP J. Wireless Commun. Netw.}, 2009.

\bibitem{infocom_8} X. Ma, J. Zhang, and T. Wu, ``Reliability analysis of one-hop safety critical broadcast services in VANETs,'' \textit{IEEE Trans. Veh. Technol.}, vol. 60, no. 8, 2011.

\bibitem{infocom_9} X. Yin, X. Ma, K. S. Trivedi, and A. Vinel, ``Performance and reliability evaluation of BSM broadcasting in DSRC with multi-channel schemes,'' \textit{IEEE Trans. Comput.}, vol. 63, no. 12, 2014.

\bibitem{infocom_10} C. Campolo, A. Vinel, A. Molinaro, and Y. Koucheryavy, ``Modeling broadcasting in IEEE 802.11p/WAVE vehicular networks,'' \textit{IEEE Commun. Lett.}, vol. 15, no. 2, 2011.

\bibitem{bennis_lett18} C-F. Liu and M. Bennis, ``Ultra-reliable and low-latency vehicular transmission: an extreme value theory approach,'' \textit{IEEE Commun. Lett.}, vol. 22, iss. 6, Jun. 2018.

\bibitem{elsevier14} R. Stanica, E. Chaput, and A.-L. Beylot, ``Reverse back-off mechanism for safety vehicular ad hoc networks,'' \textit{Elsevier Ad Hoc Netw.}, vol. 16, 2014.

\bibitem{vnc10} M. I. Hassan, H. L. Vu, T. Sakurai, L. L. Andrew, and M. Zukerman, ``Effect of retransmission on the performance of the IEEE 802.11 MAC protocol for DSRC,'' in \textit{Proc. IEEE Veh. Netw. Conf. 2010}.

\bibitem{eurasip19} X. Lei and S. H. Rhee, ``Performance analysis and enhancement of IEEE 802.11p beaconing,'' \textit{EURASIP J. Wireless Commun. Netw.}, vol. 61, 2019.

\bibitem{irt_vanet06} T. ElBatt, S. K. Goel, G. Holland, H. Krishnan, and J. Parikh, ``Cooperative collision warning using dedicated short range wireless communications,'' in \textit{Proc. ACM VANET 2006}.

\bibitem{irt_elsevier16} M. Renda, G. Resta, P. Santi, F. Martelli, and A. Franchini, ``IEEE 802.11p VANets: experimental evaluation of packet inter-reception time,'' \textit{Elsevier Comput. Commn.}, vol. 75, 2016.

\bibitem{infocom_12} J. Lansford, J. B. Kenney, and P. Ecclesine, ``Coexistence of unlicensed devices with DSRC systems in the 5.9 GHz band,'' in \textit{Proc. IEEE Veh. Netw. Conf. 2013}.

\bibitem{infocom_13} K.-H. Chang, ``Wireless communications for vehicular safety,'' \textit{IEEE Wireless Commun.}, vol. 22, no. 1, 2015.

\bibitem{infocom_14} National Telecommunications and Information Administration (NTIA), \textit{Evaluation of the 5350-5470 MHz and 5850-5925 MHz bands}, Jan. 2013.

\bibitem{mag14} Y. Park and H. Kim, ``On the coexistence of IEEE 802.11ac and WAVE in the 5.9 GHz band,'' \textit{IEEE Commun. Mag.}, vol. 52, no. 6, 2014.

\bibitem{gaurang17} G. Naik, J. Liu, and J. Park ``Coexistence of dedicated short range communication (DSRC) and Wi-Fi: implications to Wi-Fi performance,'' in \textit{Proc. IEEE INFOCOM 2017}.

\bibitem{daley} D. Daley and D. Vere-Jones, \textit{An Introduction to the Theory of Point Processes: Volume I: Elementary Theory and Methods}, Springer Probability and its Applications, Second edition, 2003.

\bibitem{bennis_jsac17} C. Perfecto, J. Del Ser, and M. Bennis, ``Millimeter-wave V2V communications: distributed association and beam alignment,'' \textit{IEEE J. Sel. Areas Commun.}, vol. 35, iss. 9, Jun. 2017.


\bibitem{distance_velocity} P. Lutus, Website of \textit{The Physics Behind Stopping a Car}, [Online]. Available: \url{https://arachnoid.com/braking_physics/index.html}


\bibitem{tcom13} X. Yin, X. Ma, and K. S. Trivedi, ``An interacting stochastic models approach for the performance evaluation of DSRC vehicular safety communication,'' \textit{IEEE Trans. Comput.}, vol. 62, no. 5, May 2013.

\bibitem{haenggi16} Z. Tong, H. Lu, M. Haenggi, and C. Poellabauer, ``A stochastic geometry approach to the modeling of DSRC for vehicular safety communication,'' \textit{IEEE Trans. Intell. Transp. Syst.}, vol. 17, iss. 5, May 2016.




\bibitem{bian18} P. Wang, H. Zhang, K. Bian, and L. song, ``Cellular V2X communications in unlicensed spectrum: harmonious coexistence with VANET in 5G systems,'' \textit{IEEE Trans. Wireless Commun.}, vol. 17, no. 8, Aug. 2018.

\bibitem{hanbat16} S.-S. Raymond, A. Abubakari, and H.-S. Jo, ``Coexistence of power-controlled cellular networks with rotating radar,'' \textit{IEEE J. Sel. Areas Commun.}, vol. PP., Iss. 99, 2016.

\bibitem{milcom19} S. Kim and T. Dessalgn, ``Mitigation of civilian-to-military interference in DSRC for urban operations,'' in \textit{Proc. IEEE MILCOM 2019}.

\bibitem{elsevier14_30} C. Campolo, A. Molinaro, A. Vinel, and Y. Zhang, ``Modeling prioritized broadcasting in multichannel vehicular networks,'' \textit{IEEE Trans. Veh. Technol.}, vol. 61, No. 2, 2012.

\bibitem{elsevier14_31} M. Di Felice, L. Bedogni, and L. Bononi, ``DySCO: a dynamic spectrum and contention control framework for enhanced broadcast communication in vehicular networks,'' in \textit{Proc. ACM MobiWac 2012}.

\bibitem{bennis_tcom19} S. Samarakoon, M. Bennis, W. Saad, and M. Debbah, ``Distributed federated learning for ultra-reliable low-latency vehicular communications,'' \textit{IEEE Trans. Commun.}, Early Access, Nov. 2019.

\bibitem{bennis_proc19} J. Park, S. Samarakoon, M. Bennis, and M. Debbah, ``Wireless network intelligence at the edge,'' \textit{Proc. IEEE}, vol. 107, iss. 11, Nov. 2019.

\bibitem{bennis_tvt19} X. Chen, C. Wu, M. Bennis, Z. Zhao and Z. Han, ``Learning to entangle radio resources in vehicular communications: an oblivious game-theoretic perspective,'' \textit{IEEE Trans. Veh. Technol.}, vol. 68, no. 5, May 2019.


\bibitem{access19} S. Kim, ``Impacts of mobility on performance of blockchain in VANET,'' \textit{IEEE Access}, vol. 7, May 2019


\bibitem{ieee80211p} IEEE 802.11p, \textit{Part 11: Wireless LAN Medium Access Control (MAC) and Physical Layer (PHY) specifications: Amendment 6: Wireless Access in Vehicular Environments}, IEEE Std., Jun. 2010.



\bibitem{globecom18} S. Kim and C. Dietrich, ``A novel method for evaluation of coexistence between DSRC and Wi-Fi at 5.9 GHz,'' in \textit{Proc. IEEE Globecom 2018}.

\bibitem{integral} \textit{Table of integrals}, [Online]. Available: \url{http://integral-table.com/}.

\bibitem{root} \textit{Linear approximation of the square root}, [Online]. Available: \url{https://socratic.org/questions/how-do-you-use-linear-approximation-to-the-square-root-function-to-estimate-squa-1}.


\bibitem{bianchi} G. Bianchi, ``IEEE 802.11--saturation throughput analysis,'' \textit{IEEE Commun. Lett.}, vol. 12, no. 2, Dec. 1998.


\bibitem{pai19} \textit{Remarks of FCC chairman Ajit Pai at the Wi-Fi world congress 2019}, May 2019. [Online]. Available: \url{https://www.fcc.gov/document/chairman-pai-remarks-wi-fi-world-congress-2019}

\bibitem{tr36300} 3GPP, ``Evolved universal terrestrial radio access (E-UTRA) and evolved universal terrestrial radio access network (E-UTRAN); overall description; stage 2 (v14.3.0, release 14),'' 3GPP, Tech. Rep. 36.300, Jun. 2017.

\bibitem{wifi6ghz} FCC, ``Promoting unlicensed use of the 6 GHz band,'' Oct. 2018. [Online]. Available: \url{https://www.fcc.gov/document/promoting-unlicensed-use-6-ghz-band}

\bibitem{icc05} J. W. Tantra, C. H. Foh, and A. B. Mnaouer, ``Throughput and delay analysis of the IEEE 802.11e EDCA saturation,'' in \textit{Proc. IEEE ICC 2015}.

\bibitem{ieee1609_4} IEEE 1609.4-2016, \textit{IEEE Standard for Wireless Access in Vehicular Environments (WAVE) -- Multi-Channel Operation}, IEEE Std., Jan. 2016.

\bibitem{etri18} Y. Kim and S. Park, ``Analytical calculation of spectrum requirements for LTE-A using the probability distribution on the scheduled resource blocks,'' \textit{IEEE Commun. Lett.}, vol. 22, no. 3, Mar. 2018.


\bibitem{haenggi05} M. Haenggi, ``On distances in uniformly random networks,'' \textit{IEEE Trans. Inf. Theory}, vol. 51, no. 10, Oct. 2005.


\bibitem{daesik17} K. Lee, J. Kim, Y. Park, H. Wang, and D. Hong, ``Latency of cellular-based V2X: perspectives on TTI-proportional latency and TTI-independent latency,'' \textit{IEEE Access}, vol. 5, Jul. 2017.

\bibitem{rcs17} B. Skorup, ``The department of transportation’s proposed vehicle-to-vehicle technology mandate is unprecedented and hasty,'' \textit{Public Interest Comment}, Apr. 2017. [Online]. Available at \url{https://www.mercatus.org/system/files/skorup-v2v-technologies-pic-v1.pdf}



\end{thebibliography}
\end{document}